\documentclass[
  journal=largetwo,
  manuscript=article-type,
  year=2024,
  volume=0,
]{cup-journal}

\usepackage{amsmath}
\usepackage[nopatch]{microtype}
\usepackage{booktabs}

\usepackage{multicol}
\usepackage{hyperref}
\usepackage{subcaption}
\usepackage{makecell}
\usepackage{graphicx}	
\usepackage{amssymb}

\def\aap{Astron.~Astrophys.}            
\def\apjl{Astrophys.~J.}                
\def\apjs{Astrophys.~J.~Suppl.~Ser.}    

\title{Revisiting the Mysterious Origin of FRB 20121102A with Machine-learning Classification}

\author{Leah Ya-Ling Lin}
\affiliation{Department of Physics, National Tsing Hua University, 101, Section 2. Kuang-Fu Road, Hsinchu, 30013, Taiwan}
\email[Leah Ya-Ling Lin]{stu109022104@gapp.nthu.edu.tw}

\author{Tetsuya Hashimoto}
\affiliation{Department of Physics, National Chung Hsing University, 145 Xingda Rd., South Dist., Taichung 40227, Taiwan}

\author{Tomotsugu Goto}
\affiliation{Department of Physics, National Tsing Hua University, 101, Section 2. Kuang-Fu Road, Hsinchu, 30013, Taiwan}\alsoaffiliation{Institute of Astronomy, National Tsing Hua University, 101, Section 2. Kuang-Fu Road, Hsinchu, 30013, Taiwan}

\author{Bjorn Jasper Raquel}
\affiliation{Department of Physics, National Chung Hsing University, 145 Xingda Rd., South Dist., Taichung 40227, Taiwan}\alsoaffiliation{Department of Earth and Space Sciences, Rizal Technological University, Boni Avenue, Mandaluyong, 1550 Metro Manila, Philippines}

\author{Simon C.-C. Ho}
\affiliation{Research School of Astronomy and Astrophysics, The Australian National University, Canberra, ACT 2611, Australia}\alsoaffiliation{Centre for Astrophysics and Supercomputing, Swinburne University of Technology, P.O. Box 218, Hawthorn, VIC 3122, Australia}\alsoaffiliation{OzGrav: The Australian Research Council Centre of Excellence for Gravitational Wave Discovery, Hawthorn, VIC 3122, Australia}\alsoaffiliation{ASTRO3D: ARC Centre of Excellence for All-sky Astrophysics in 3D, ACT 2611, Australia}

\author{Bo-Han Chen}
\affiliation{Graduate School of Data Science, Seoul National University, 1, Gwanak-ro, Gwanak-gu, Seoul 08826, Korea}

\author{Seong Jin Kim}
\affiliation{Department of Physics, National Tsing Hua University, 101, Section 2. Kuang-Fu Road, Hsinchu, 30013, Taiwan}\alsoaffiliation{Institute of Astronomy, National Tsing Hua University, 101, Section 2. Kuang-Fu Road, Hsinchu, 30013, Taiwan}

\author{Chih-Teng Ling}
\affiliation{Institute of Astronomy, National Tsing Hua University, 101, Section 2. Kuang-Fu Road, Hsinchu, 30013, Taiwan}


\begin{document}

\begin{abstract}
    Fast radio bursts (FRBs) are millisecond-duration radio waves from the Universe. Even though more than 50 physical models have been proposed, the origin and physical mechanism of FRB emissions are still unknown. The classification of FRBs is one of the primary approaches to understanding their mechanisms, but previous studies classified conventionally using only a few observational parameters, such as fluence and duration, which might be incomplete. To overcome this problem, we use an unsupervised machine-learning model, the Uniform Manifold Approximation and Projection (UMAP) to handle seven parameters simultaneously, including amplitude, linear temporal drift, time duration, central frequency, bandwidth, scaled energy, and fluence.
    We test the method for homogeneous 977 sub-bursts of FRB 20121102A detected by the Arecibo telescope. 
    Our machine-learning analysis identified five distinct clusters, suggesting the possible existence of multiple different physical mechanisms responsible for the observed FRBs from the FRB 20121102A source. 
    \textcolor{black}{The geometry of the emission region and the propagation effect of FRB signals could also make such distinct clusters. }
    This research will be a benchmark for future FRB classifications when dedicated radio telescopes such as the Square Kilometer Array (SKA) or Bustling Universe Radio Survey Telescope in Taiwan (BURSTT) discover more FRBs than before.
\end{abstract}

\section{Introduction}

Fast radio bursts (FRBs) are a type of highly energetic astrophysical transient that last only a few milliseconds \citep[e.g.,][]{lorimer2007}. 
Many FRBs have dispersion measures (DMs) that exceed the expected maximum of the Galactic electron density, indicating their extragalactic origins. 
DM represents the column density of free electrons traversed along the propagation path of an FRB.
Despite their discovery over a decade ago \citep{lorimer2007}, the origin of FRBs remains a mystery. Recently, the detection of repeating FRBs \citep[e.g.,][]{spitler2014,niu2022} has opened up new avenues of research into the origin of these phenomena.

With the emergence of a large number of FRBs samples in recent years, repeated FRBs (referred to as \lq repeating bursts\rq\ for simplicity) have also been noticed by astronomers, especially FRB 20121102A, which has been observed to have a very high burst rate \citep[e.g.,][]{Li2021,Jahns2018}. 
FRB 20121102A is the first-discovered repeating FRB source \citep{scholz2016}. 
This source was first recorded in 2012 and was detected again in the same spatial location in 2015 with the same dispersion measure \citep{scholz2016}. 
In subsequent observations, FRB 20121102A exhibited an extremely high repetition rate compared to other FRBs \citep[e.g.,][]{Li2021,Jahns2018} and became the first repeating burst to be localized \citep{Chatterjee2017}.

\textcolor{black}{Given the large sample size of recent FRB detections (e.g., \cite{Li2021}), machine-learning approaches have been becoming important. Applying deep learning to single-pulse classification was proposed in a pioneering paper by \citet{Connor2018}. They trained a deep neural network using single pulses and false-positive triggers from real telescopes to develop a framework for ranking events. The ranking was ordered by their probability of being astrophysical transients with high accuracy, recall, and quick computational time, indicating the power of deep learning.}

\textcolor{black}{Since then, unsupervised machine learning has been applied to the Canadian Hydrogen Intensity Mapping Experiment (CHIME) data \citep[e.g.,][]{chen2022,zhuge2023}. \cite{chen2022} and \cite{zhuge2023} found distinct physical properties (i.e., the ratio of the highest frequency to the peak frequency by \cite{chen2023}, brightness temperature and rest-frame frequency bandwidth by \cite{zhuge2023}) between repeaters and one-off events, which allows the machine to predict the repetitiveness of FRBs. Based on the unsupervised machine learning approaches, both studies identified some potentially repeating FRBs currently reported as one-off FRBs. A few active repeaters, including FRB 20201124A \citep{chen2023} and FRB 20121102A \citep{Raquel2023}, were also classified by unsupervised machine algorithm. Some distinct clusters were commonly identified for these active repeating FRB sources, suggesting multiple radiation mechanisms of active repeaters or distinct physical environments of emission regions. These approaches to the FRB classification used catalogs, including measured physical properties of individual FRBs. In addition to such catalog-based classifications, the UMAP algorithm was used for the image data (i.e., waterfall) of the CHIME FRBs \citep{yang2023}. They found that the UMAP algorithm using image data produced more accurate results in predicting the repetitiveness.}

In this paper, we revisit the repeating FRB 20121102A using Arecibo samples \citep{Jahns2018} with a machine-learning classification, being free of human bias, an approach to understanding its properties and origin. The Uniform Manifold Approximation and Projection (UMAP) \citep{McInnes2018,McInnesUMAP2018} is an algorithm that utilizes manifold learning techniques and incorporates concepts from topological data analysis to achieve dimension reduction. It offers a versatile framework for approaching manifold learning and dimension reduction, providing both a broad scope and specific practical implementations. 
This paper aims to explain the practical workings of the UMAP algorithm. 
UMAP is useful because it allows the two-dimensional projection of higher-dimensional data points, which can be handled easily. Previous studies demonstrate the effectiveness of UMAP and the practical usage of a follow-up science case. \cite{2022MNRAS.514.5987K} 
\cite{chen2022}

After the classification, we make a comparison between this work and the previous machine learning classification result using Five hundred meter Aperture Spherical Telescope (FAST) data \citep{Raquel2023} to mitigate a possible observational bias.
We note that the Arecibo samples include relatively brighter FRB populations \cite[$\gtrsim$0.095 Jy ms;][]{Jahns2018} than those in the FAST samples \citep[$\gtrsim$0.02 Jy ms;][]{Li2021}, making this work independent of \cite{Raquel2023}.
We investigate whether there are groups with common features between the FAST and Arecibo data so that we can corroborate the previous classification result with conjectures about the origin.

\section{Data Pre-processing}
We use the FRB catalogue detected in the Arecibo archival data \citep{Jahns2018}. 
The catalogue includes a total of 849 FRBs from the identical source of FRB 20121102A.
Each FRB can contain multiple sub-bursts.
There are 988 sub-bursts in total in the catalogue (classified by visual inspection). 
In this work, we treat sub-bursts independently, following \citet{Raquel2023}.
To ensure adherence to physical principles, all data points with negative amplitudes were removed, resulting in the final samples of 977 FRB sub-bursts.
The catalogue contains the following parameters:
        \begin{itemize}
            \item Time of arrival (ms)
            \item Amplitude (A)
            \item Bandwidth (sig$\_$nu) (MHz)
            \item Central frequency (nu$\_$0) (MHz)
            \item Dispersion Measure (pc $\cdot$ cm$^{-3}$)
            \item Linear temporal drift (d) (ms $\cdot$ MHz$^{-1}$)
            \item Fluence (Jy $\cdot$ ms)
            \item Time duration (sig$\_$t) (ms)
            \item Scaled energy (erg).
        \end{itemize}
        
We exclude the time of arrival in this work because it can only convey the sequence of arrival of various FRBs, and its correlation with physical properties is limited. 
In other words, the time of arrival alone would not be closely related to the distinct physical characteristics of each FRB.\\

Equation (2) in \citep{Jahns2018} fits two-dimensional, elliptical Gaussians to each sub-burst in
the burst spectra. The exact form depending on time $t$ and radio frequency $\nu$ is

\begin{equation}
    \mathcal{G}_{2D}(t,\nu)= A \exp\left(-\frac{(t - t_0 - d_t (\nu - \nu_0))^2}{2{\sigma_t}^2}- \frac{(\nu - \nu_0)^2}{2{\sigma_\nu}^2}\right).
    \label{equation_1}
\end{equation}

The variable $A$ represents the amplitude of the fitting function.

Following \cite{Raquel2023}, we also exclude DM from the classification process since the repeating FRBs from the FRB 20121102A source have almost the same DMs. 
In other words, each burst in FRB 20121102A exhibits an almost identical DM. 
Therefore, including it in the classification process would not provide significant and meaningful information. 
The fluence is a readily quantifiable property of a transient that remains less affected by the time resolution of the observation \citep[e.g.,][]{macquart2018,hashimoto2022}.
Therefore, we use fluences provided by \cite{Jahns2018} rather than using flux densities.

In summary, we utilise seven parameters \citep{Jahns2018}, including: Amplitude (corresponding to the fitting relation, as seen in Eqn. \ref{equation_1}), Linear temporal drift (temporal change of the peak frequency), Time duration (temporal burst width), Central frequency (spectral peak), Bandwidth (width in the frequency domain), Scaled energy (the isotropic equivalent energy that is scaled
from the fluence and the 2D Gaussian fits), Fluence (the flux of FRBs integrated over the time duration). 

These parameters collectively contribute to the analysis presented in this study.


\section{Data Processing/Methodology}
\subsection{Unsupervised machine learning}
\label{Unsupervised machine learning}

UMAP is an innovative manifold learning technique used for dimension reduction. It is built on a theoretical foundation rooted in Riemannian geometry and algebraic topology \citep{McInnes2018,McInnesUMAP2018}.
After pre-processing, our data consists of 977 rows and 7 columns. 
To facilitate data visualization and conduct unsupervised learning, we employ the UMAP algorithm.
Here are our processing steps:
\begin{enumerate}
\item Embedding the data with the following hyperparameter of {\ttfamily n\_neighbors}. Embedding refers to the process of mapping high-dimensional data points to a lower-dimensional space while preserving certain structural relationships and patterns present in the data. The goal of embedding is to represent complex and high-dimensional data in a more visually interpretable form, typically in two or three dimensions, without losing important information \citep[e.g.,][]{McInnes2018,McInnesUMAP2018,chen2022}.
{\ttfamily n\_neighbors} is one of UMAP's basic hyperparameters, which significantly affect the resulting embedding of the data \citep[e.g.,][]{Raquel2023}.

\item Clustering analysis with Hierarchical Density-Based Spatial Clustering of Applications with Noise (HDBSCAN) \citep{campello2013} to identify a group(s) in the embedded data.

\item Testing the processes 1 and 2 with different values of {\ttfamily n\_neighbors}. We investigate {\ttfamily n\_neighbors}$=$5,6,7,8, and 9 in this work.
The embedding and clustering results of {\ttfamily n\_neighbors}$=$5,6,7, and 9 are presented in APPENDIX A (Fig.\ref{nn9parcol2}), whereas we adopt {\ttfamily n\_neighbors}$=$8 as the fiducial result (see Section \ref{rand_score_sec} for details). 

\item Determining the optimised {\ttfamily n\_neighbors}, which maximizes the \lq rand score\rq. The concept of the rand score is introduced in the next section, which delves into the evaluation of similarities among different {\ttfamily n\_neighbors} outcomes. This evaluation aims to identify which results exhibit the highest degree of agreement with others.

\item Parameter colouring and histograms to investigate the characteristics of each cluster based on the optimised 

{\ttfamily n\_neighbors}.
\end{enumerate}

\subsection{Rand score for Clustering performance}
\label{rand_score_sec}
\begin{figure}
            \centering
            \includegraphics[width=1.0\linewidth]{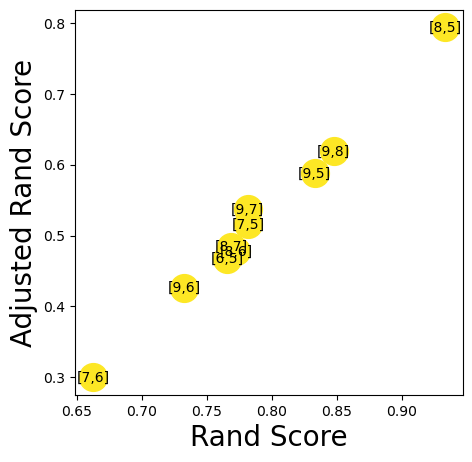}
            \caption{Adjusted Rand Score as a function of Rand Score. 
             Higher values of Adjusted Rand Score and Rand Score indicate a greater similarity of the classification results between the pair of {\ttfamily n\_neighbors} values.
             The pair of {\ttfamily n\_neighbors}$=8$ and $5$ shows the highest Adjusted Rand Score and Rand Score, indicating that {\ttfamily n\_neighbors}$=8$ clustering result is most similar to that of {\ttfamily n\_neighbors}$=5$.
             {\ttfamily n\_neighbors}$=8$ is commonly included in the two highest cases.
             }
            \label{rand_score}
        \end{figure}

Because this is an unsupervised ML, we need Rand Score to make the comparison. 
A clustering performance metric, namely the Rand Index \citep{HubertArabie1985}, and its adjusted form provide us with a Rand score and Adjusted Rand score for each pair of compared different {\ttfamily n\_neighbors} clustering results. 
A high score indicates that the two clustering results are in excellent agreement \citep[e.g.,][]{HubertArabie1985,Raquel2023}.
A higher Rand score indicates a greater similarity with the classification results of other {\ttfamily n\_neighbors} values, i.e., a high Rand Score (high agreement) agrees with another result.
To find the most suitable {\ttfamily n\_neighbors}, we need to find the rand score of each pair of {\ttfamily n\_neighbors}.
In Fig.\ref{rand_score}, the adjusted rand score is compared with the rand score with each point annotated with the pair of {\ttfamily n\_neighbors}.
In this work, we choose {\ttfamily n\_neighbors}$=8$, which has the highest rand score compared with the other values of {\ttfamily n\_neighbors}, and is included in the 2nd highest rand score. 
This way, even by considering only one of these results for discussion, we could extract the common groups for the different values of {\ttfamily n\_neighbors}.
The details of {\ttfamily n\_neighbors} and rand score arguments are presented in \citet{chen2022} and \citet{Raquel2023}, respectively.

\subsection{Hyperparameters}
\textcolor{black}{
There are two sets of hyperparameters in this study. The first set is of the UMAP and the second set is the HDBSCAN. UMAP hyperparameters that are considered in this study are \texttt{min\_dist}, \texttt{metric}, \texttt{n\_components}, and \texttt{n\_neighbours}. 
}

\textcolor{black}{
\texttt{min\_dist} controls the clumping of the embedded data points which means that the smaller the value we assign to this hyperparameter the clumpier the resulting embedding would be. Thus, in this study, we set \texttt{min\_dist}$=0.01$. 
}

\textcolor{black}{
\texttt{metric} is essentially the way distance is defined on the resulting embedding. Since using other metrics is not intuitive or straightforward, we set \texttt{metric} $=$ Euclidean.
}

\textcolor{black}{
\texttt{n\_components} dictates the spatial dimension of the resulting embedding. Thus, for simplicity and ease of visualization, we set \texttt{n\_components} $= 2$.
}

\textcolor{black}{
\texttt{n\_neighbours} is the most important hyperparameter in this stage. This hyperparameter estimates the manifold structure by controlling the size of the local neighborhood. This suggests that a lower value would emphasize the regional structure compared to a higher value which then emphasizes the global structure. Thus in this study, we set \texttt{n\_neighbors} $= 8$ (see also Section \ref{rand_score_sec} for details).
}

\textcolor{black}{
HDBSCAN compared to UMAP has a larger number of hyperparameters, however only four major parameters significantly affect the resulting clustering. These hyperparameters are \texttt{min\_cluster\_size}, \texttt{min\_samples}, \\
\texttt{cluster\_selection\_epsilon}, and \texttt{alpha}.
}

\textcolor{black}{
\texttt{min\_cluster\_size} is the size of the grouping that can be considered a cluster. This in return affects the number of clusters that can be identified by HDBSCAN. In this study the value of \texttt{min\_cluster\_size} $= 80$ because after numerous trials with different values ranging between 30 and 100, we found that setting \texttt{min\_cluster\_size} $= 80$ resulted in more significant differences between the parameters of the clusters. Also, it can be classified clearly between clusters and noises.
}

\textcolor{black}{
\texttt{min\_samples} is also an important hyperparameter and should be considered depending on the resulting embedding. When a large value is used for this hyperparameter, a large number of points will then be considered Noise. Thus, the researchers set a conservative value of \texttt{min\_samples} $= 15$ since the default setting of \texttt{min\_samples} was found to be the most appropriate value after thorough examinations. This process involved checking whether some data points were mistakenly considered as noise, despite their values and errors conforming to reasonable physical interpretations.
}

\textcolor{black}{
\texttt{cluster\_selection\_epsilon} controls the merging of microclusters located in high-concentration regions when tuned correctly. However, adjusting this hyperparameter to merge microclusters will not provide further insight into the clustering aside from the fact that they are considered a group or a cluster. Therefore, we set \texttt{cluster\_selection\_epsilon} $= 0$ which is its default value.
}

\textcolor{black}{
\texttt{alpha} is a hyperparameter that is rarely adjusted, if not avoided, and only acts as a last resort when tuning \texttt{min\_samples} or \texttt{cluster\_selection\_epsilon}, which does not provide any useful changes to the clustering. This hyperparameter is used to determine how conservative the clustering will become but since its adjustment is not necessary, we set it to its default value of \texttt{alpha} $=1.0$.
}

\section{Results}

\subsection{Embedding and Clustering results}
\begin{figure}
        \centering
        \includegraphics[width=1\linewidth]{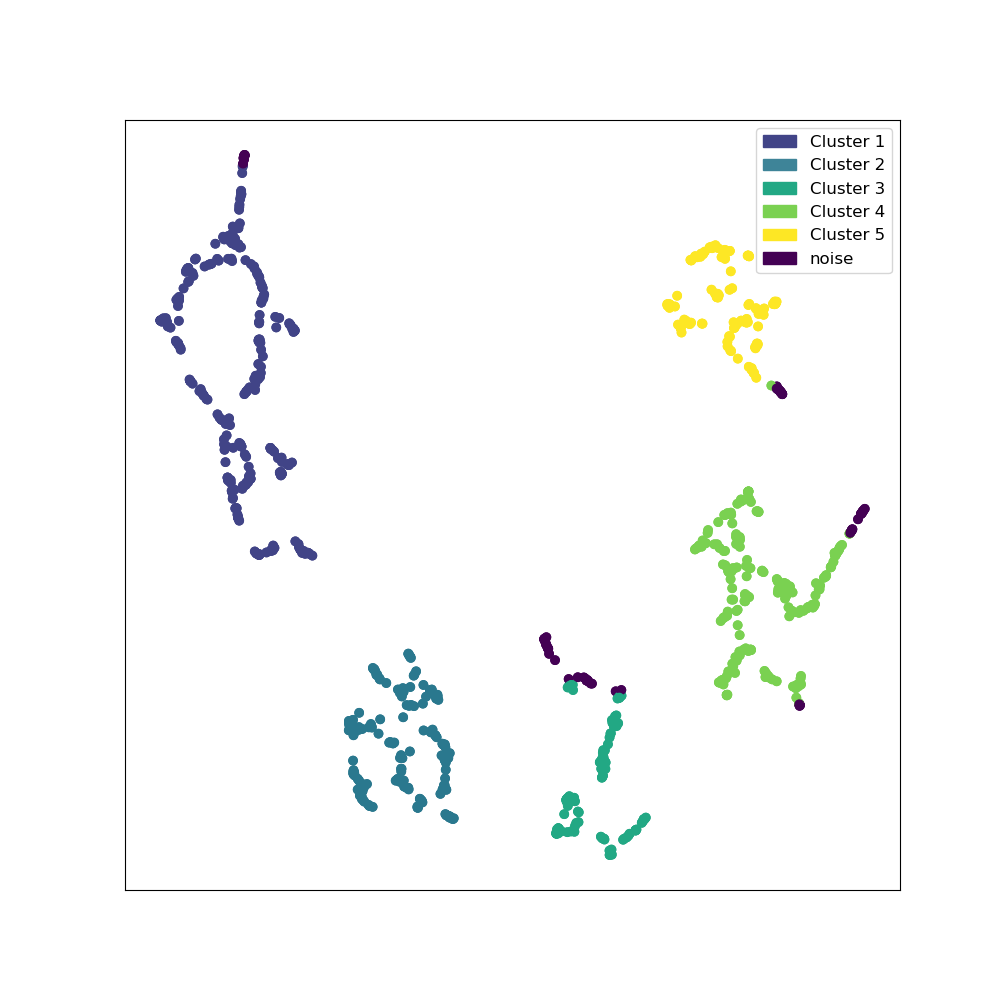}
        \caption{Two-dimensional UMAP embedding for {\ttfamily n\_neighbors=8}. The 977 FRB data are classified into five clusters with {\ttfamily n\_neighbors=8}. 
        After \lq projection\rq, HDBSCAN \citep{campello2013} is utilised to identify individual groups.}
        \label{test_8} 
    \end{figure}

The embedding result with {\ttfamily n\_neighbors}$=8$ is shown as Fig. \ref{test_8}.
Fig. \ref{test_8} shows distinct data distributions, indicating the existence of multiple clusters in the data.
The clustering algorithm described in Section \ref{Unsupervised machine learning} is applied to Fig. \ref{test_8} to identify clusters.
The embedded data are classified into five clusters as shown in Fig.\ref{test_8}. 
The clusters are clearly separated from each other, demonstrating the distinct distribution of each cluster. 
Distinct clusters are assigned unique colours to represent groups of data points (Fig. \ref{test_8}).
The distinct characteristics of these clusters are elaborated in Section \ref{characteristics}.

\subsection{Identifying characteristic properties of each cluster}
\label{characteristics}
\begin{figure*}
                \begin{multicols}{2}
                    \subcaptionbox{\label{a8}}{\includegraphics[width=1.03\linewidth, height=0.85\columnwidth]{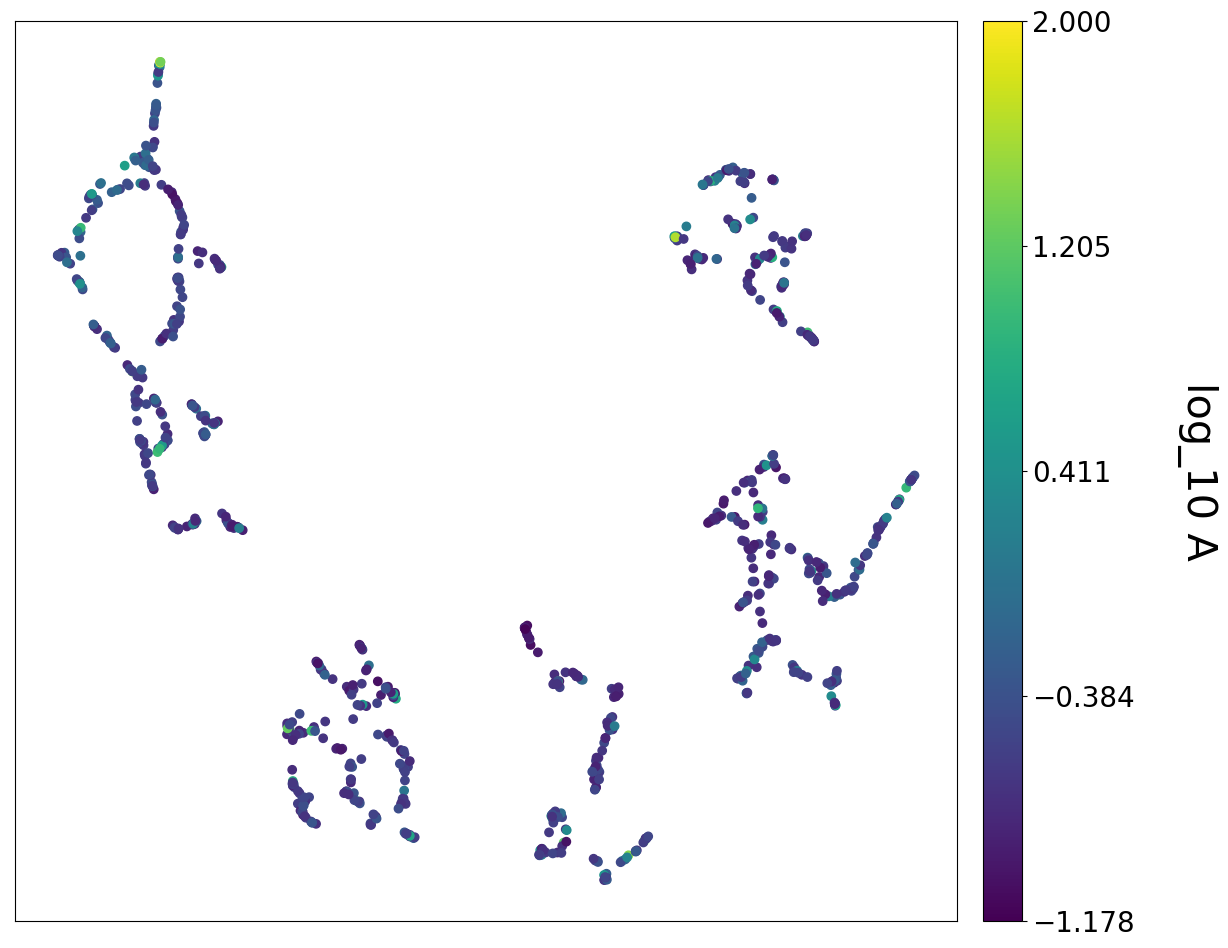}}\par 
                    \subcaptionbox{\label{b8}}{\includegraphics[width=1\linewidth, height=0.85\columnwidth]{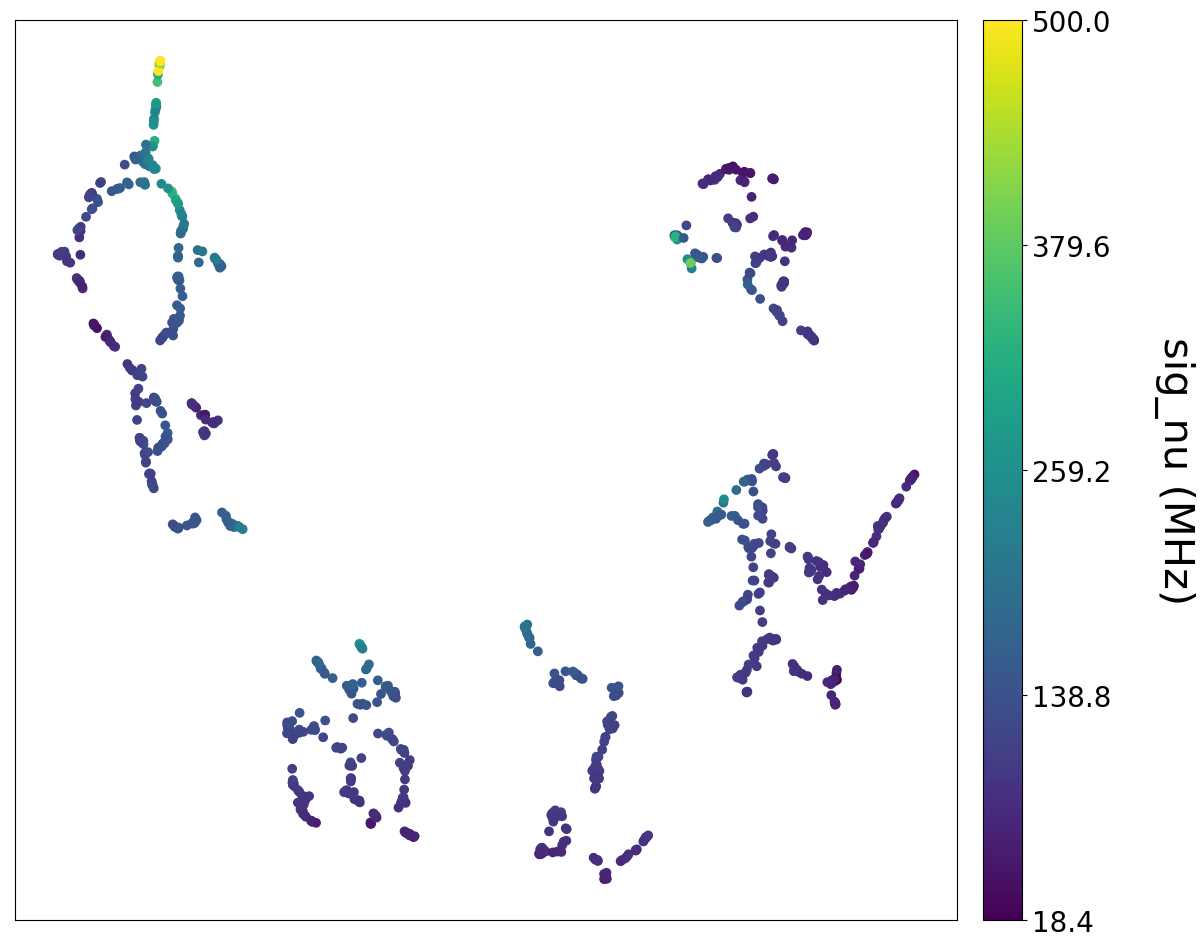}}\par 
                \end{multicols}
                \begin{multicols}{2}
                    \subcaptionbox{\label{c8}}{\includegraphics[width=1\linewidth, height=0.85\columnwidth]{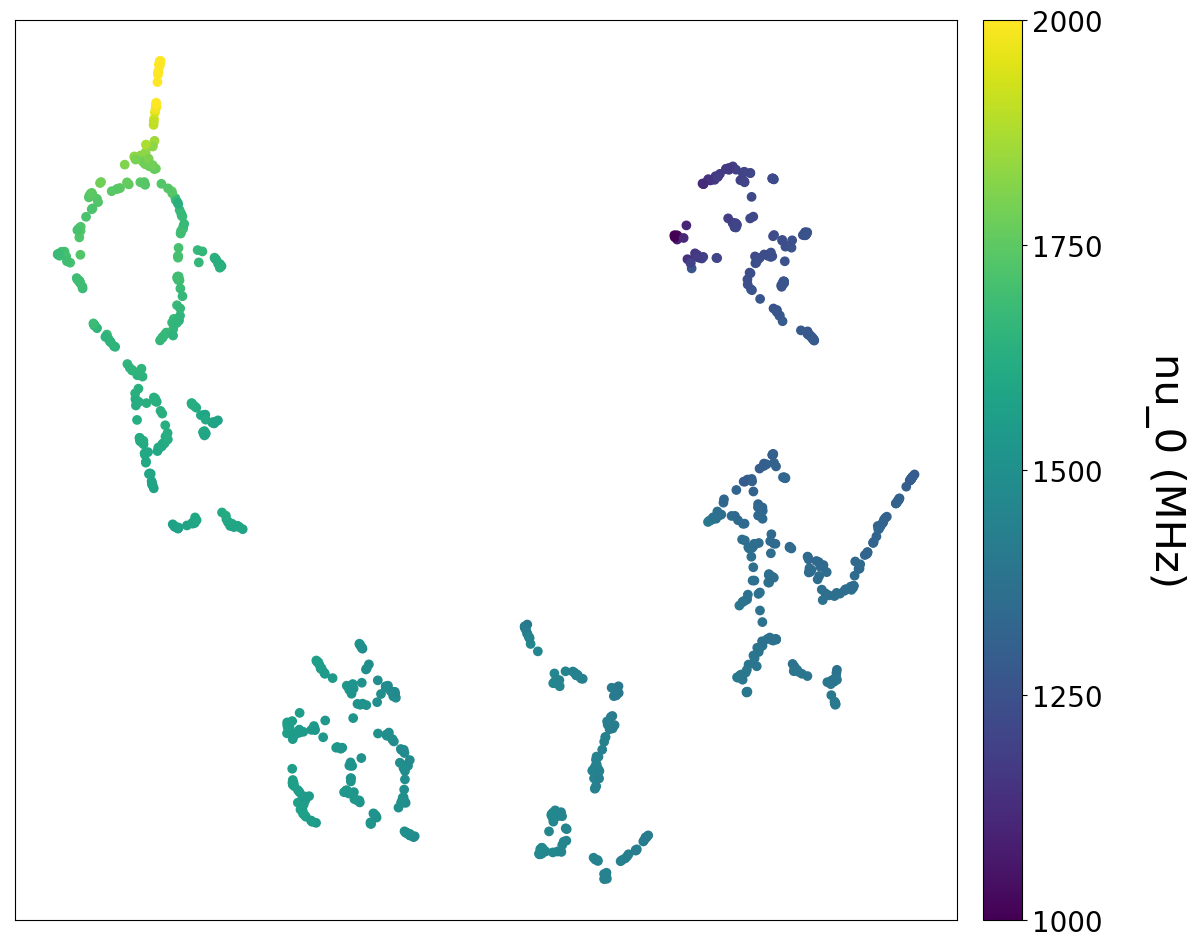}}\par
                    \subcaptionbox{\label{d8}}{\includegraphics[width=1.05\linewidth, height=0.85\columnwidth]{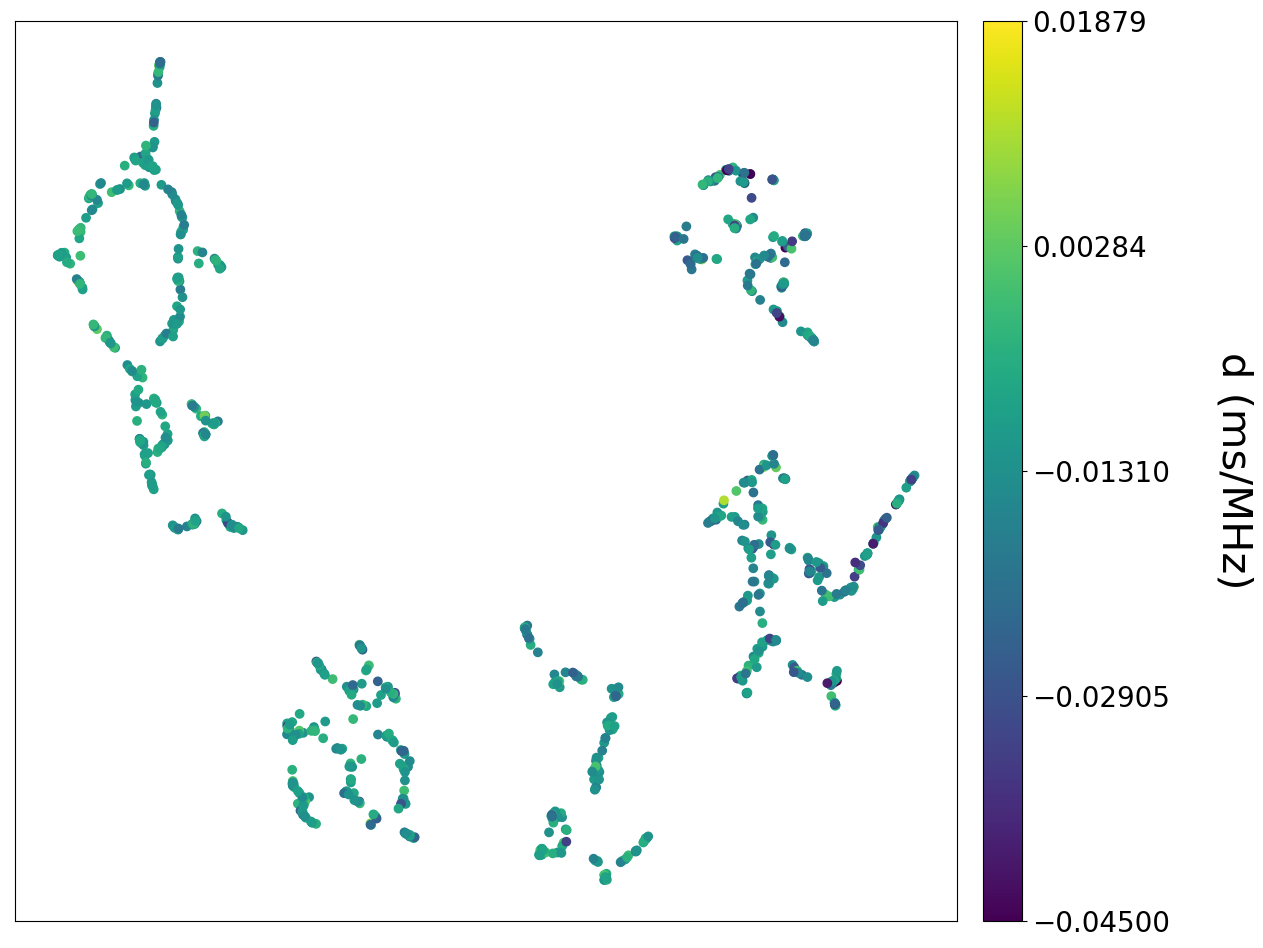}}\par
                \end{multicols}    
            \caption{Parameter colouring of the clustering result for {\ttfamily n\_neighbors = 8}. 
            From (a) to (d), the amplitude, bandwidth, central frequency, and linear temporal drift are shown, respectively.
            For amplitude Fig.\ref{a8}, the colour is shown in the logarithmic scale for visualization purposes.}
            
            \label{colouring_1}
            \end{figure*}

\begin{figure*}
                    \begin{multicols}{2}
                    \subcaptionbox{\label{tw9}}{\includegraphics[width=1\linewidth, height=0.85\columnwidth]{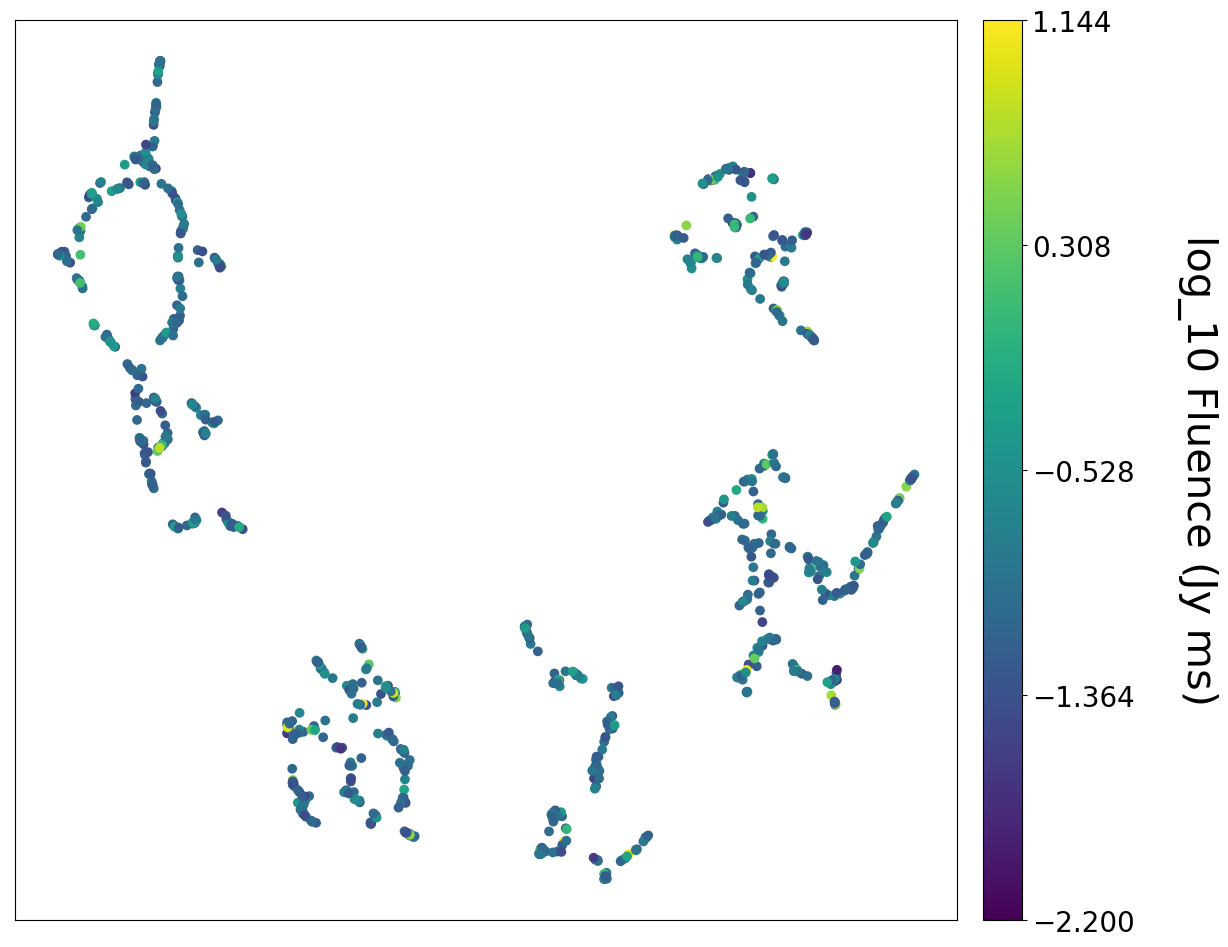}}\par
                    \subcaptionbox{\label{scaled_energy_8new}}{\includegraphics[width=1\linewidth, height=0.85\columnwidth]{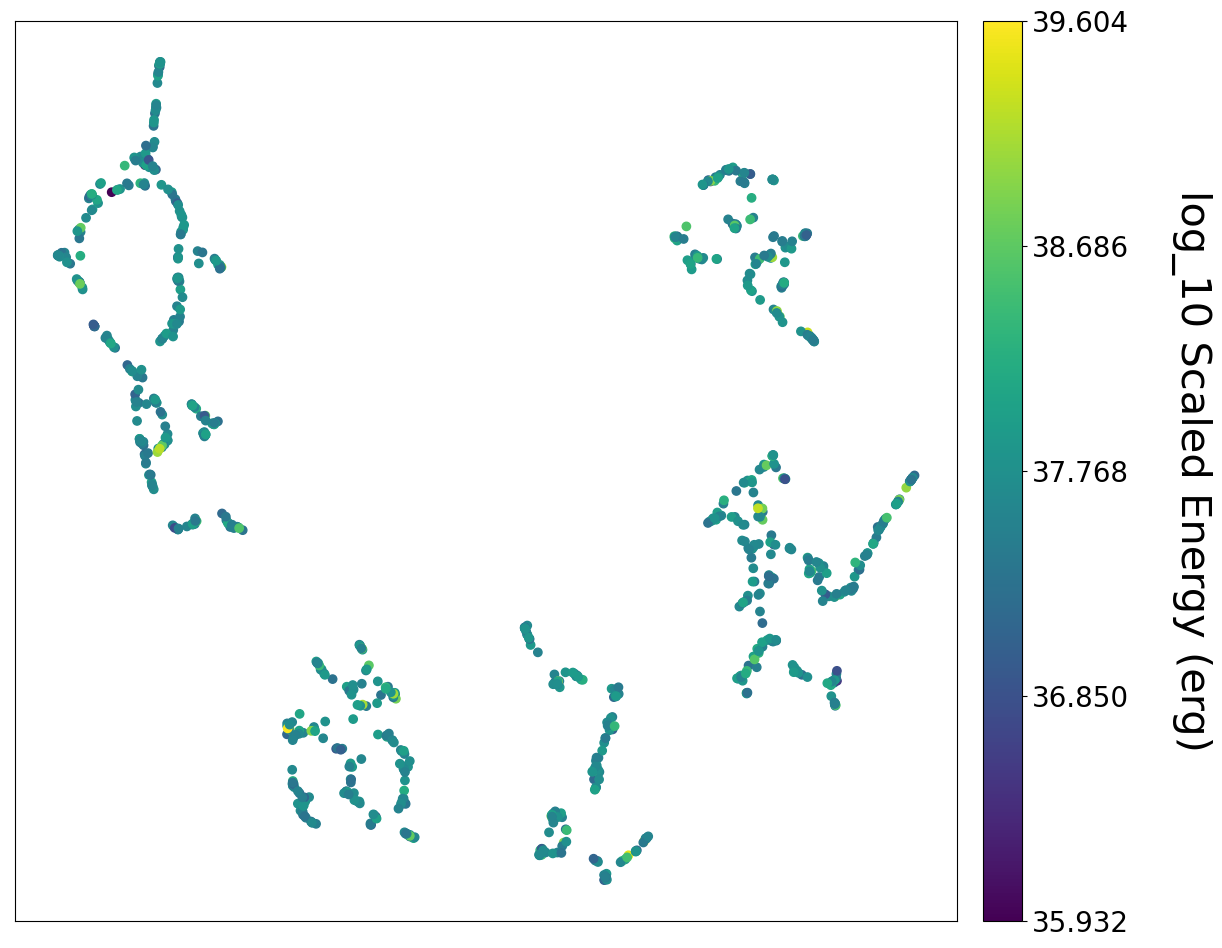}}\par
                \end{multicols}
                \begin{multicols}{2}
                    \subcaptionbox{\label{xd9}}{\includegraphics[width=1\linewidth, height=0.85\columnwidth]{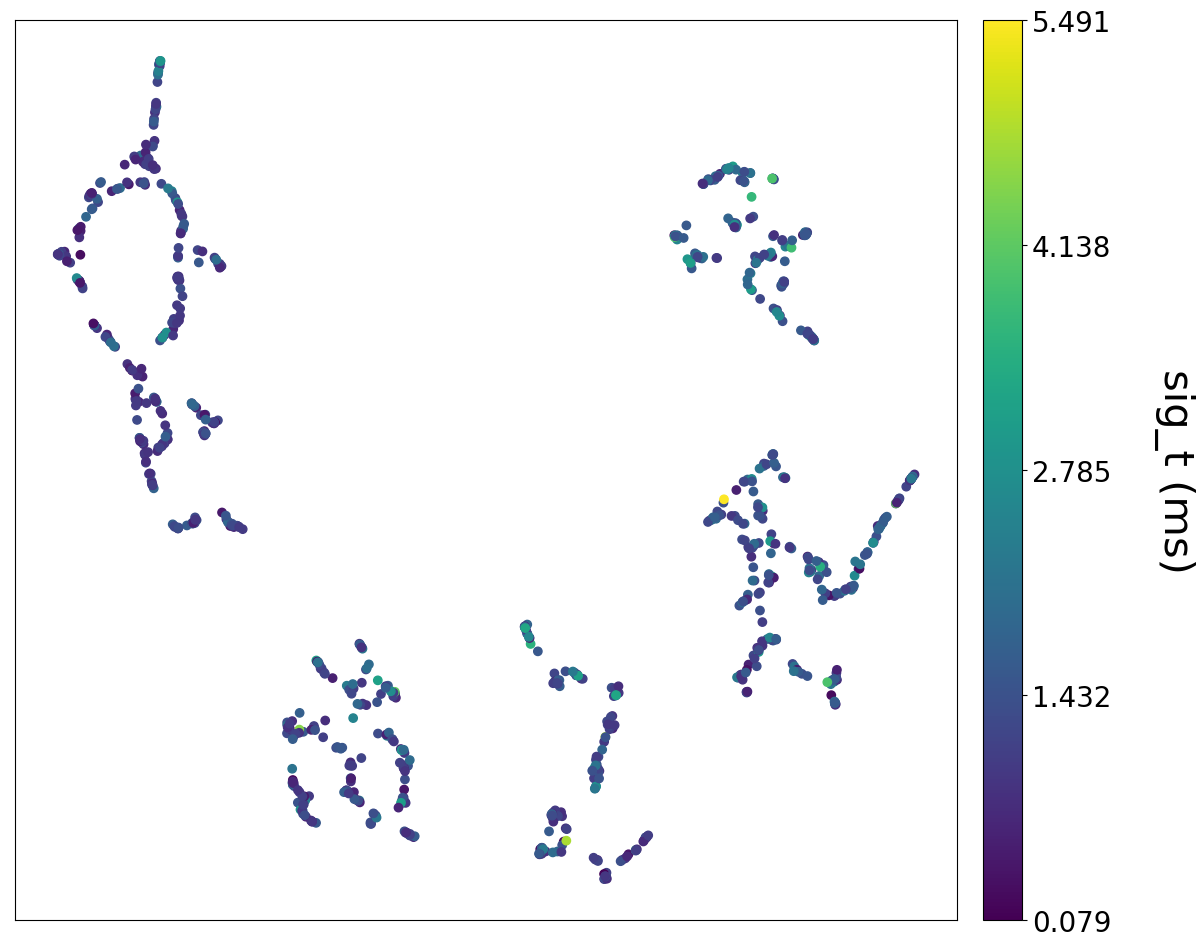}}\par
                \end{multicols}
            \caption{Parameter colouring of the clustering result for {\ttfamily n\_neighbors = 8.}
            From (a) to (c), the fluence, scaled energy, and time duration are shown, respectively.}
            \label{colouring_2}
            \end{figure*}

Because there are seven parameters in our analysis, we show seven plots of embedded data with colouring for each of the seven parameters in Fig. \ref{colouring_1} and Fig. \ref{colouring_2}.
These colouring plots allow us to infer the distinct characteristics of each cluster.

         \begin{figure*}
                \centering
                \begin{multicols}{2}
                    \subcaptionbox{\label{a8_his}}{\includegraphics[width=1\linewidth, height=0.8\columnwidth]{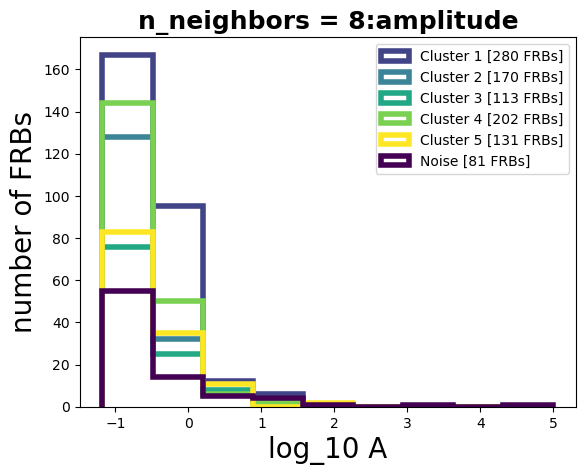}}\par 
                    \subcaptionbox{\label{b8_his}}{\includegraphics[width=1\linewidth, height=0.8\columnwidth]{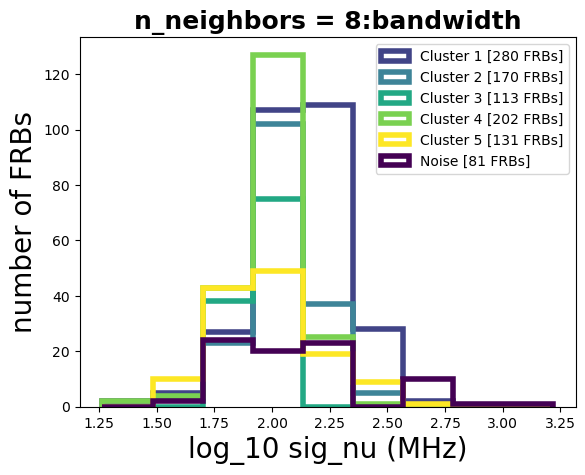}}\par 
                \end{multicols}
                \begin{multicols}{2}
                    \subcaptionbox{\label{c8_his}}{\includegraphics[width=1\linewidth, height=0.8\columnwidth]{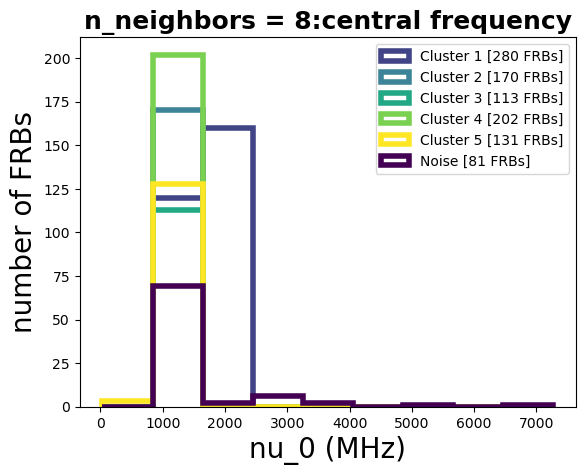}}\par
                    \subcaptionbox{\label{d8_his}}
                    {\includegraphics[width=1\linewidth, height=0.8\columnwidth]{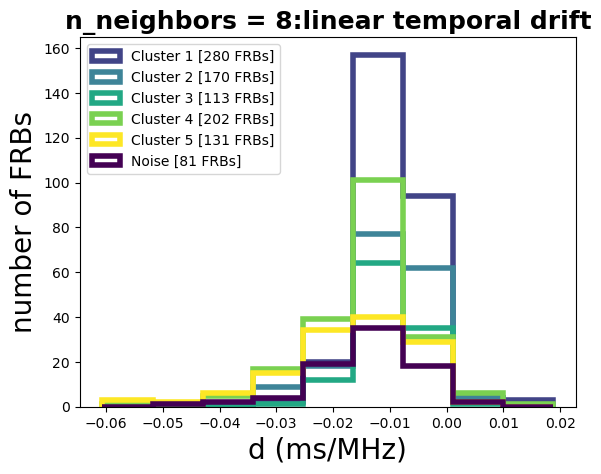}}\par
                \end{multicols}   

                \caption{ 
                Histograms for each parameter with {\ttfamily n\_neighbors = 8}. }
            \label{his_8}
            \end{figure*}

            \begin{figure*}
                \begin{multicols}{2}
                    \subcaptionbox{\label{f8_his}}{\includegraphics[width=1\linewidth, height=0.8\columnwidth]{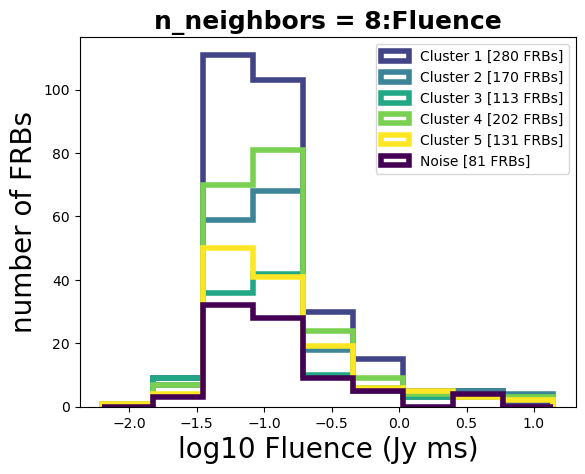}}\par
                    \subcaptionbox{\label{s8_his}}{\includegraphics[width=1\linewidth, height=0.8\columnwidth]{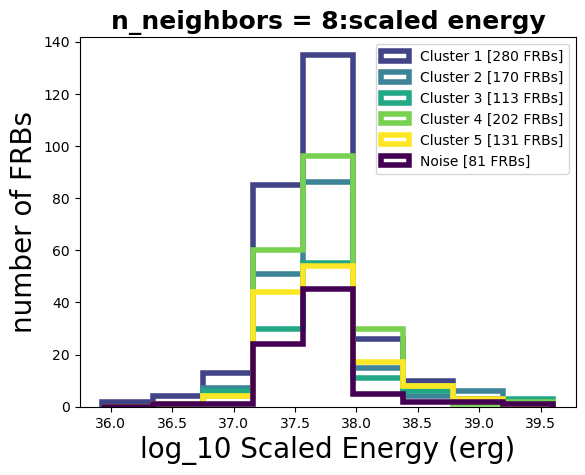}}\par
                \end{multicols}
                \begin{multicols}{2}
                    \subcaptionbox{\label{t8_his}}{\includegraphics[width=1\linewidth, height=0.8\columnwidth]{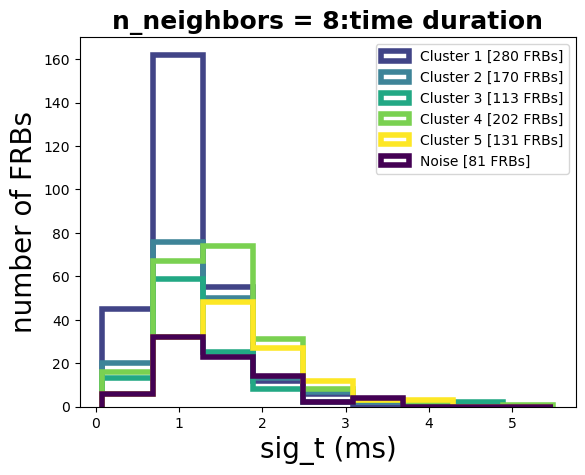}}\par
                \end{multicols}
                \caption{Parameter colouring of the clustering result for {\ttfamily n\_neighbors = 8}. 
                }
            \label{his_8_2}
            \end{figure*}

            \begin{table*}
            \centering
            \begin{tabular}{|c|c|c|c|c|c|c|}
                    \hline
                    
                 Average Value &  &  & & & &  \\
                 of each Parameter &  &  &  & & &  \\
                 in each Cluster & Cluster 1 & Cluster 2 & Cluster 3 & Cluster 4 & Cluster 5 & Noise   \\
                 
                 \hline
                 Amplitude & 0.8$\pm$2.2 & 0.7$\pm$2.0 & 1.1$\pm$3.3 & 0.5$\pm$1.5 & 1.5$\pm$6.9 & 1000$\pm$11000 \\

                 \hline
                 Bandwidth (MHz) &146$\pm$62 &118$\pm$40 &96$\pm$20 &102$\pm$32 &110$\pm$65 &190$\pm$220 \\
                 \hline
                 Central Frequency (MHz) & 1676$\pm$96 & 1519$\pm$30 & 1441$\pm$17
                & 1355$\pm$30
                & 1180$\pm$160 & 1660$\pm$930 \\
                 \hline
                 Linear Temporal Drift (ms MHz$^{-1}$) & -0.0090$\pm$0.0056 & -0.0104$\pm$0.0071 & -0.0099$\pm$0.0055 & -0.0141$\pm$0.0094  & -0.017$\pm$0.012 &  -0.0142$\pm$0.0091 \\
                 \hline
                 Fluence (Jy ms) & 0.24$\pm$0.61 & 0.5$\pm$1.9 & 0.7$\pm$2.1 & 0.4$\pm$1.3& 0.5$\pm$1.8 & 0.33$\pm$0.85 \\
                 \hline 
                 Scaled Energy ($\log _{10}$ erg) &37.66$\pm$ 0.40& 37.73$\pm$0.39 &37.76$\pm$0.46 &37.73$\pm$0.37 &37.74$\pm$0.39 &37.71$\pm$0.38 \\
                 \hline 
                 Time Duration (ms) & 1.10$\pm$0.47 & 1.32$\pm$0.67 & 1.33$\pm$0.78 & 1.49$\pm$0.69 & 1.73$\pm$0.72 &1.52$\pm$0.67 \\ 
                 \hline
                 BT (K) &             \makecell{$3.3\times10^{33}$ \\ $\pm 3.9\times10^{33}$} & \makecell{$5.1\times10^{33}$ \\ $\pm 3.3\times10^{33}$} & \makecell{$6.9\times10^{33}$ \\ $\pm 3.5\times10^{33}$} & \makecell{$3.2\times10^{33}$ \\ $\pm 2.6\times10^{33}$} & \makecell{$3.4\times10^{33}$ \\ $\pm 2.2\times10^{33}$} & \makecell{$1.7\times10^{33}$ \\ $\pm 3.2\times10^{33}$} \\
                 \hline
            \end{tabular}
            \caption{Average value of each parameter in each cluster with {\ttfamily n\_neighbors=8}. The errors include two significant figures. For the purpose of comparing with the Critical Temperature \citep{xiao2022}, we computed the Brightness Temperature (BT) using the average values of each parameter, as presented in the last row.}
            \label{tab1}
          \end{table*}
           
        \begin{table*}
            \centering
            \begin{tabular}{|c|c|c|c|c|c|c|}
                    \hline
                    
                 Invariant &  &  & &  & & \\
                 Cluster 
                 & Cluster 1 & Cluster 2 & Cluster 3 & Cluster 4 & Cluster 5 & Noise  \\
                 Properties &  &  &  & & &   \\                 
  
                 \hline
                 Amplitude & Low & Low & Low & Low & Low & High \\
                 \hline
                 Bandwidth & Wide & Wide & Narrow & Medium & Medium & Diverse  \\
                 \hline
                 Central Frequency & High & High & Medium & Medium & Low & Diverse \\
                 \hline
                 Linear Temporal Drift & Uniform & Uniform & Uniform & Diverse & Diverse & Diverse  \\
                 \hline
                 Fluence & Uniform & Uniform & Uniform & Diverse & Diverse & Diverse  \\
                 \hline 
                 Scaled Energy & Uniform & Diverse & Diverse & Uniform & Diverse & Diverse \\ 
                 \hline
                 Time Duration & Very Short & Short & Short & Long & Very Long & Long \\
                 \hline
            \end{tabular}
            \caption{Cluster properties that remain constant with {\ttfamily n\_neighbors=8}. The qualitative description of the clusters is based on the range of values for each parameter of a given cluster.
            }
            \label{tab2}
            \end{table*}
To validate the characteristics of clusters, we construct histograms, followed by analysis and summarization in the form of tables, Table \ref{tab1} and Table \ref{tab2}.
In the histograms provided in Fig.\ref{his_8}, we conducted an examination of the Amplitude Fig.\ref{a8_his}, Bandwidth Fig.\ref{b8_his}, Central Frequency Fig.\ref{c8_his}, Linear Temporal Drift Fig.\ref{d8_his}, Fluence Fig.\ref{f8_his}, Scaled Energy Fig.\ref{s8_his}, and Time Duration Fig.\ref{t8_his} histograms. 
Notably, the bandwidth distributions exhibited unique patterns in all clusters, supporting that the resulting clusters are significantly different.

We combine the results of histograms and colouring figures for discussion. Some parameters clearly show distinct differences among each cluster, especially central frequency and bandwidth. 
While others may appear less distinguishable, we carefully examine their distribution patterns, noting some are wider in the frequency domain while others are narrower. This allows us to identify the unique characteristics of each cluster.

The result of the analysis is summarized in Table \ref{tab1} and Table \ref{tab2}. As shown in Table \ref{tab2}, each cluster encompasses a distinct set of attributes, as illustrated in the Appendix (Amplitude Fig. \ref{A_3}, Bandwidth Fig. \ref{A_4}, Central Frequency Fig. \ref{A_5}, Fluence Fig. \ref{A_6}, Scaled Energy Fig. \ref{A_7}, and Time Duration Fig. \ref{A_8}). Even with varying {\ttfamily n$\_$neighbor} values, each cluster exhibits similar distributions, as seen in Table.\ref{tab2}. We might refer to these attributes as "invariant" cluster properties. 
Although these attributes do not immediately pin point us to specific physical mechanisms, the classyfying is an important step advance, because now we can try to theoretically understand each cluster one by one, instead of understanding them all at once while mixed.

\section{Discussion}   

\subsection{Relationships between different variables}

       \begin{figure*}
       \centering
       \includegraphics[width=0.9\linewidth]{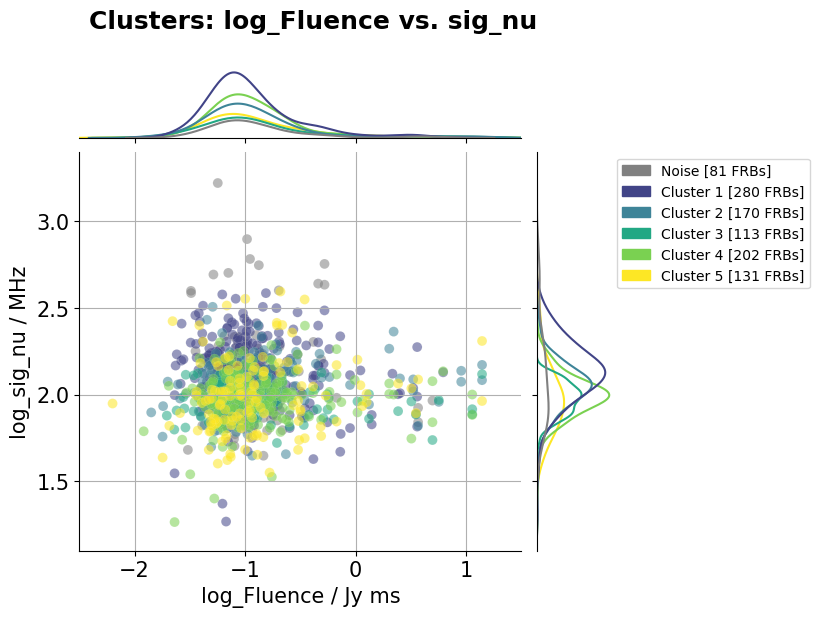}
       \caption{
       \textcolor{black}{
       Mapping plot of Bandwidth and Fluence with {\ttfamily n\_neighbors=8}.
       Different colors correspond to different clusters. 
       The histograms on the vertical and horizontal axes represent the data distributions of Bandwidth and Fluence, respectively. 
       }
       }
       \label{com_his} 
       \end{figure*}

\textcolor{black}{After performing dimensionality reduction on the clusters, we aimed to map these clusters onto joint distribution plots of the variables. To identify significant differences, we selected physical parameters that showed notable variations between clusters. We will discuss these differences following the analysis of histograms (Fig. \ref{his_8}, Fig. \ref{his_8_2}) and tables (Tab. \ref{tab1} and Tab. \ref{tab2}). We observe that Fluence and Bandwidth exhibit the most significant differences among clusters, as shown in the histograms in Figures \ref{his_8} and \ref{his_8_2}, and the data in Tables \ref{tab1} and \ref{tab2}. To further analyze these parameters, we combined and mapped them into a distribution plot (Fig. \ref{com_his}). In Fig. \ref{com_his}, the distribution of Fluence in Cluster 1 appears more concentrated compared to the other clusters, and so does the distribution of Bandwidth, indicating that most of the FRBs in Cluster 1 are similar.}

\textcolor{black}{Additionally, there is a subtle secondary peak beside the main peak in the distribution curve of Cluster 1. This raises an interesting question: could there be a physical mechanism that generates two extremums of Fluence, unlike other mechanisms that result in a single peak, as observed in the other Clusters?}

\textcolor{black}{As for the Bandwidth of each cluster, there are noticeable differences between their peaks, especially between Cluster 1 and Cluster 4. There is also something interesting that the Bandwidth distribution of Cluster 3 has two significant peaks. It would be valuable to explore what causes these significant differences in future work.} 

\subsection{Comparison with the classification by the previous work}
\begin{figure*}
            \begin{multicols}{2}
                    \subcaptionbox{\label{hdbscan_L}}{\includegraphics[width=1.4\linewidth]{test/test_8_new.png}}\par 
            \end{multicols}
            \begin{multicols}{2}
                \subcaptionbox{\label{hdbscan_J}}
                {\includegraphics[width=1.3\linewidth]{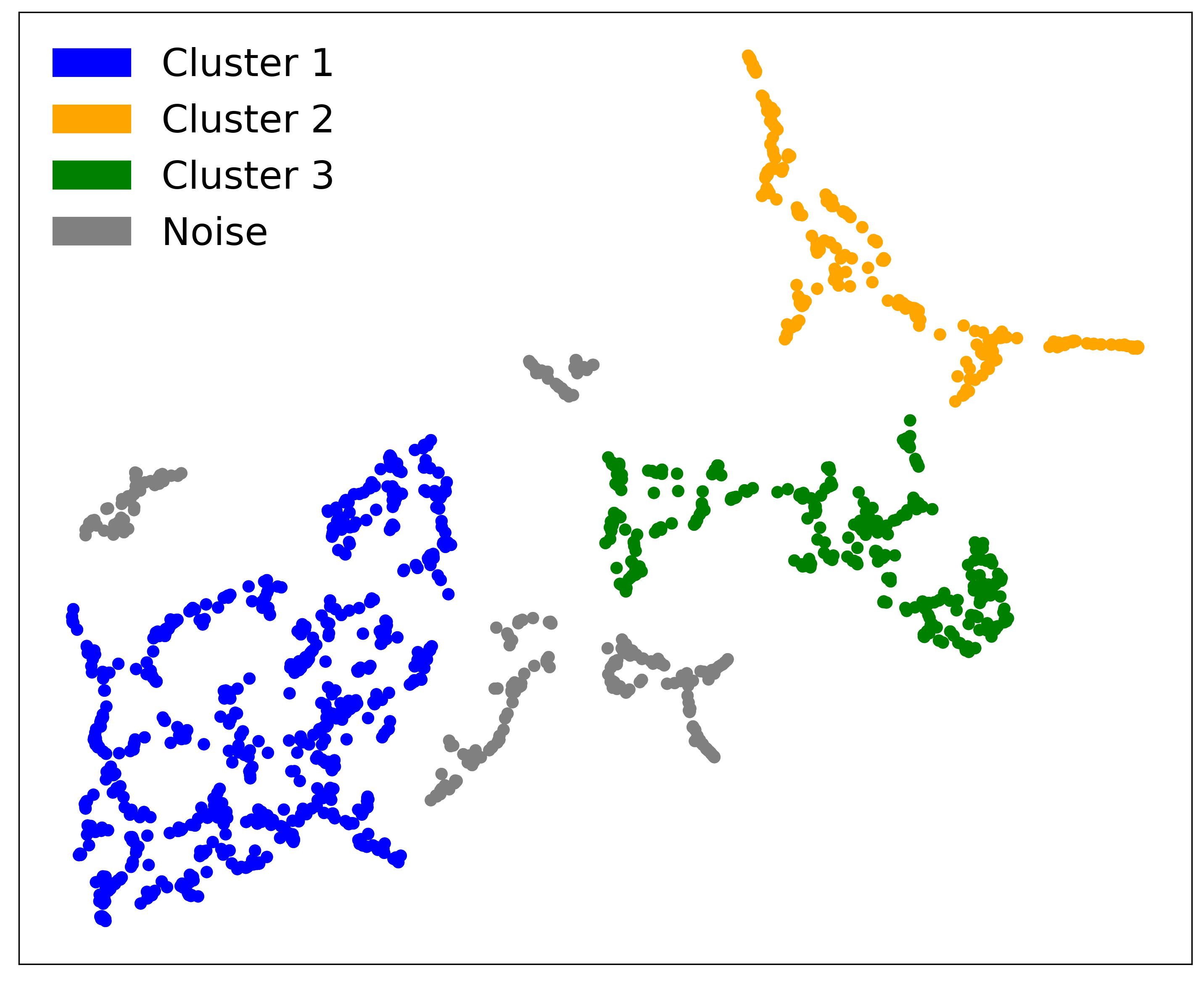}}\par 
            \end{multicols}
            \caption{The comparison of the clustering of our result (a) and \citet{Raquel2023} (b).}
        \label{adjrandvrand}
        \end{figure*}

We compare our classification with another classification result of FRB 20121102A using the FAST data \citep{Raquel2023}. 
Our result, presented in Fig.  \ref{hdbscan_L}, shows five clusters, while Fig.  \ref{hdbscan_J} from \cite{Raquel2023} shows three clusters. Noises are also plotted in the figures.
        
\begin{figure*}
                \centering
                \begin{multicols}{2}
                    \subcaptionbox{\label{bw9}}{\includegraphics[width=\linewidth, height=0.8\columnwidth]{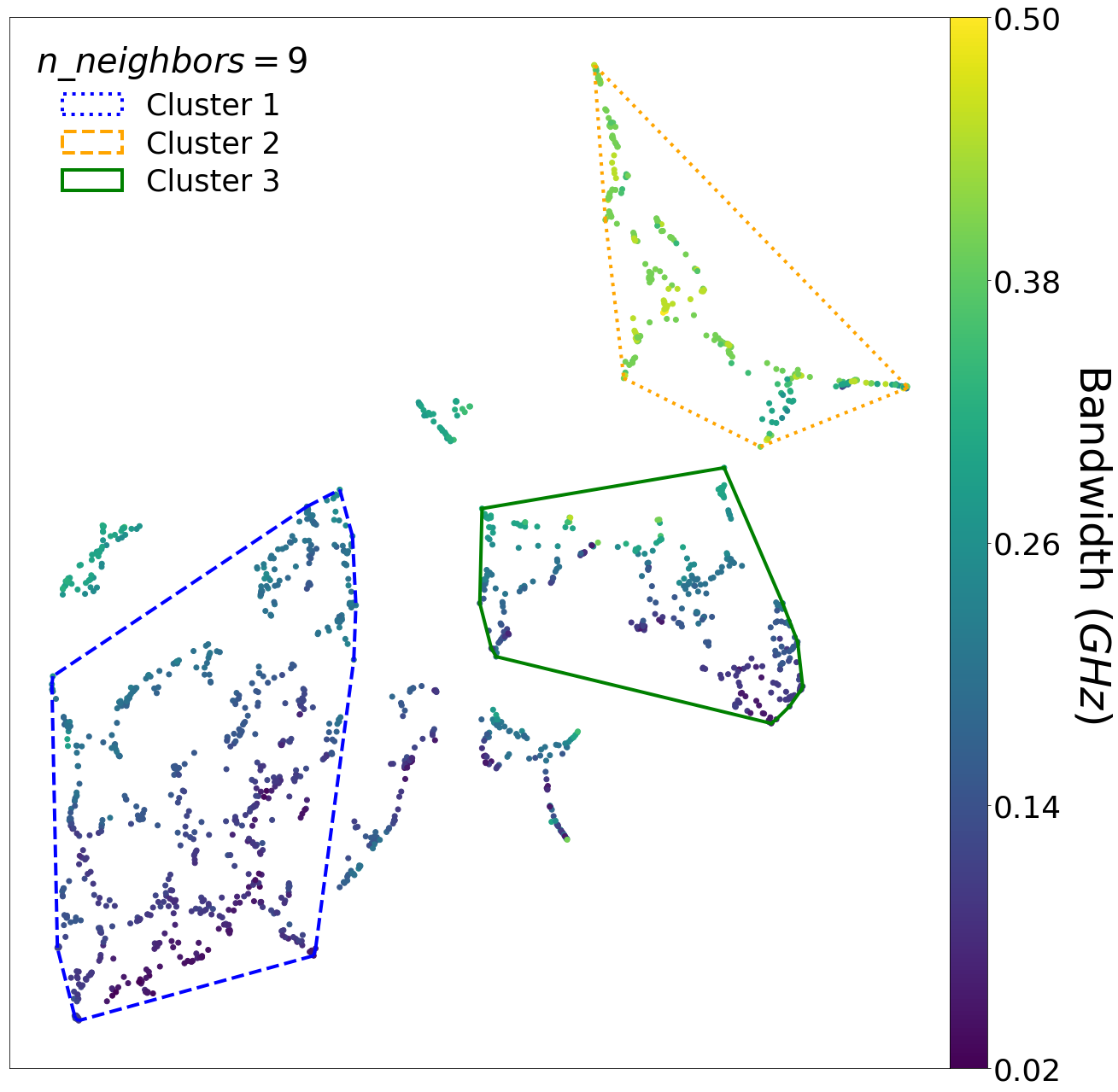}}\par 
                    \subcaptionbox{\label{bw8}}{\includegraphics[width=\linewidth, height=0.8\columnwidth]{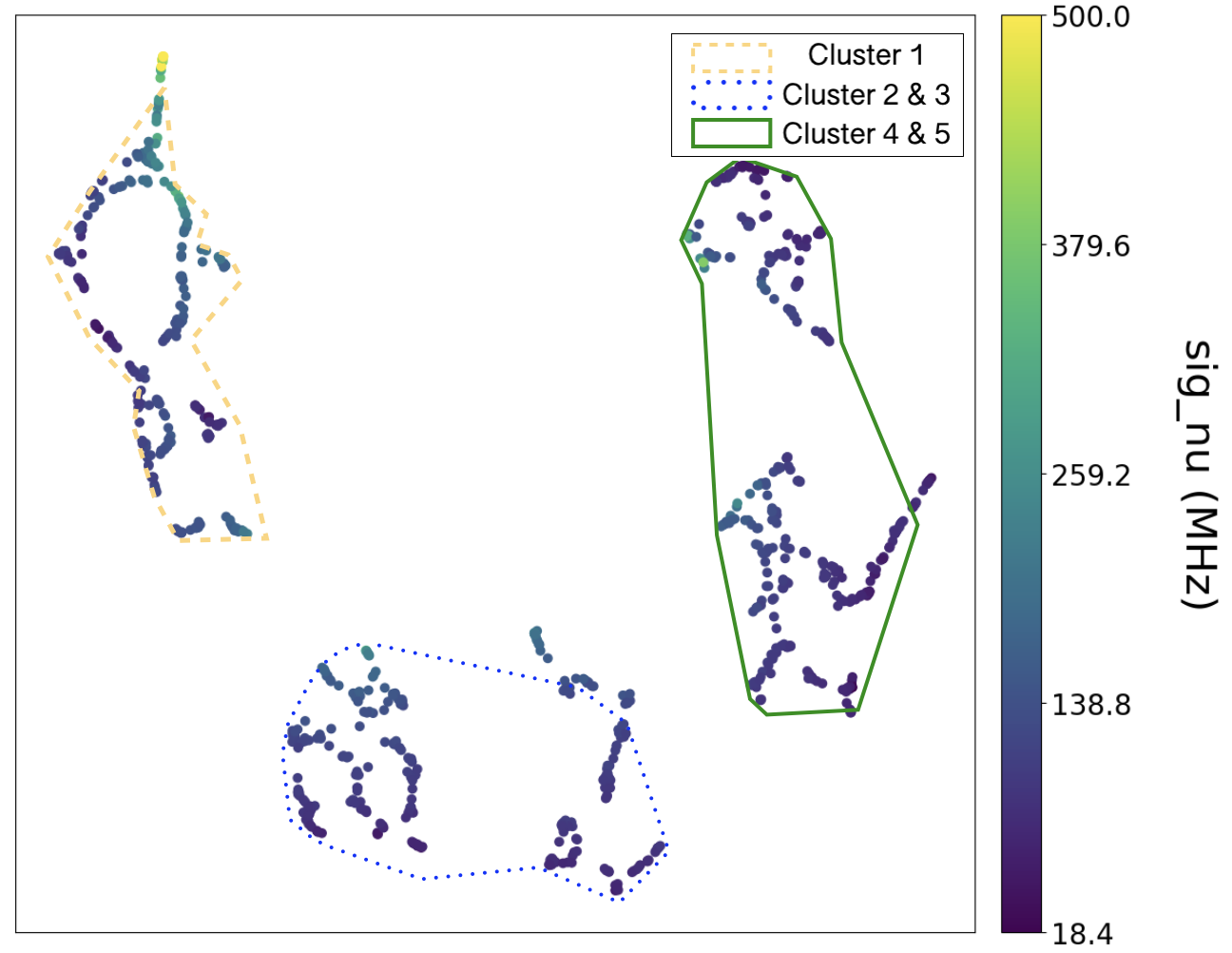}}\par 
                \end{multicols}
                \begin{multicols}{2}
                    \subcaptionbox{\label{fl9}}{\includegraphics[width=\linewidth, height=0.8\columnwidth]{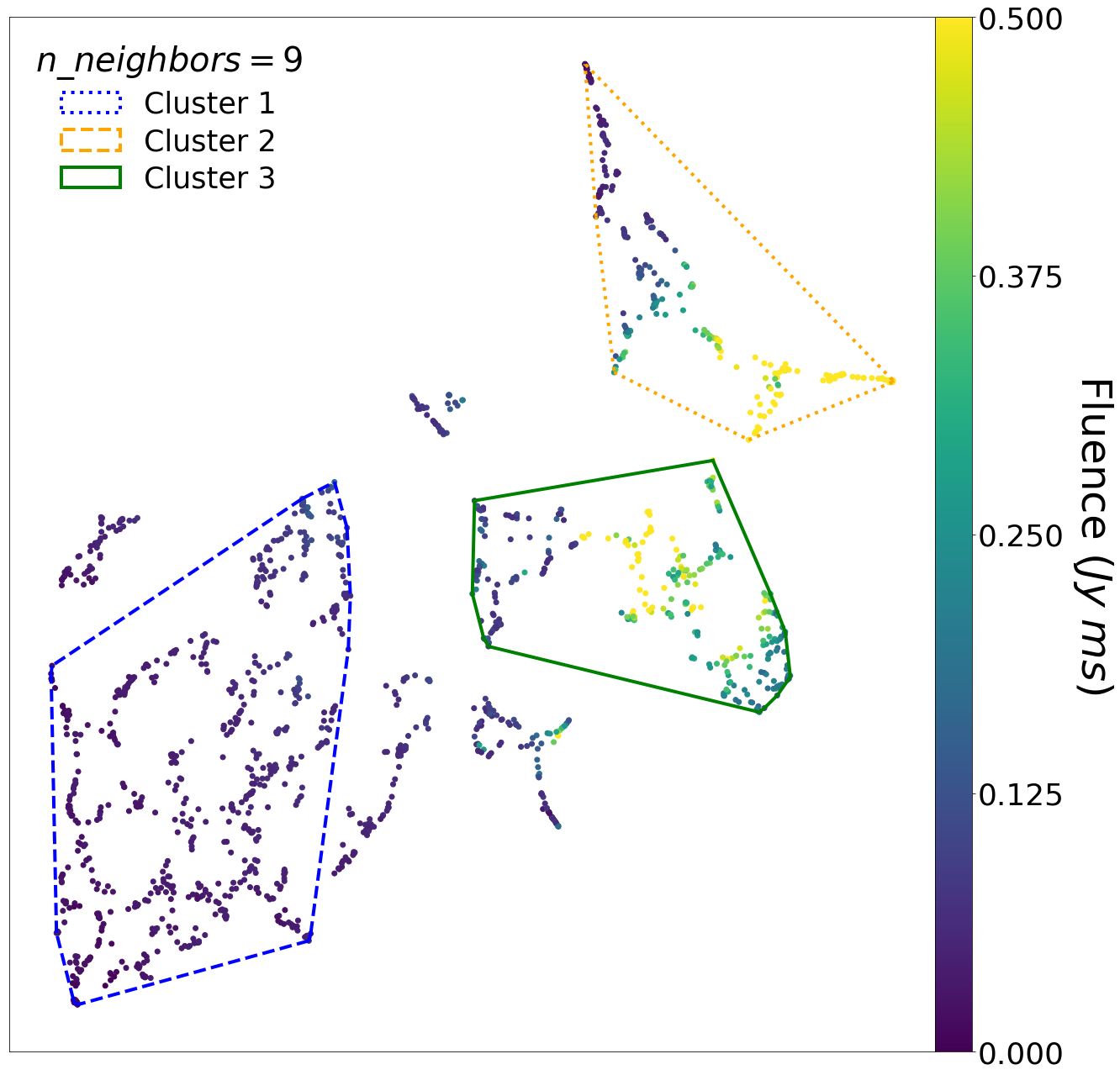}}\par
                    \subcaptionbox{\label{pf8}}{\includegraphics[width=\linewidth, height=0.8\columnwidth]{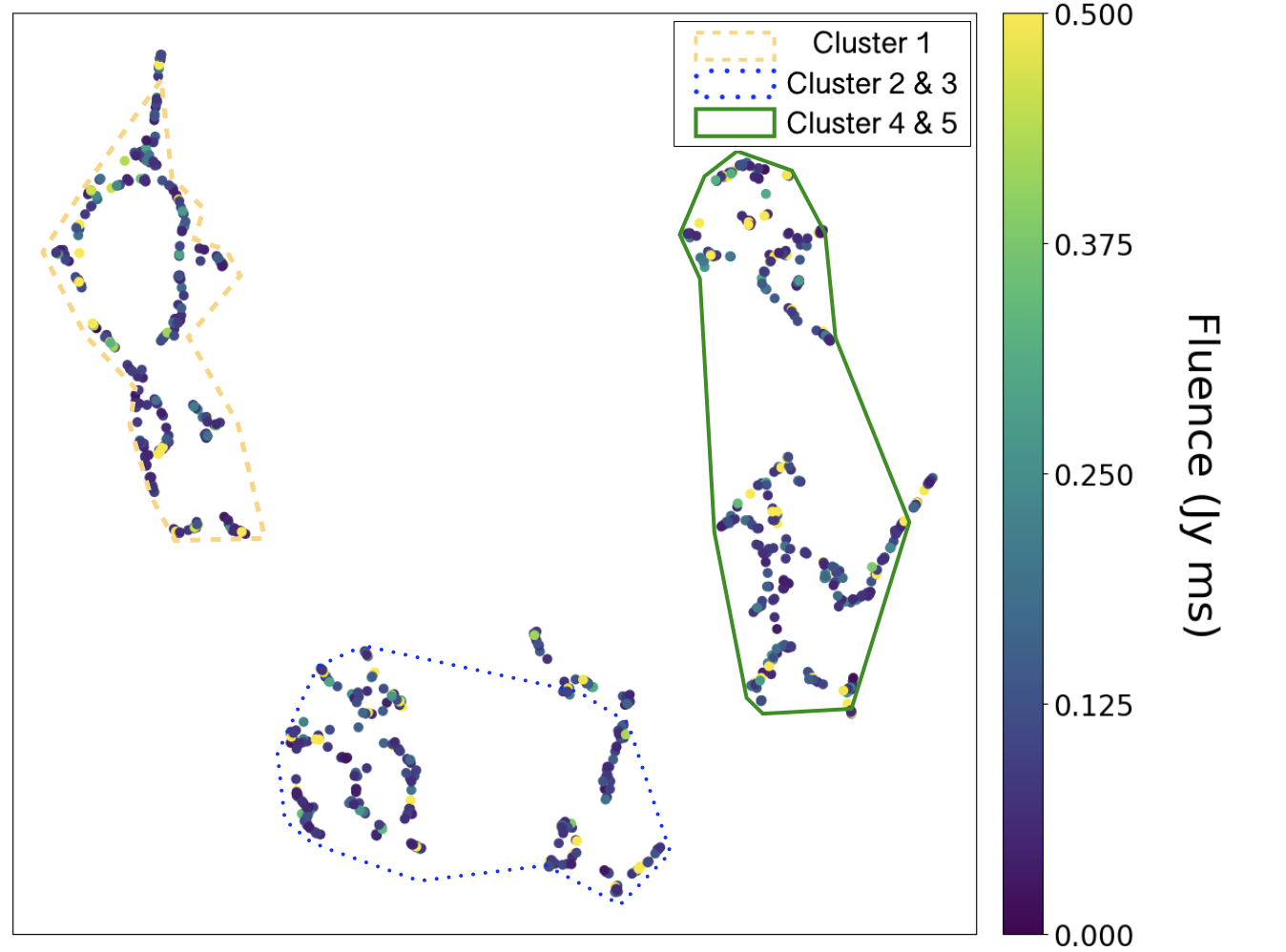}}\par
                \end{multicols}               
                \caption{(Left) Classification results from FAST data \citep{Raquel2023}, there are three clusters, Cluster 2 has the highest value of Bandwidth, same as Fluence, Cluster 2 has the higher value in general, and so does Cluster 3. (Right) Classification results from this work. While our result shows that there are five clusters,  Fluence looks more uniform in each cluster. As for Bandwidth(sig$\_$nu), same as Cluster 2 in another classification result \citep{Raquel2023}, our Cluster 1 has the highest value in general.}
            \label{Fig_9}
            \end{figure*}

\begin{figure*}
                \begin{multicols}{2}
                    \subcaptionbox{\label{ee9}}{\includegraphics[width=\linewidth, height=0.8\columnwidth]{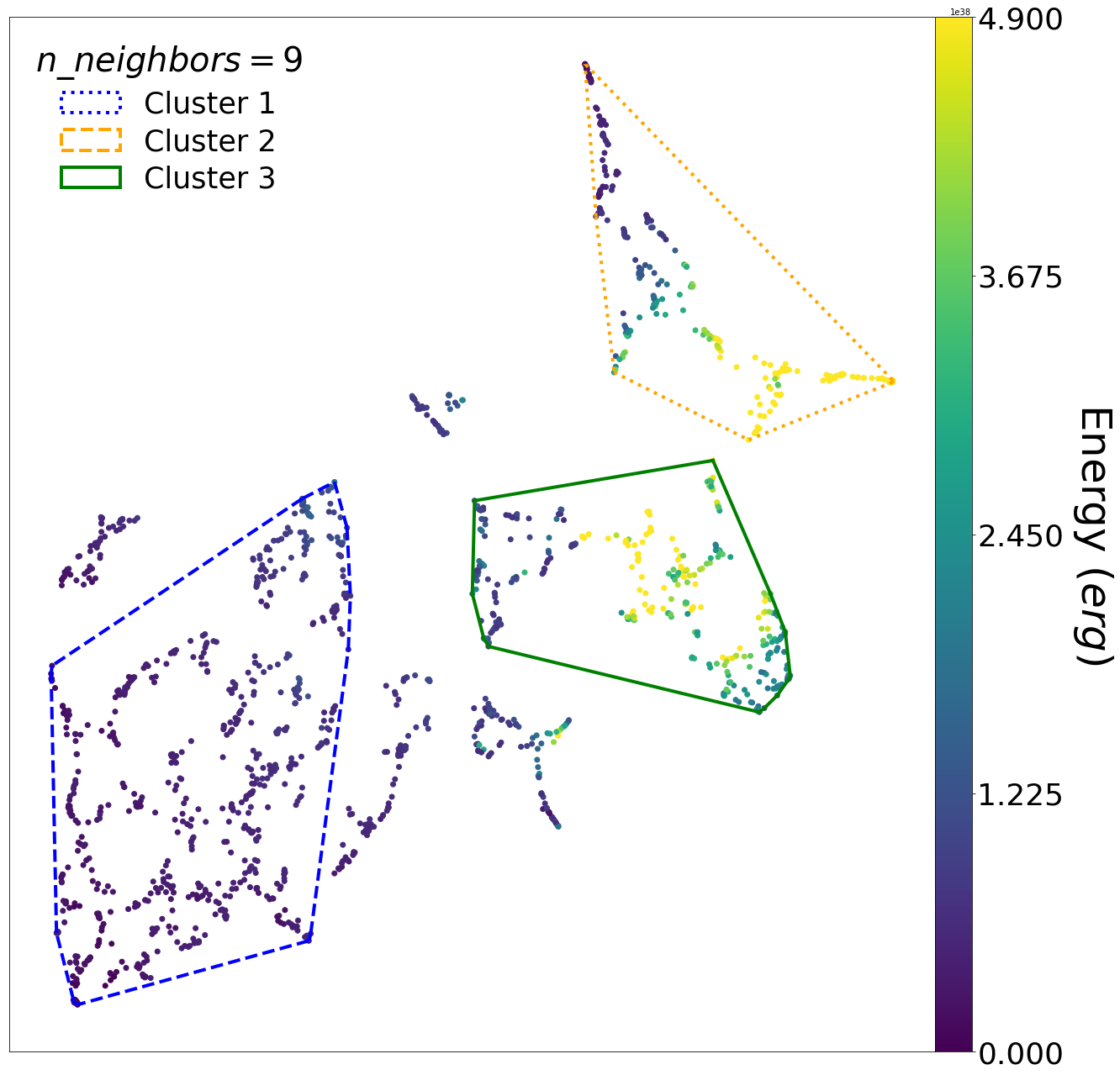}}\par
                    \subcaptionbox{\label{se8}}{\includegraphics[width=\linewidth, height=0.8\columnwidth]{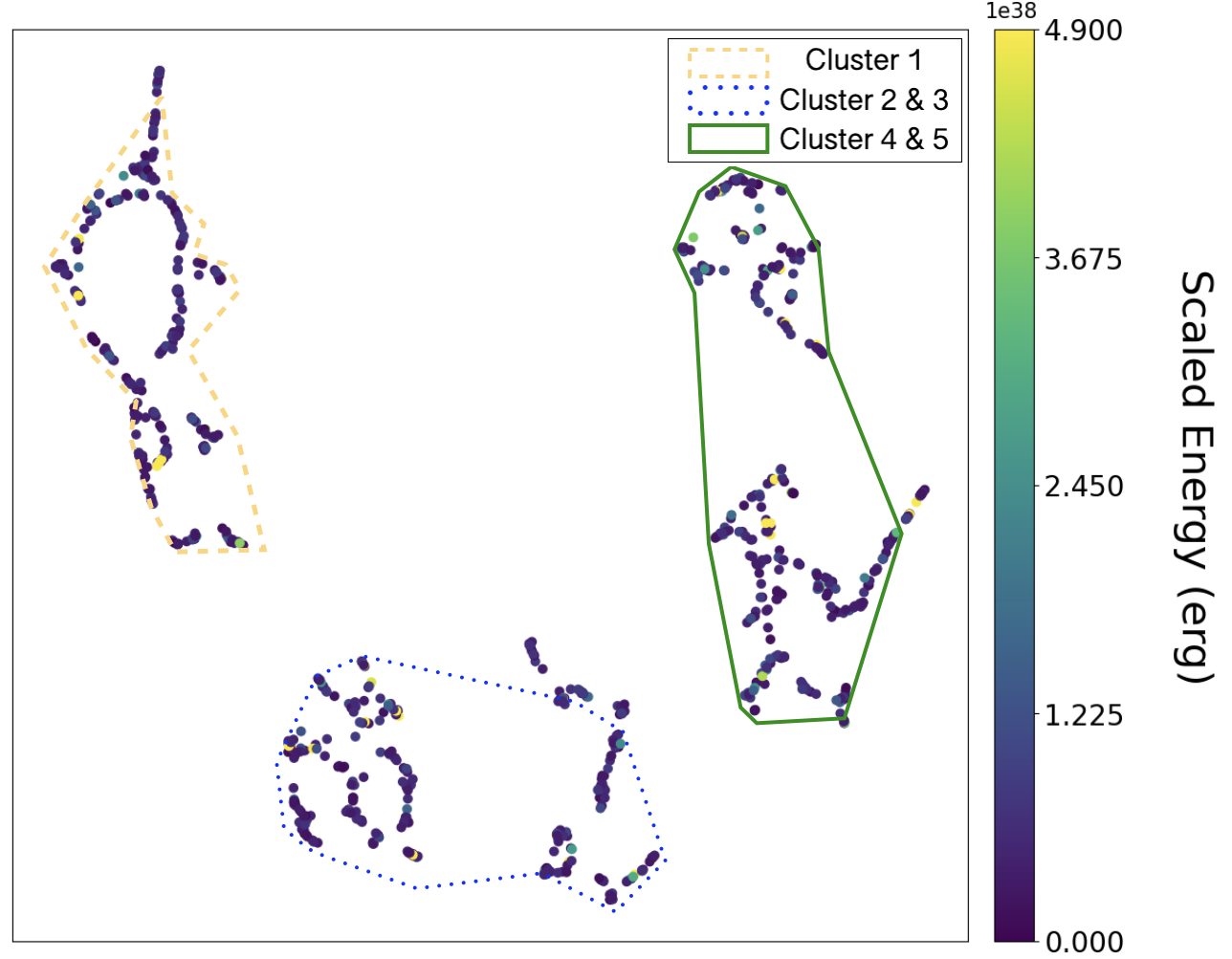}}\par
                \end{multicols}
                \begin{multicols}{2}
                    \subcaptionbox{\label{wt9}}{\includegraphics[width=\linewidth, height=0.8\columnwidth]{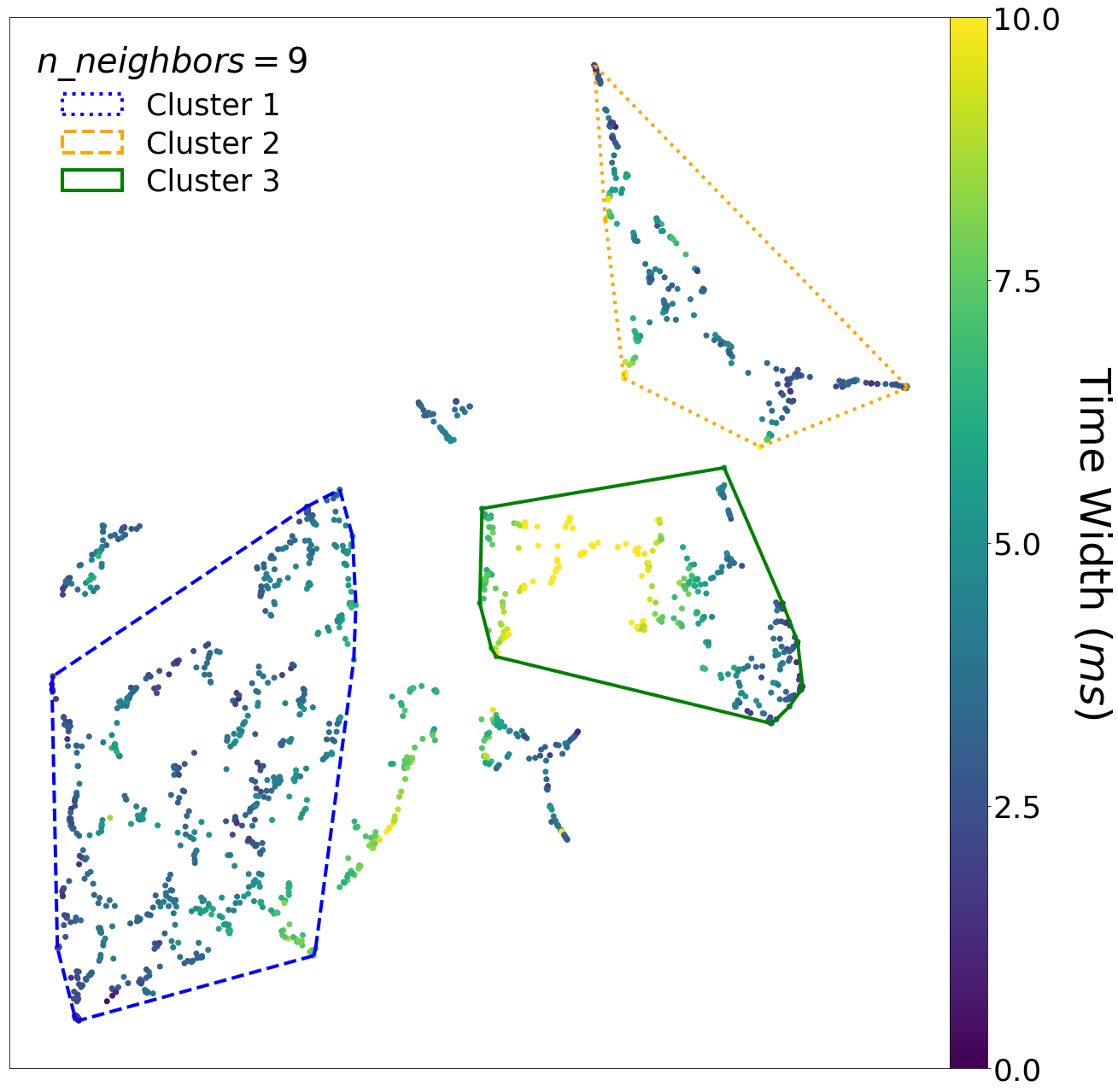}}\par
                    \subcaptionbox{\label{td8}}{\includegraphics[width=\linewidth, height=0.8\columnwidth]{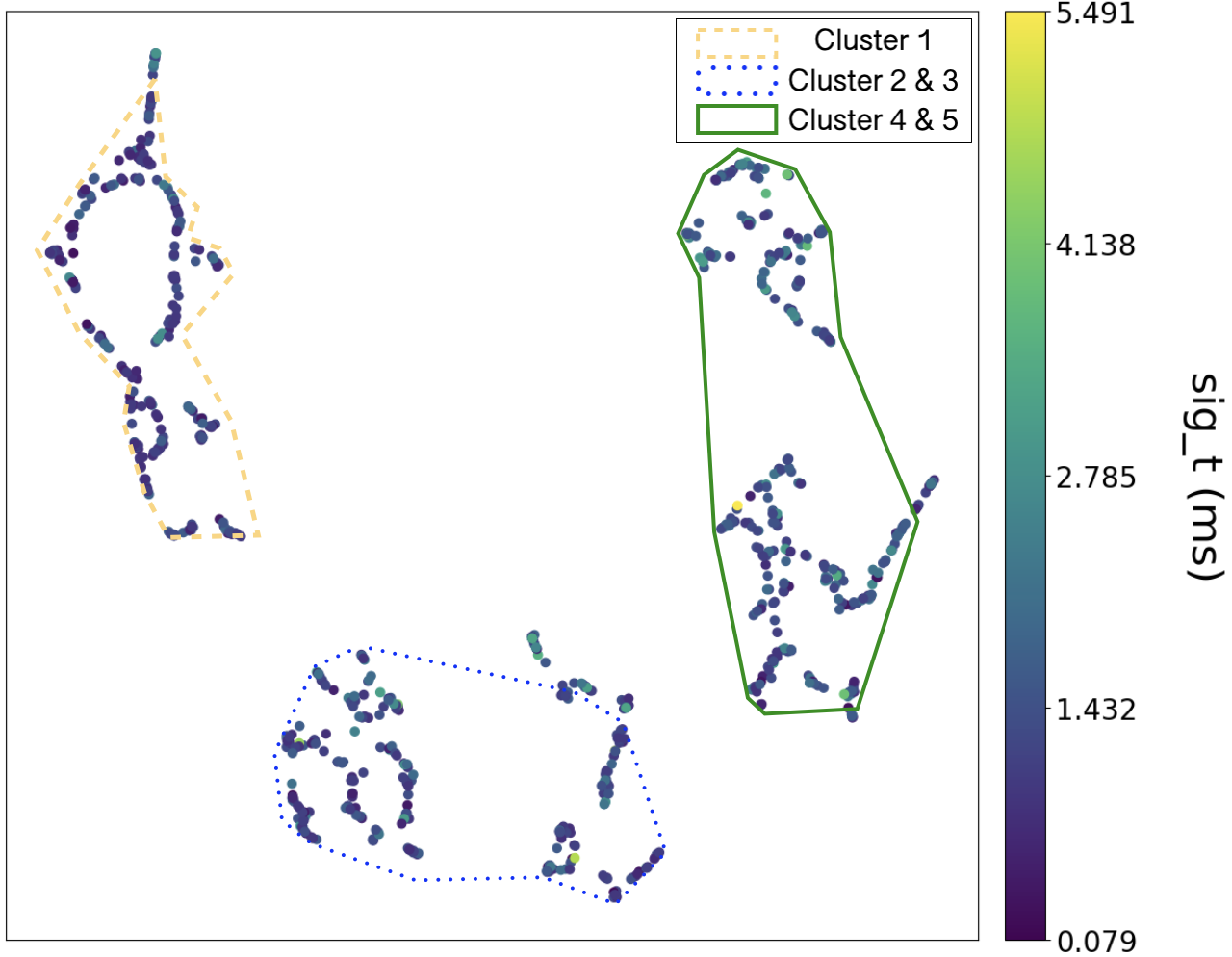}}\par
                \end{multicols}                
                \caption{(Left) Classification results from FAST data \citep{Raquel2023}, there are three clusters, both Cluster 2 and Cluster 3 have the higher values of Energy, Cluster 1 has the lowest, most of the data points show most of the values lower than $1.225\times10^{38}$ erg. On the other hand, Cluster 3 has the longest time width (also duration), most of them are longer than 5.0 milliseconds according to Fig. \ref{wt9}. (Right) Classification results from this work. While our result shows that there are five clusters, Scaled Energy looks more uniform to each cluster, however, Cluster 4 and Cluster 5 seem to have higher values of data points. As for Time Duration (sig$\_$t), Cluster 4 and Cluster 5 have the longer time duration, followed by Cluster 2 and Cluster 3.}
            \label{fig_10}
            \end{figure*}

Fig. \ref{Fig_9} and Fig. \ref{fig_10} compare the colouring results between \citet{Raquel2023} and this work for parameters commonly used in both studies.
Fig.\ref{ee9} and Fig.\ref{se8} are the energy \citep{Raquel2023} and scaled energy (this work) colouring of the clustering results, respectively. 
The scaled energy represents the isotropic equivalent energy, derived by scaling the fluence and the 2D Gaussian fits using Equation 2 in \cite{Jahns2018}.
In Fig. \ref{Fig_9}, both Cluster 2 in \citet{Raquel2023} and Cluster 1 in this work show higher Bandwidths than the other clusters.
These clusters also include FRBs with high Fluence (Fig. \ref{Fig_9}c and d) and high Energy/Scaled Energy (Fig. \ref{fig_10}a and b).
Therefore, we speculate that Cluster 2 in \citet{Raquel2023} is a similar population to Cluster 1 in this work.

Cluster 3 in \citet{Raquel2023} includes FRBs with two distinct properties with low and high values of Bandwidth (Fig. \ref{Fig_9}a), Fluence (Fig. \ref{Fig_9}c), Energy (Fig. \ref{fig_10}a), and Time Width (Fig. \ref{fig_10}c).  
These properties of Cluster 3 in \citet{Raquel2023} would correspond to a combination of Cluster 4 and 5 in this work.
The remaining Cluster 1 in \citet{Raquel2023} shows similar physical properties to a combination of Cluster 2 and 3 in this work in terms of Bandwidth/sig\_nu, Fluence, Energy/Scaled Energy and Time Width/sig\_t (Fig. \ref{Fig_9} and Fig. \ref{fig_10}).
Therefore, we circle borders with similar colours and shapes to individual clusters that might correspond to each other (e.g., their Cluster 2 might correspond to our Cluster 1. Therefore, we encircle Cluster 1 with a yellow dashed line, just as they encircled their Cluster 2 with a yellow dashed line).

We found five clusters with noise, each of which possesses distinct physical properties.
This suggests that FRBs might involve multiple different physical mechanisms, leading to individual sets of radio emissions with unique characteristics. 
\textcolor{black}{The geometry of the emission region and the propagation effect of FRB signals could also make such distinct clusters.}
However, the previous analysis by \cite{Raquel2023} identified three different clusters, whereas our result includes five. 
We speculate the following reasons for the different numbers of clusters between \citet{Raquel2023} and this work:
\begin{enumerate}
\item Their analysis yielded three clusters \citep{Raquel2023}, but this does not necessarily mean there are only three distinct groups. FAST telescope is larger and more sensitive than Arecibo. Therefore, their data are dominated by fainter bursts than ours. This may have led them to miss clusters dominated by brighter bursts. For example, our cluster 1 and 4 include brighter bursts.


\item Differences in the parameters used in our study compared to theirs \citep{Raquel2023} may lead to variations in the machine learning outcomes.  One contributing factor may be the omission of noise during their analysis.  

\end{enumerate}

\subsection{Critical Temperature}

The Critical Temperature serves as a criterion for distinguishing between \lq Classical\rq\ and \lq Atypical\rq\ bursts proposed by \cite{xiao2022} using FAST data of FRB 20121102A. 
Following \citet{xiao2022}, we investigate the Critical Temperature of the Arecibo data \citep{Jahns2018} in this work.
We compute the average Brightness Temperature (BT) values for each cluster, which are presented as the bottom row in Table \ref{tab1}. 
BTs of Cluster 2, 3, and 5 exceed the Critical BT of $10^{33}$ K proposed by \citet{xiao2022}.
The errors of BTs in Cluster 1, 4, and Noise are too large to determine whether their BTs exceed the Critical BT.
The average BT in Cluster 3 is significantly higher than $10^{33}$ K. 
This is probably because \citet{xiao2022} used FAST to detect fainter populations of FRBs, whereas we use Arecibo data, which include relatively brighter populations than those of FAST.

When we calculated the BT for each cluster using the average parameter values, we found that most clusters either aligned with or exceeded the Critical Temperature. 
However, it is worth noting that applying the Critical Brightness Temperature \citep{xiao2022} may not be entirely suitable for interpreting Arecibo data, given that the Critical Temperature was empirically proposed by using properties of FAST FRBs derived by a particular pulse-fitting algorithm \citep{Li2021}.
We note that, as shown in Table \ref{tab1}, the errors associated with BTs are significantly large, making it challenging to discern the distinctive BT of each cluster. The Critical Temperature criterion proposed by \citep{xiao2022} may not be a suitable approach for identifying the underlying physical mechanisms in this work.
However, classification is an important step forward in theoretically modeling FRB physical mechanisms, because it allows us to tackle the mechanisms one by one, rather than mixed mechanisms at the same time.

\subsection{Physical Interpretation of Clusters}

\textcolor{black}{The geometry of the emission region could make the distinct clusters identified in this work. For instance, the concept of radius-frequency mapping \citep[e.g.,][]{Manchester1977, Phillips1992} is broadly discussed in pulsar search, where higher-frequency radio is emitted at a shorter distance to the progenitor, corresponding to a shorter pulse duration. Clusters 1 and 5 show higher and lower frequencies with shorter and longer pulse durations, respectively (see Tab.\ref{tab1}). Therefore, Clusters 1 and 5 might have different emission radii from the progenitors. 
The pulse duration could be affected by propagation effects, including scattering. The line-broadening effect by scattering is proportional to $\nu^{-4}$ (e.g., \cite{2022ApJ...931...88C}). Because Cluster 1 shows higher frequency than Cluster 5, Cluster 1 should be less affected by the scattering effect. Therefore, scattering might make the pulse duration of Cluster 1 shorter than that of Cluster 5, making distinct clusters. }

\textcolor{black}{\citet{Li2021} found that a two-component fit was required to describe the energy distribution of FRB 20121102A, suggesting more than one radiation mechanism or emitting region. \citet{xiao2022} found double components in the distribution of brightness temperature of FRB 20121102A bursts. They suggested two different radiation mechanisms corresponding to the double components. In this context, the different clusters identified in this work might be attributed to different radiation mechanisms. There are two major scenarios of the FRB progenitor models, pulsar-like and gamma-ray burst-like (GRB-like) models \citep[e.g.,][]{2022ApJ...931...88C}.} 
 
\textcolor{black}{One of the major emission mechanisms of the pulsar-like model is curvature radiation by bunches \citep[e.g.,][]{Wang2012}. The bunches, particles that are clustered in both position and momentum spaces, slide along the magnetic field lines in a curved trajectory. This can emit coherent radio pulses, including FRBs. In general, the curvature radiation shows a broad spectrum (e.g., \cite{2018ApJ...868...31Y}), whereas all of the clusters in this work show narrow spectra confined within $<$200 MHz. Such narrow spectra could be explained by spatially separated bunches (e.g., \cite{2020ApJ...901L..13Y}). Therefore, the broader and narrower bandwidths of Cluster 1 and 3 (see Tab. \ref{tab1} and Fig. \ref{com_his}), respectively, might be due to different spatial distributions of the emitting regions.} 

\textcolor{black}{Cherenkov radiation might be another candidate for the pulsar-like model (e.g.,\cite{1999MNRAS.305..338L}). However, \citet{2018MNRAS.477.2470L} argued that they might not be favored for the FRB scenario because the required condition cannot be satisfied or the growth rate of the instability is too slow to explain FRBs. Therefore, we leave this subject for future work.}
 
\textcolor{black}{One of the major emission mechanisms of the GRB-like model is the maser radiation by external shocks (e.g., \cite{Metzger2019}). An ejecta from the central engine, e.g., magnetar, can interact with the ambient medium, invoking relativistic shocks. As the relativistic shocks propagate, particles coherently gyrate around magnetic field lines to generate coherent radio emissions, including FRBs. This scenario is characterized by a bulk Lorentz factor ($\Gamma$) of charged particles. The observed frequency corresponds to the gyration frequency boosted by $\Gamma$ (e.g., \cite{2023RvMP...95c5005Z}). The bulk Lorentz factor also governs the pulse duration which is inversely proportional to $\Gamma^2$ (e.g., \cite{2023RvMP...95c5005Z}). In this framework, the higher frequency and shorter pulse duration of Cluster 1 might be qualitatively explained by a larger $\Gamma$ value. The smaller $\Gamma$ might be the case for Cluster 5 with lower frequency and longer duration.}

\subsection{Advantage of the machine-learning approach}

The classification of \lq Classical\rq\ and \lq Atypical\rq\ bursts based solely on the BT might be an arbitrary choice.
In contrast, we simultaneously treat seven parameters, which include ones used to compute the BT.
This is where the potential advantages of ML come into play. 
ML models possess the capability to process vast amounts of data and discern complex patterns that may elude human bias. 
This could potentially lead to a more comprehensive understanding of the classification of FRBs.

Our utilization of UMAP effectively categorized FRBs into five distinct clusters, alongside noise, hinting at the possibility of multiple physical mechanisms responsible for generating FRBs, though not exclusively \citep{Jahns2018}. To mitigate the potential impact of telescope bias, we corroborated our findings with an alternative ML classification method utilizing FAST data \citep{Raquel2023}. The striking alignment between these two approaches provides intriguing insights for future investigations.



While machine learning can significantly reduce human bias in the analysis process, a complete elimination of human bias remains challenging when interpreting and comprehending the results. Nevertheless, machine learning methods tend to introduce far less human bias compared to traditional manual analysis techniques.


\section{CONCLUSIONS}\label{sec6}
    With the above underpinnings, this paper concludes the following: 
    \begin{itemize}
        \item Using machine learning classification methods, we identified five clusters among the seven parameters. Each cluster exhibits distinct characteristics in histograms and parameter colouring, which might suggest the existence of multiple mechanisms of FRB emissions. 
        \textcolor{black}{The geometry of the emission region and the propagation effect of FRB signals could also make such distinct clusters.}
        \item With parameter colouring, we have determined the invariant properties of each cluster regardless of the {\ttfamily n\_neighbors} value, which demonstrates that describing FRB subtypes without relying on the {\ttfamily n\_neighbors} setting facilitates comparison with other studies aimed at classifying FRBs.
        \item Classifying and confirming the actual physical mechanisms of the clusters in this work are challenging. Consequently, the Critical Temperature criterion may not be applicable to this work.
        \item Nevertheless, a certain degree of agreement with other results (e.g., being able to recover the FRB classification used by \citealt{Raquel2023}) exhibits consistency and foundation on physical parameters of the clusters.
    \end{itemize}

Looking ahead, we expect even more promising outcomes in the future, thanks to enhanced telescope capabilities provided by future projects like the Square Kilometre Array (SKA) \citep[e.g.,][]{dewdney2009, hashimoto2020} and the Bustling Universe Radio Survey Telescope in Taiwan (BURSTT) \citep[e.g.,][]{lin2022, ho2023}. 
We maintain optimism that these advancements will unveil the enigmatic nature of FRBs. This research serves as a benchmark for future FRB classifications, particularly as dedicated radio telescopes like SKA and BURSTT continue to detect a growing number of FRBs.

\section{Acknowledgments}
\textcolor{black}{
We appreciate the referee's insightful comments, which have greatly enhanced the quality of the manuscript. 
}
LL acknowledges the Taiwan Astronomical ObserVatory Alliance (TAOvA)  grant NSTC~111-2740-M-008-003 for the summer student internship in partial financial support of this research. 
The authors would like to express our gratitude to our collaborators in the NTHU \& NCHU Cosmology Group, including Yu-Wei Lin, Tzu-Yin Hsu, Poya Wang, Shotaro Yamasaki and many others for their invaluable contributions and support throughout this project.
TG acknowledges the support of the National Science and Technology Council of Taiwan through grants 108-2628-M-007-004-MY3, 110-2112-M-005-013-MY3, 112-2112-M-007-013, and 112-2123-M-001-004-. 
TH acknowledges the support of the National Science and Technology Council of Taiwan through grants 110-2112-M-005-013-MY3, 110-2112-M-007-034-, 113-2112-M-005-009-MY3, and 112-2123-M-001-004-. 
SH acknowledges the support of The Australian Research Council Centre of Excellence for Gravitational Wave Discovery (OzGrav) and the Australian Research Council Centre of Excellence for All Sky Astrophysics in 3 Dimensions (ASTRO 3D), through project number CE17010000 and CE170100013, respectively.
This work is based on observations made with the Arecibo Telescope. The Arecibo Observatory is a facility of the National Science Foundation operated under a cooperative agreement by the University of Central Florida and in alliance with Universidad Ana G. Mendez, and Yang Enterprises.

\section{Data availability}
The data underlying this article is available in the work of \citep{Jahns2018}. 
The dataset was derived from \url{https://academic.oup.com/view-large/389045964} and \url{https://academic.oup.com/view-large/389045965}. Other data described in this article will be shared upon reasonable request to the corresponding author.

\printbibliography


\appendix
\section{Testing n\_neighbors parameters}
Here, we present embedding, clustering, and colouring results with different assumptions on {\ttfamily n\_neighbors} values.

         \begin{figure*}
                \centering
                \begin{multicols}{2}
                    \subcaptionbox{\label{test5}}{\includegraphics[width=1.1\linewidth, height=1.1\columnwidth]{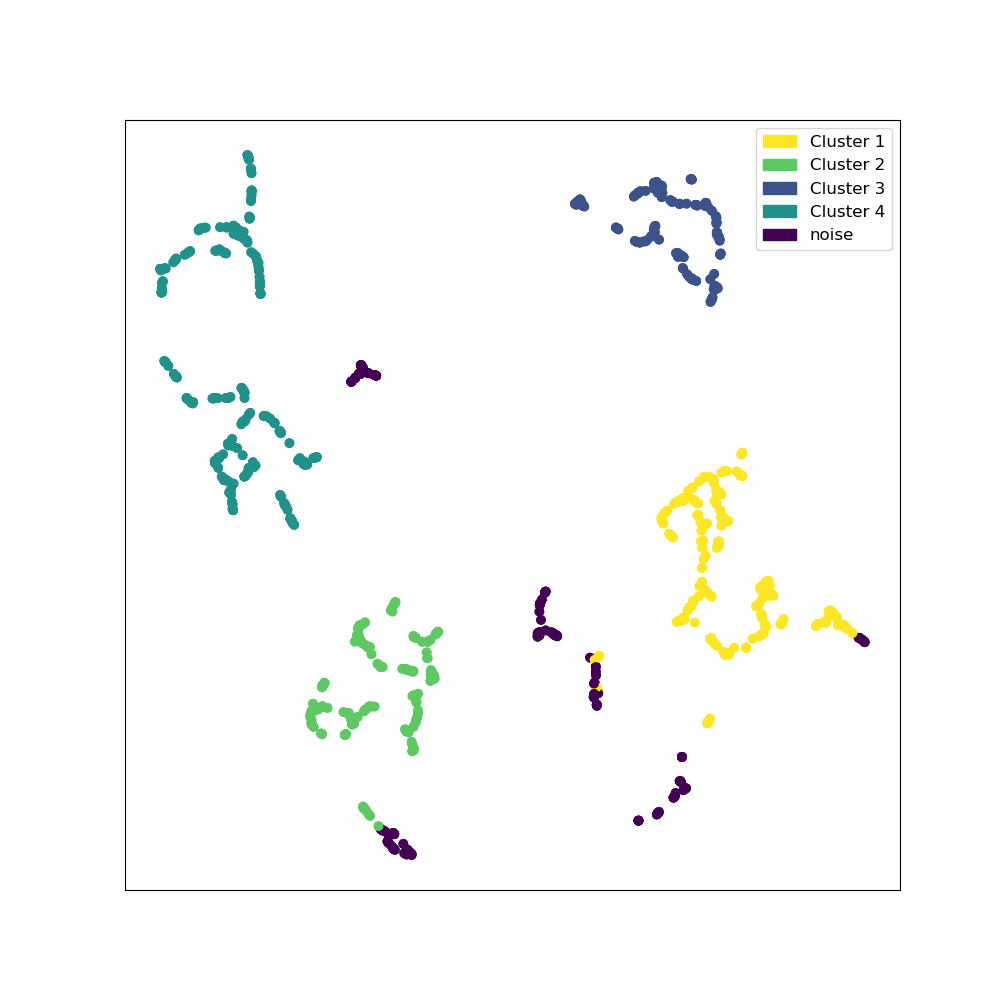}}\par 
                    \subcaptionbox{\label{test6}}{\includegraphics[width=1.1\linewidth, height=1.1\columnwidth]{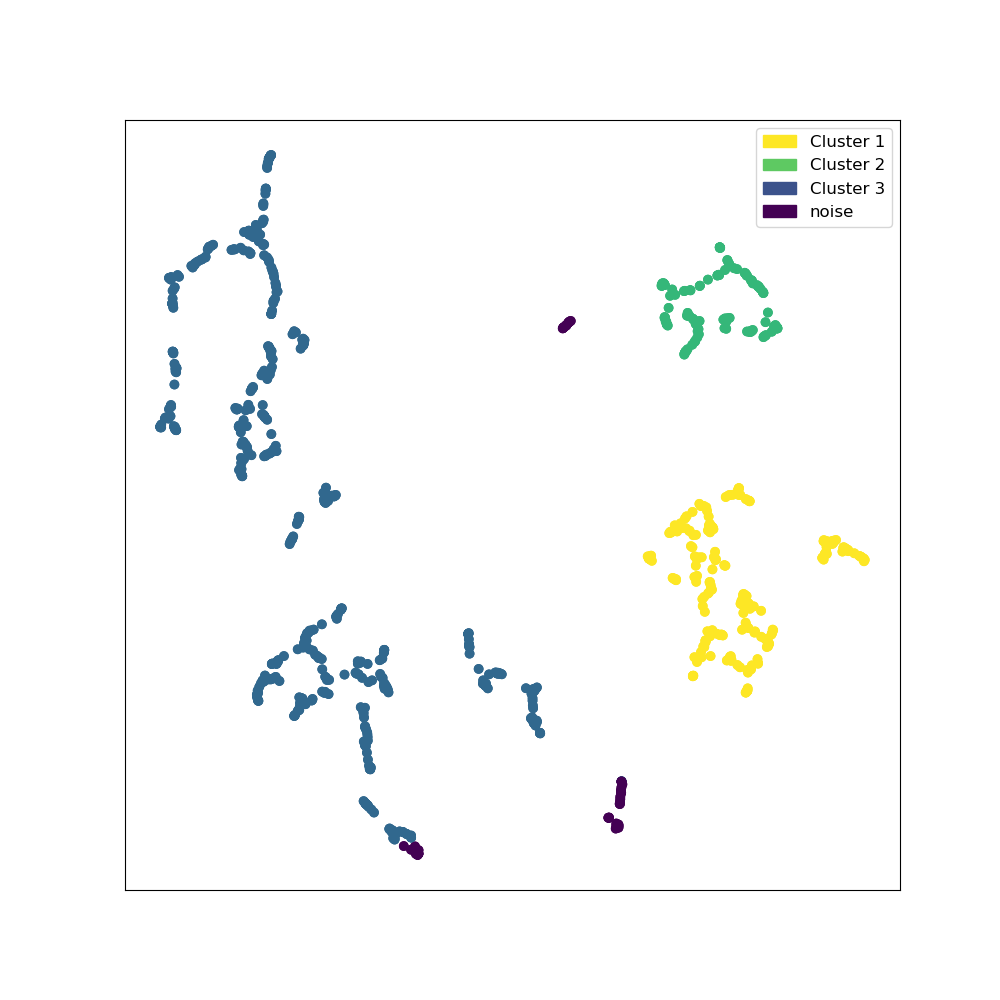}}\par 
                \end{multicols}
                \begin{multicols}{2}
                    \subcaptionbox{\label{test7}}{\includegraphics[width=1.1\linewidth, height=1.1\columnwidth]{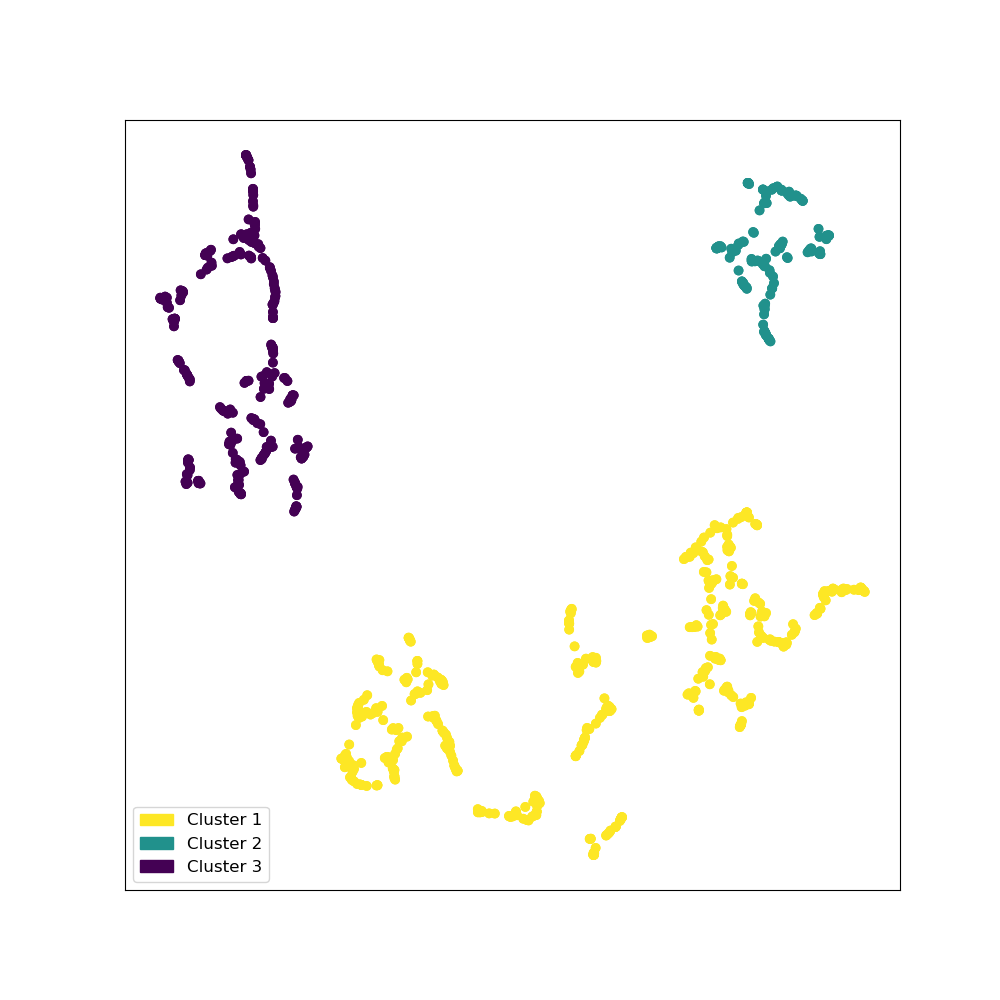}}\par
                    \subcaptionbox{\label{test9}}{\includegraphics[width=1.1\linewidth, height=1.1\columnwidth]{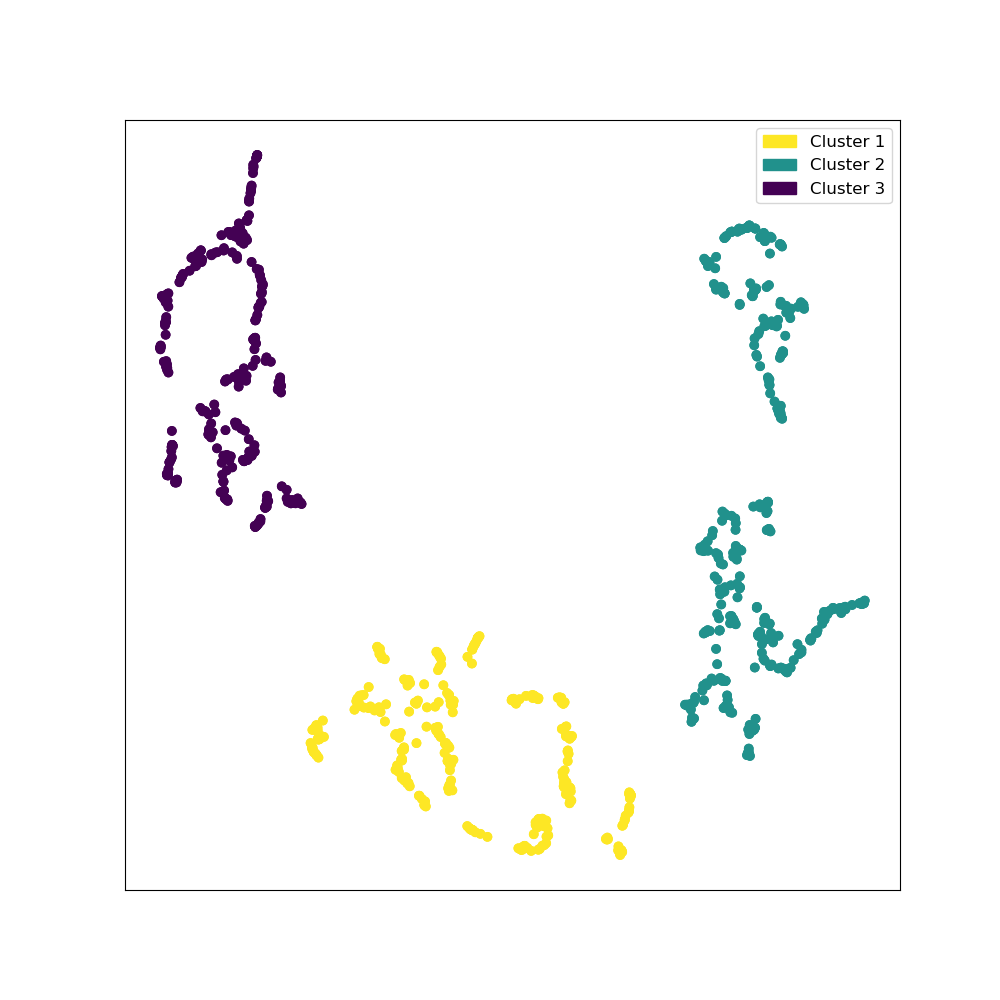}}\par
                \end{multicols}
                \caption{HDBSCAN Clustering result for Fig. \ref{test5} {\ttfamily n\_neighbors = 5}, Fig. \ref{test6} {\ttfamily n\_neighbors = 6}, Fig. \ref{test7} {\ttfamily n\_neighbors = 7}, and Fig. \ref{test9} {\ttfamily n\_neighbors = 9}. 
                }
            \label{nn9parcol2}
            \end{figure*}

         \begin{figure*}
                \centering
                \begin{multicols}{2}
                    \subcaptionbox{\label{a5}}{\includegraphics[width=1\linewidth, height=0.8\columnwidth]{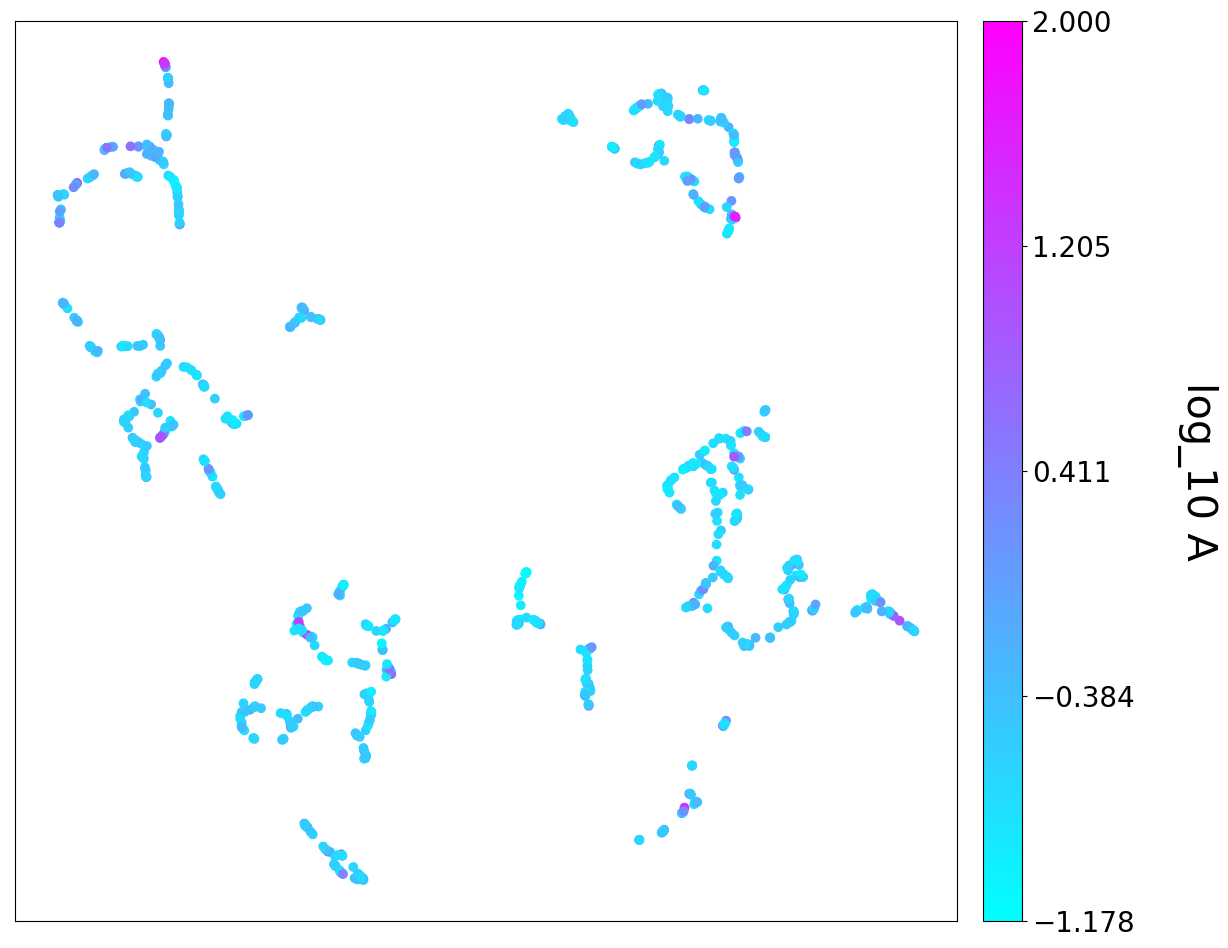}}\par 
                    \subcaptionbox{\label{a6}}{\includegraphics[width=1\linewidth, height=0.8\columnwidth]{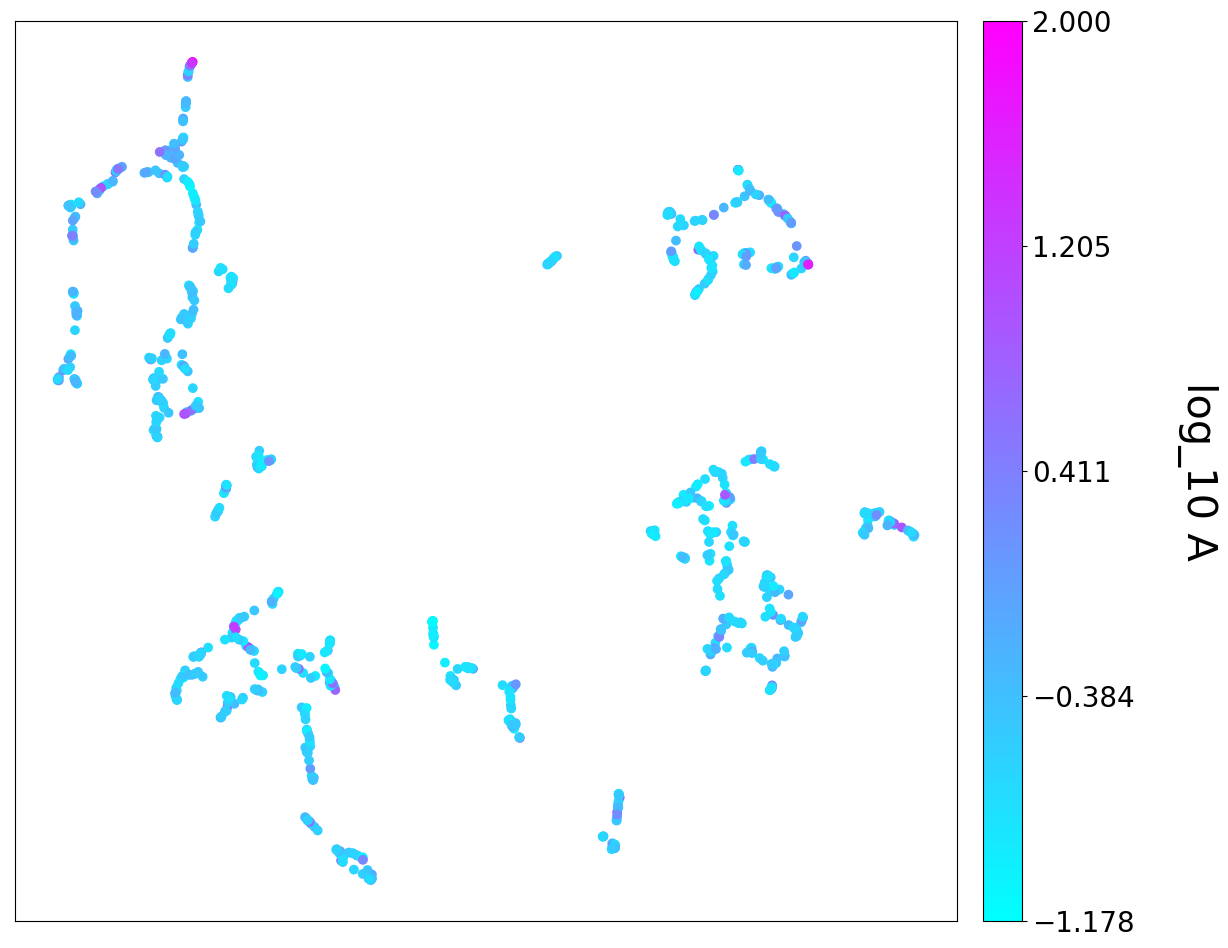}}\par 
                \end{multicols}
                \begin{multicols}{2}
                    \subcaptionbox{\label{a7}}{\includegraphics[width=1\linewidth, height=0.8\columnwidth]{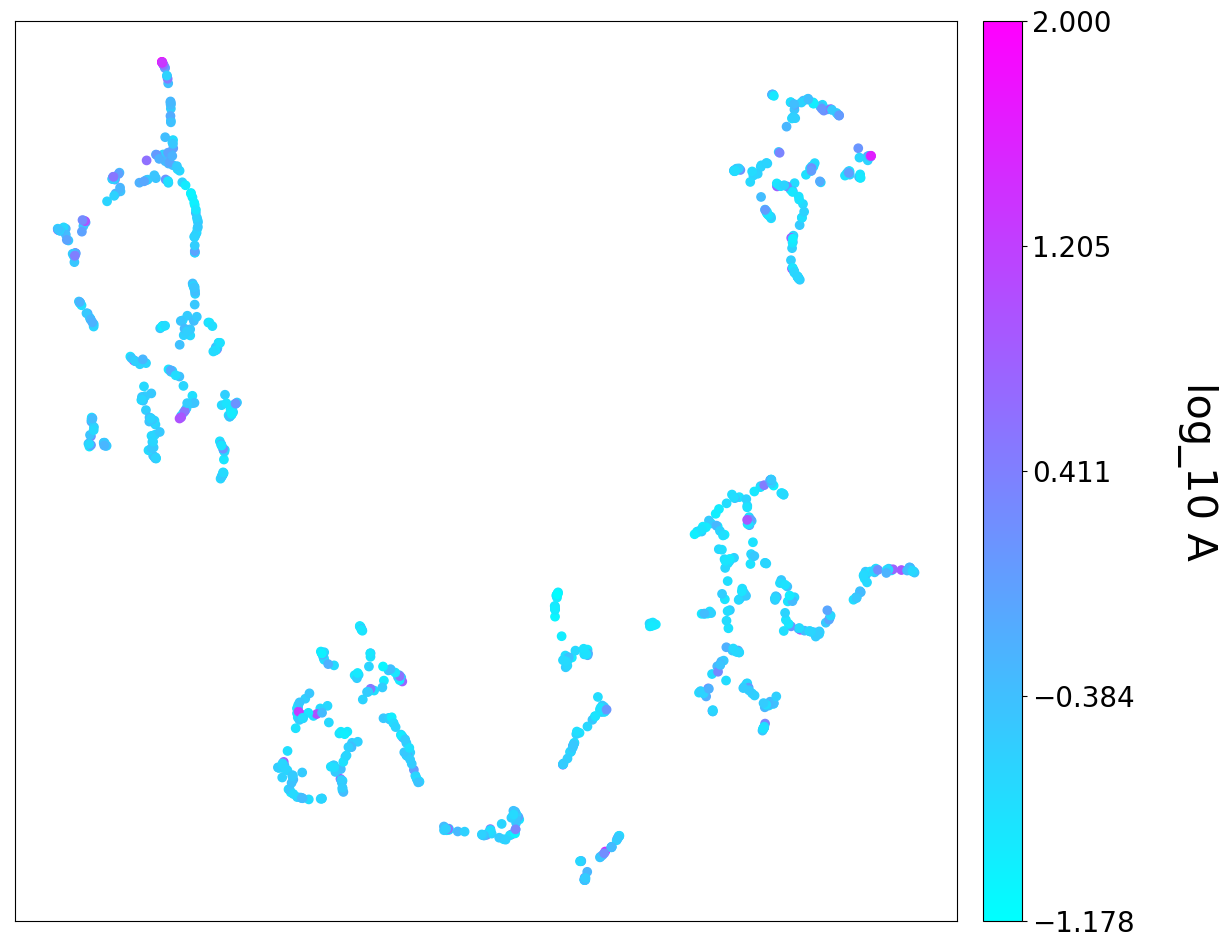}}\par
                    \subcaptionbox{\label{a9}}{\includegraphics[width=1\linewidth, height=0.8\columnwidth]{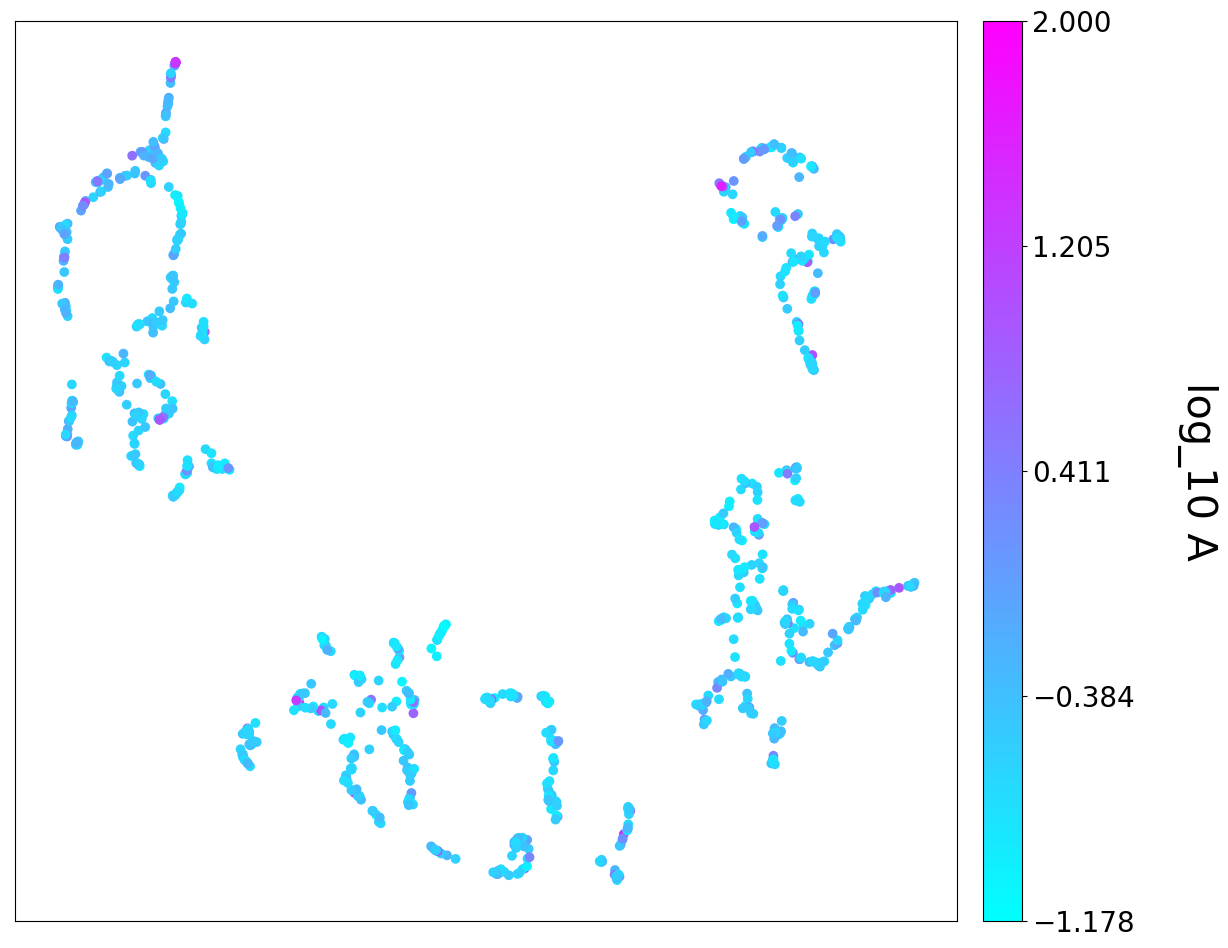}}\par
                \end{multicols}
                \caption{Amplitude colouring of the clustering results for Fig. \ref{a5} {\ttfamily n\_neighbors = 5}, Fig. \ref{a6} {\ttfamily n\_neighbors = 6}, Fig. \ref{a7} {\ttfamily n\_neighbors = 7}, and Fig. \ref{a9} {\ttfamily n\_neighbors = 9}. 
                }
            \label{A_3}
            \end{figure*}
         \begin{figure*}
                \centering
                \begin{multicols}{2}
                    \subcaptionbox{\label{b5}}{\includegraphics[width=1\linewidth, height=0.8\columnwidth]{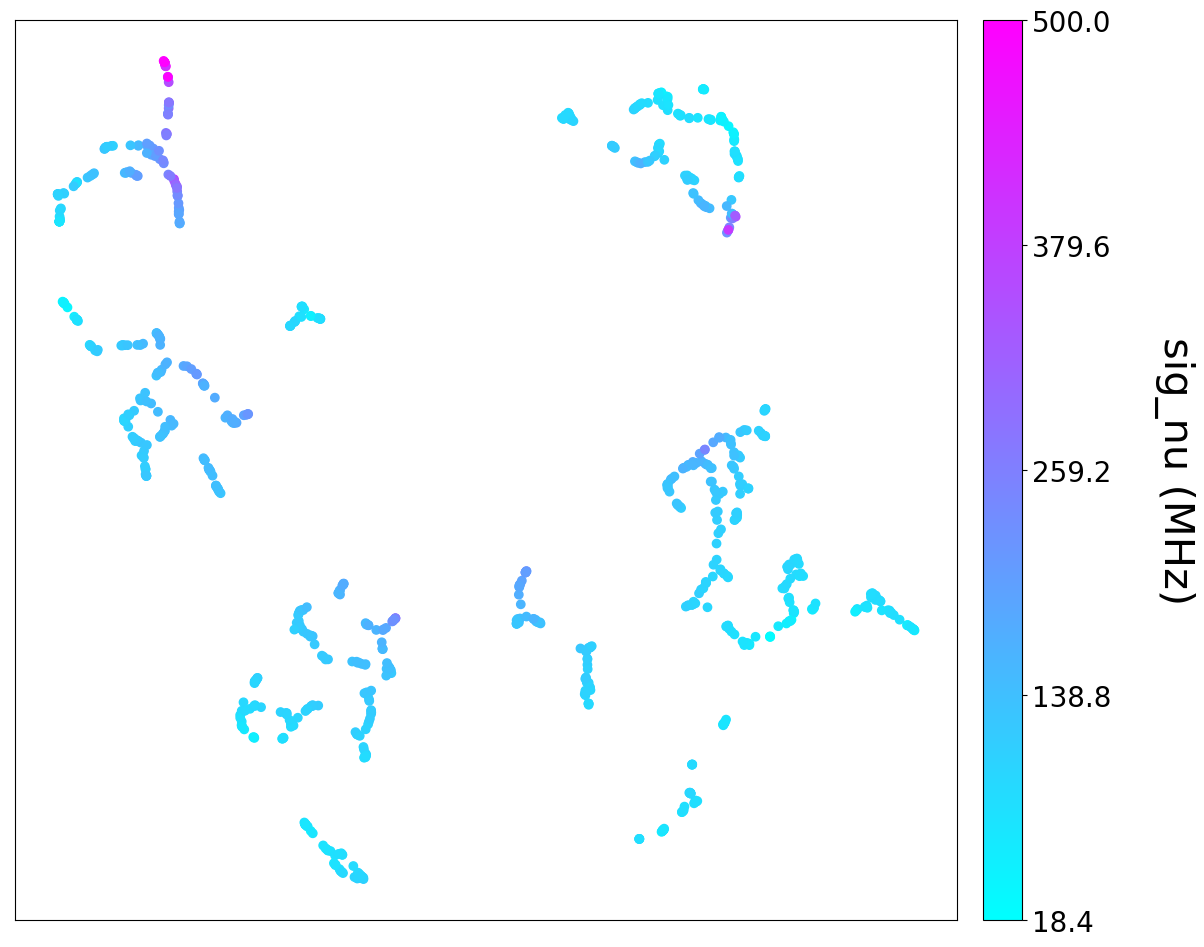}}\par 
                    \subcaptionbox{\label{b6}}{\includegraphics[width=1\linewidth, height=0.8\columnwidth]{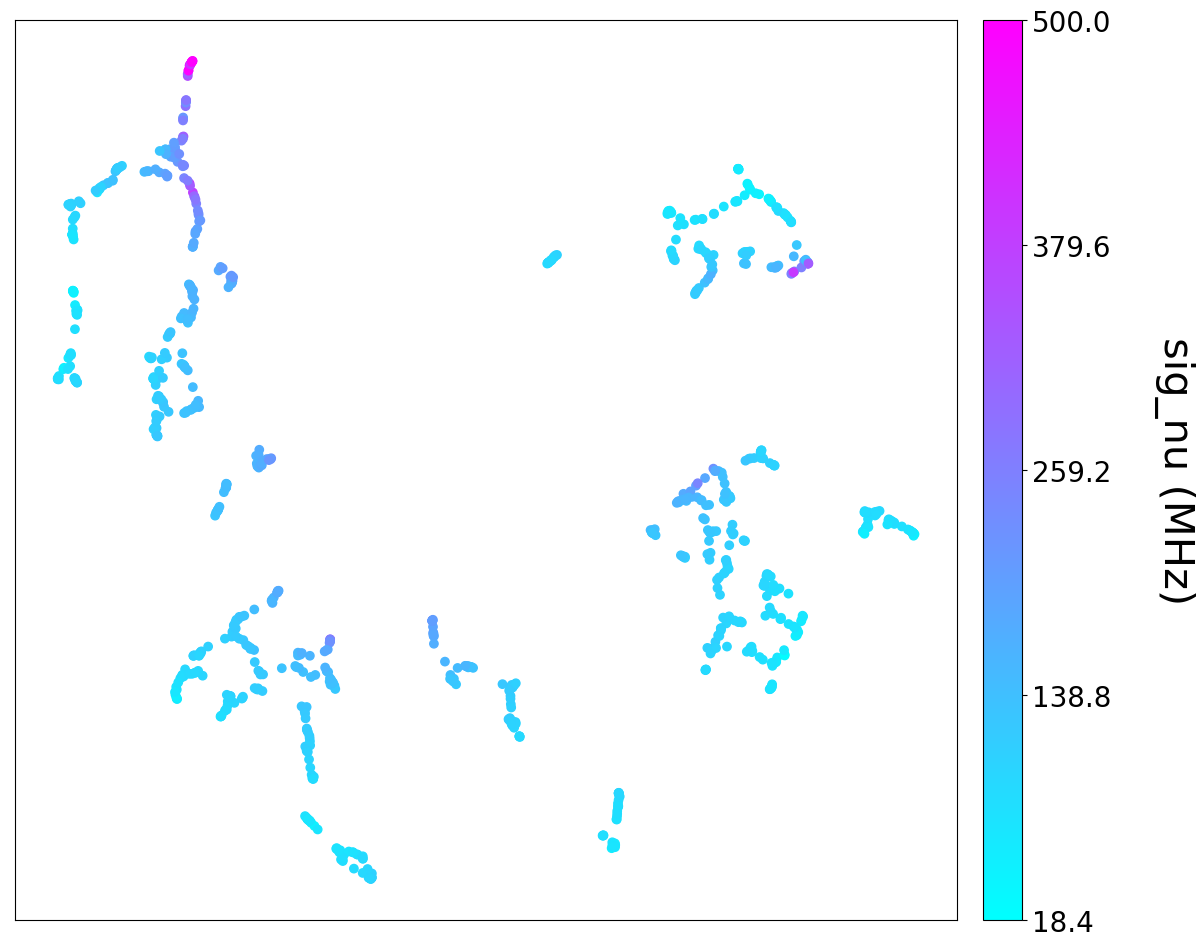}}\par 
                \end{multicols}
                \begin{multicols}{2}
                    \subcaptionbox{\label{b7}}{\includegraphics[width=1\linewidth, height=0.8\columnwidth]{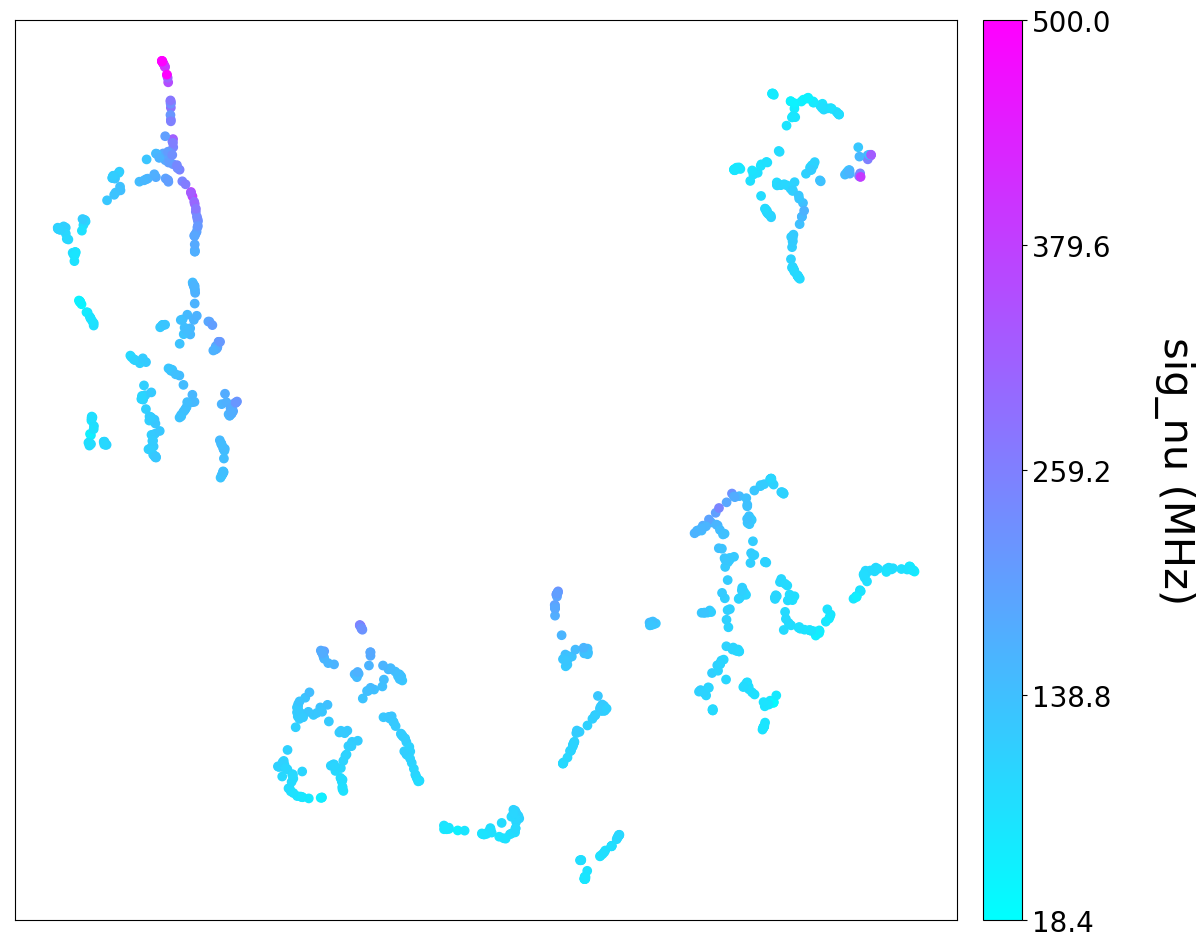}}\par
                    \subcaptionbox{\label{b9}}{\includegraphics[width=1\linewidth, height=0.8\columnwidth]{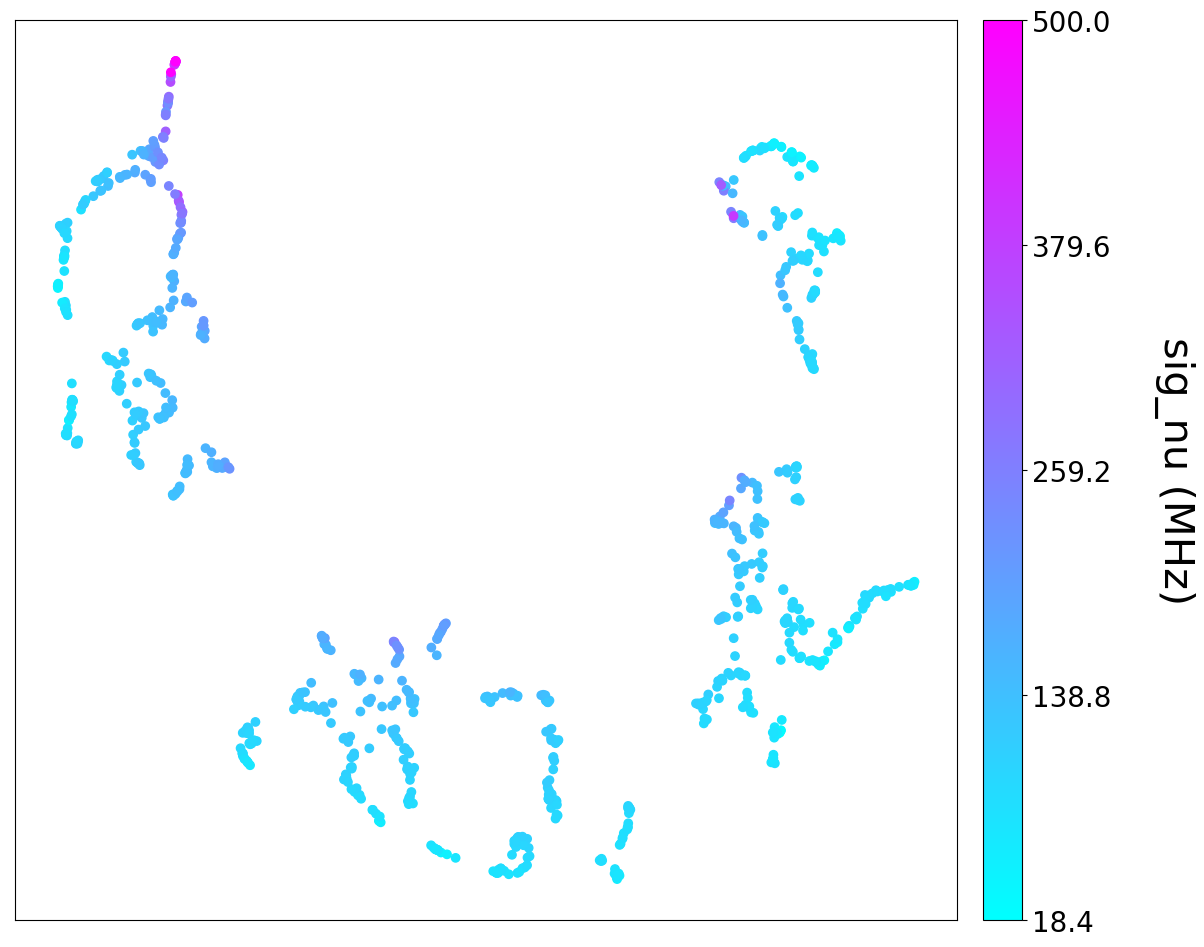}}\par
                \end{multicols}
                \caption{Bandwidth colouring of the clustering results for Fig. \ref{b5} {\ttfamily n\_neighbors = 5}, Fig. \ref{b6} {\ttfamily n\_neighbors = 6}, Fig. \ref{b7} {\ttfamily n\_neighbors = 7}, and Fig. \ref{b9} {\ttfamily n\_neighbors = 9}. 
                }
            \label{A_4}
            \end{figure*}

         \begin{figure*}
                \centering
                \begin{multicols}{2}
                    \subcaptionbox{\label{c5}}{\includegraphics[width=1\linewidth, height=0.8\columnwidth]{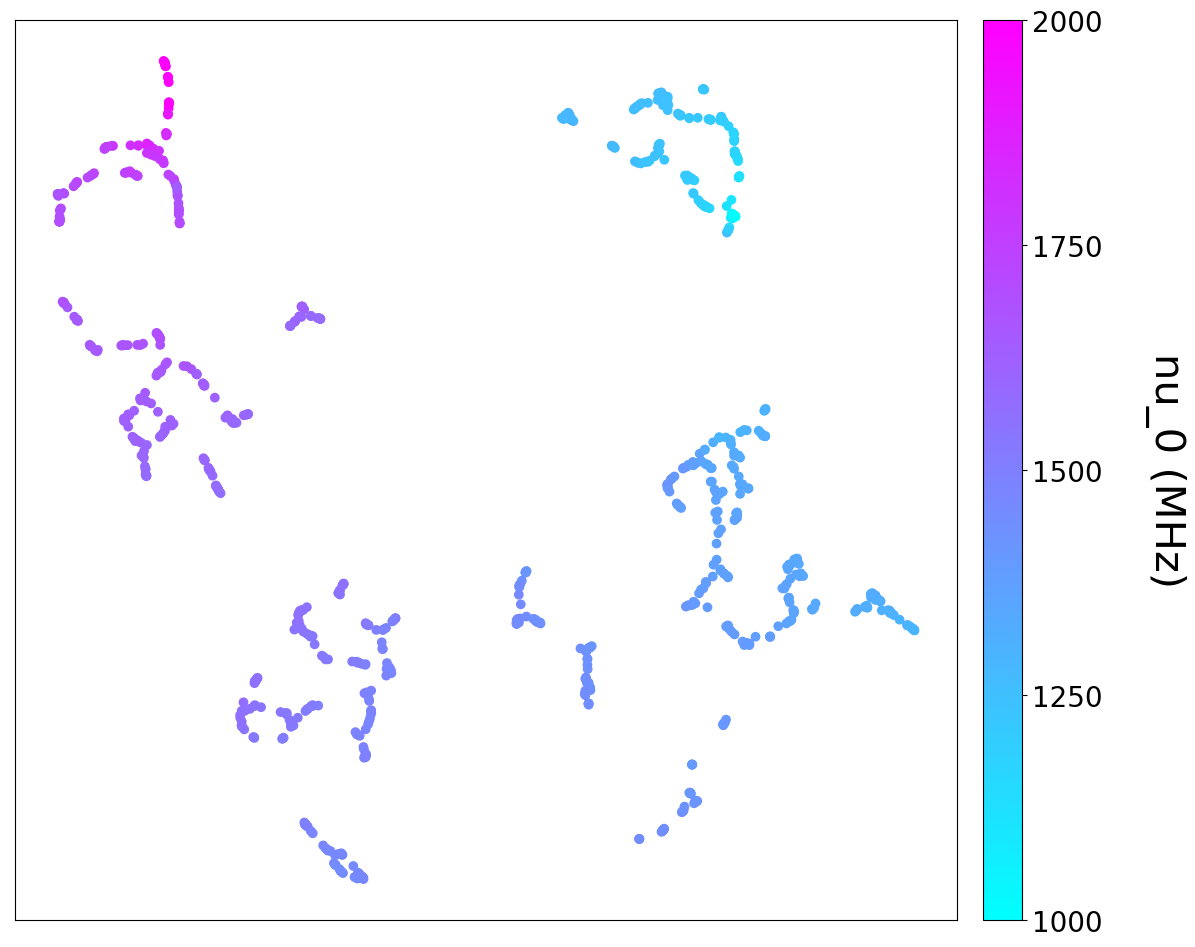}}\par 
                    \subcaptionbox{\label{c6}}{\includegraphics[width=1\linewidth, height=0.8\columnwidth]{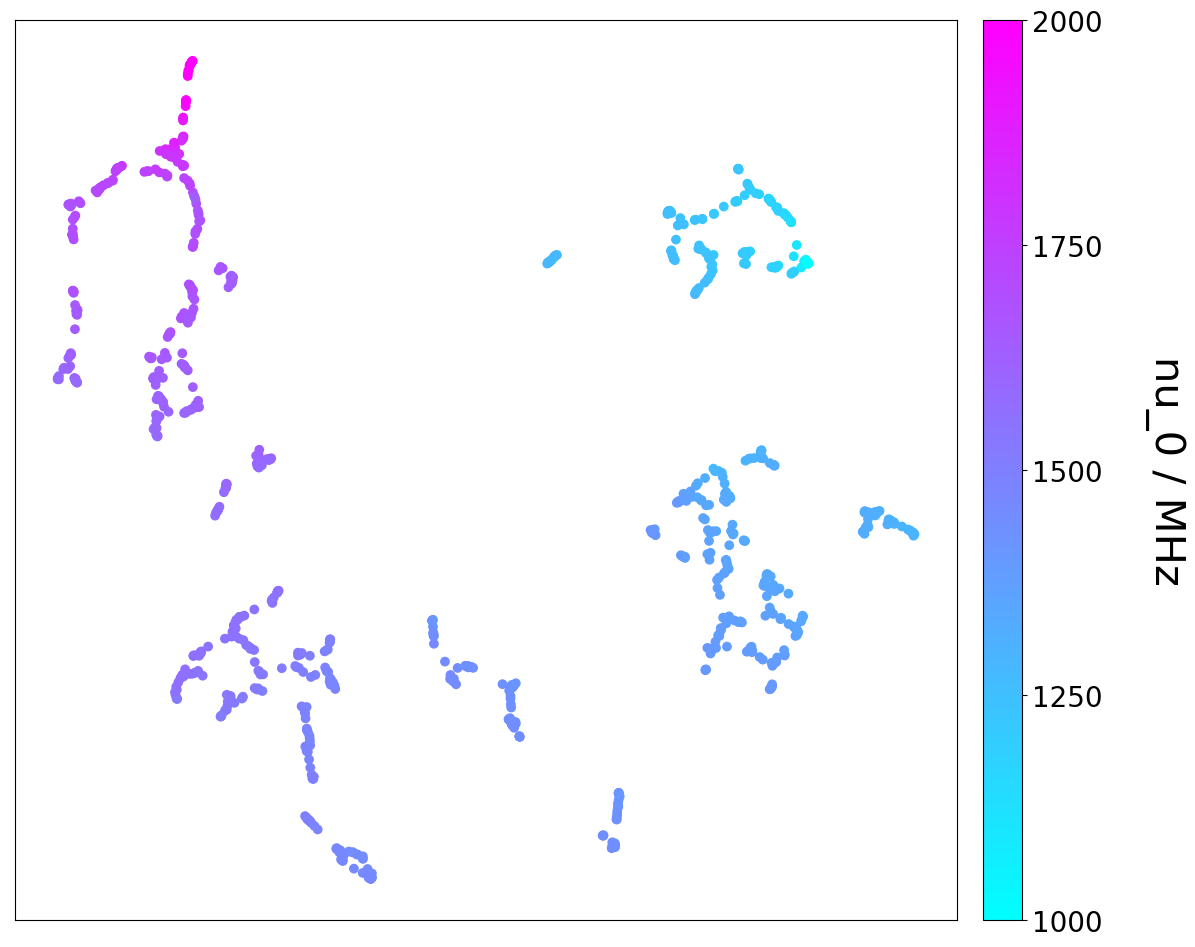}}\par 
                \end{multicols}
                \begin{multicols}{2}
                    \subcaptionbox{\label{c7}}{\includegraphics[width=1\linewidth, height=0.8\columnwidth]{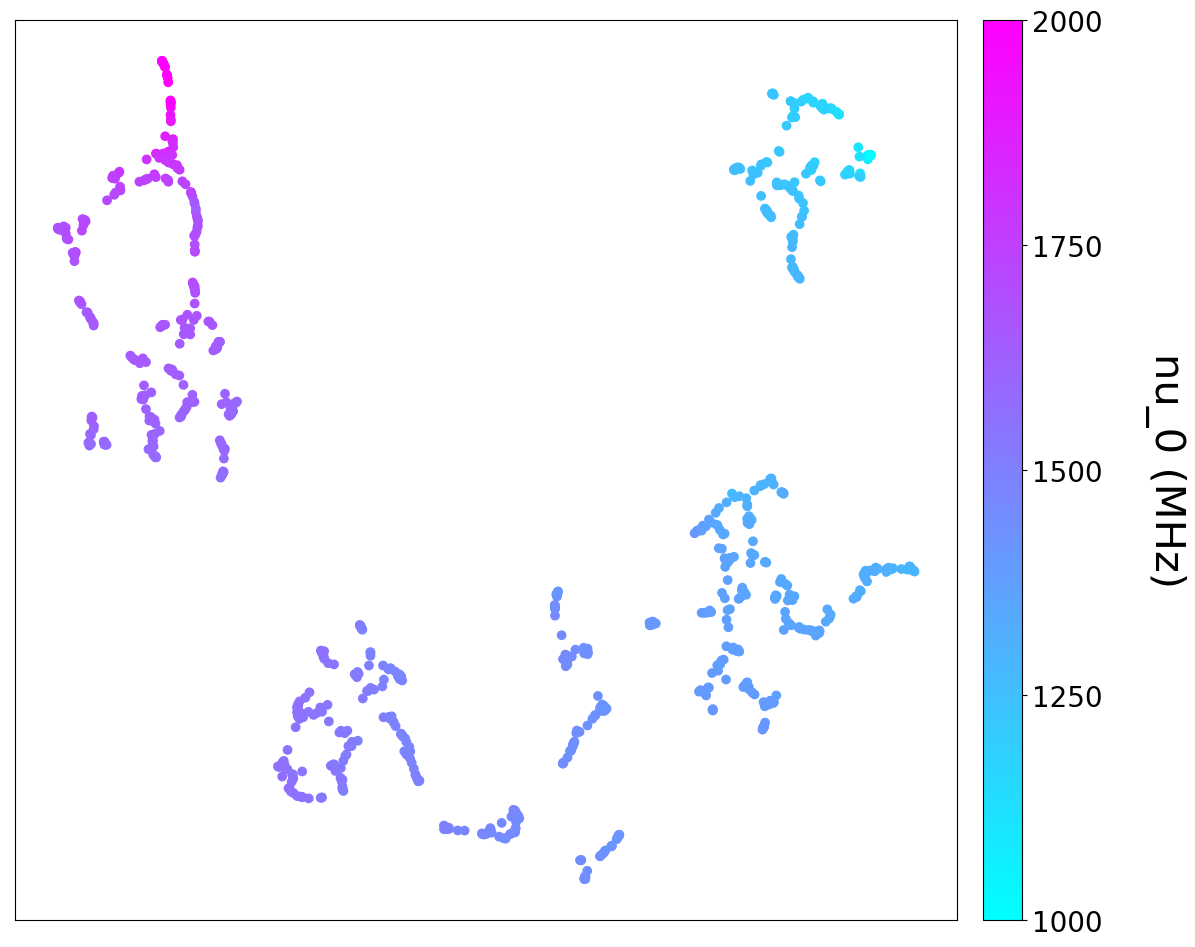}}\par
                    \subcaptionbox{\label{c9}}{\includegraphics[width=1\linewidth, height=0.8\columnwidth]{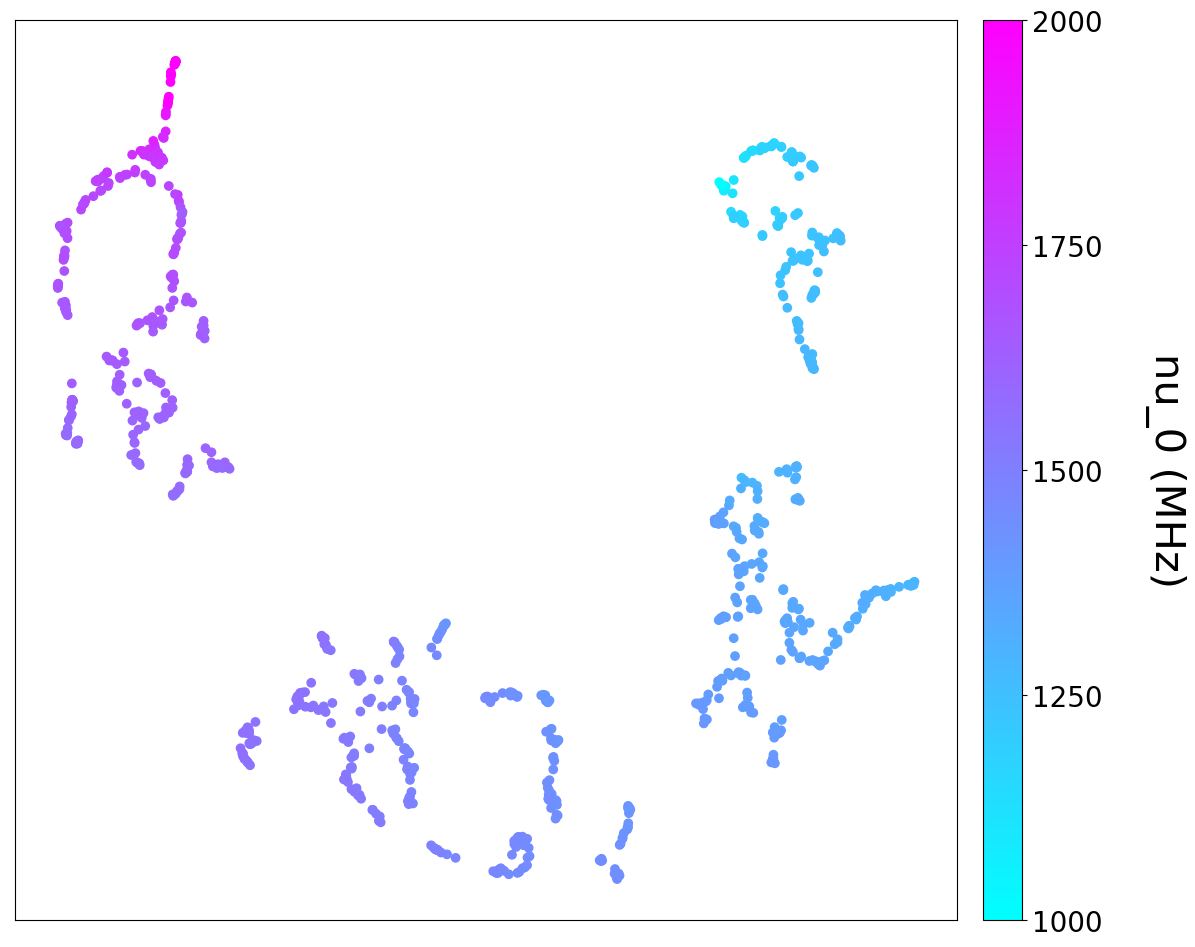}}\par
                \end{multicols}
                \caption{Central Frequency colouring of the clustering results for Fig. \ref{c5} {\ttfamily n\_neighbors = 5}, Fig. \ref{c6} {\ttfamily n\_neighbors = 6}, Fig. \ref{c7} {\ttfamily n\_neighbors = 7}, and Fig. \ref{c9} {\ttfamily n\_neighbors = 9}. 
                }
            \label{A_5}
            \end{figure*}

         \begin{figure*}
                \centering
                \begin{multicols}{2}
                    \subcaptionbox{\label{d5}}{\includegraphics[width=1\linewidth, height=0.8\columnwidth]{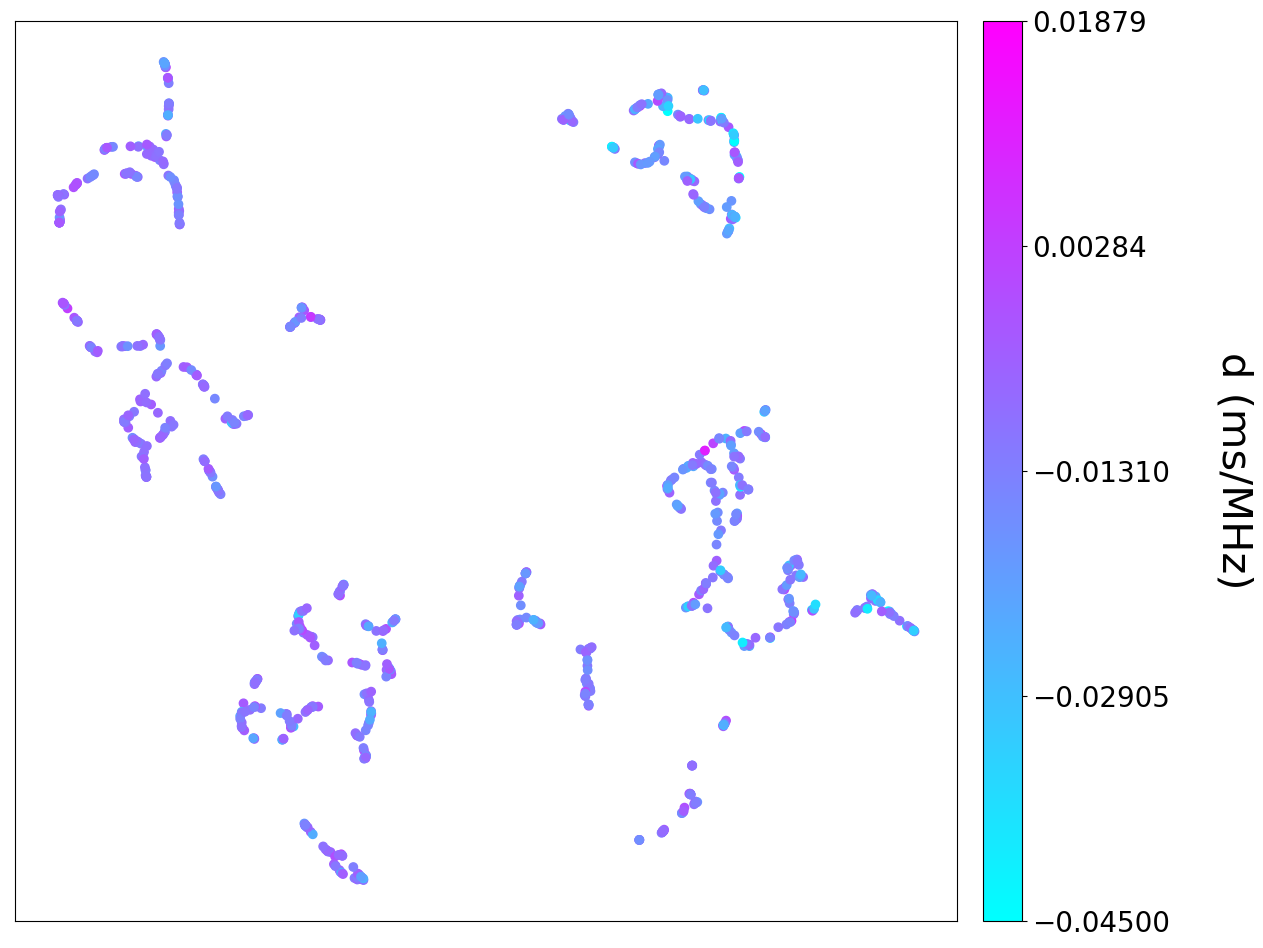}}\par 
                    \subcaptionbox{\label{d6}}{\includegraphics[width=1\linewidth, height=0.8\columnwidth]{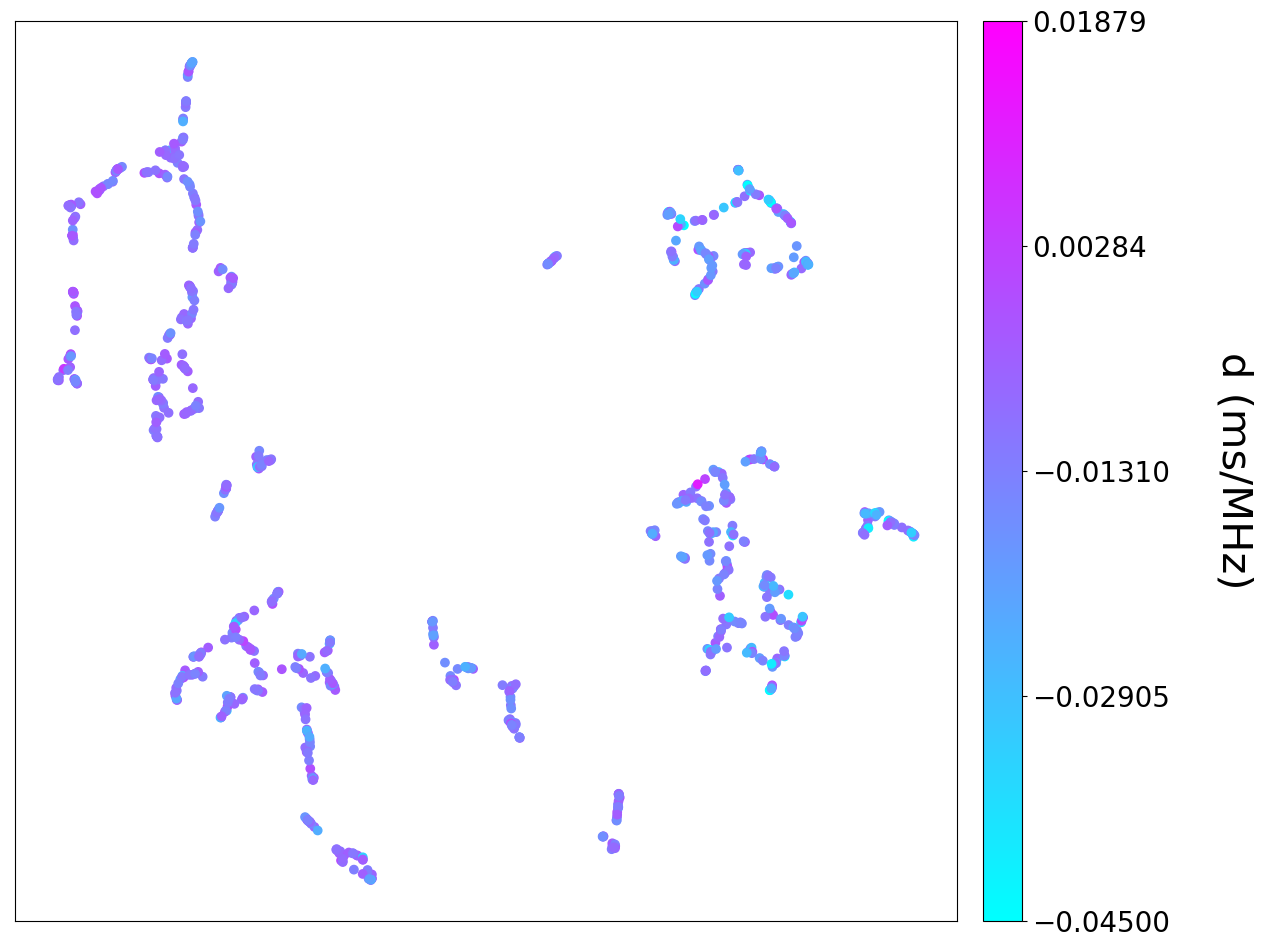}}\par 
                \end{multicols}
                \begin{multicols}{2}
                    \subcaptionbox{\label{d7}}{\includegraphics[width=1\linewidth, height=0.8\columnwidth]{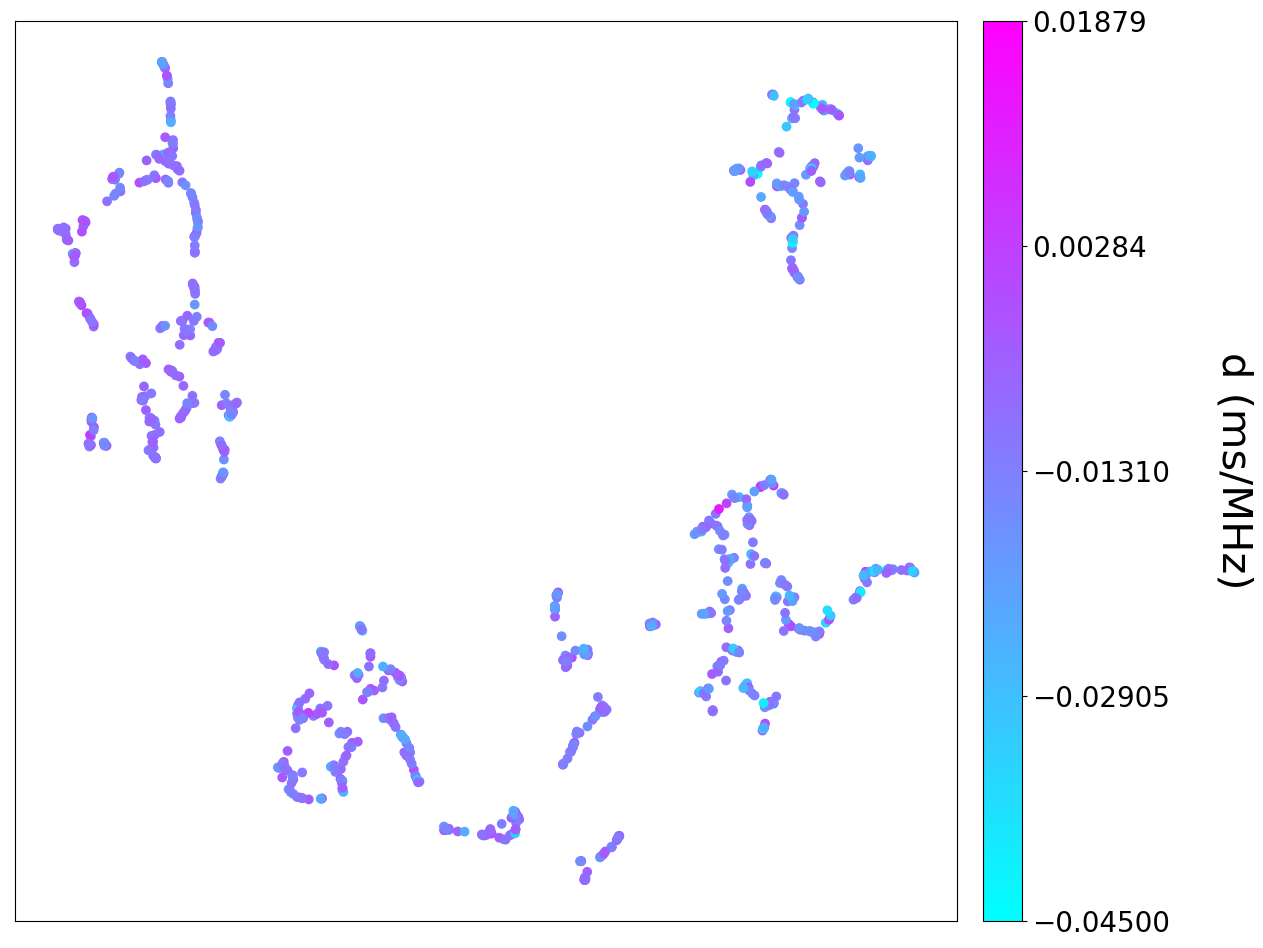}}\par
                    \subcaptionbox{\label{d9}}{\includegraphics[width=1\linewidth, height=0.8\columnwidth]{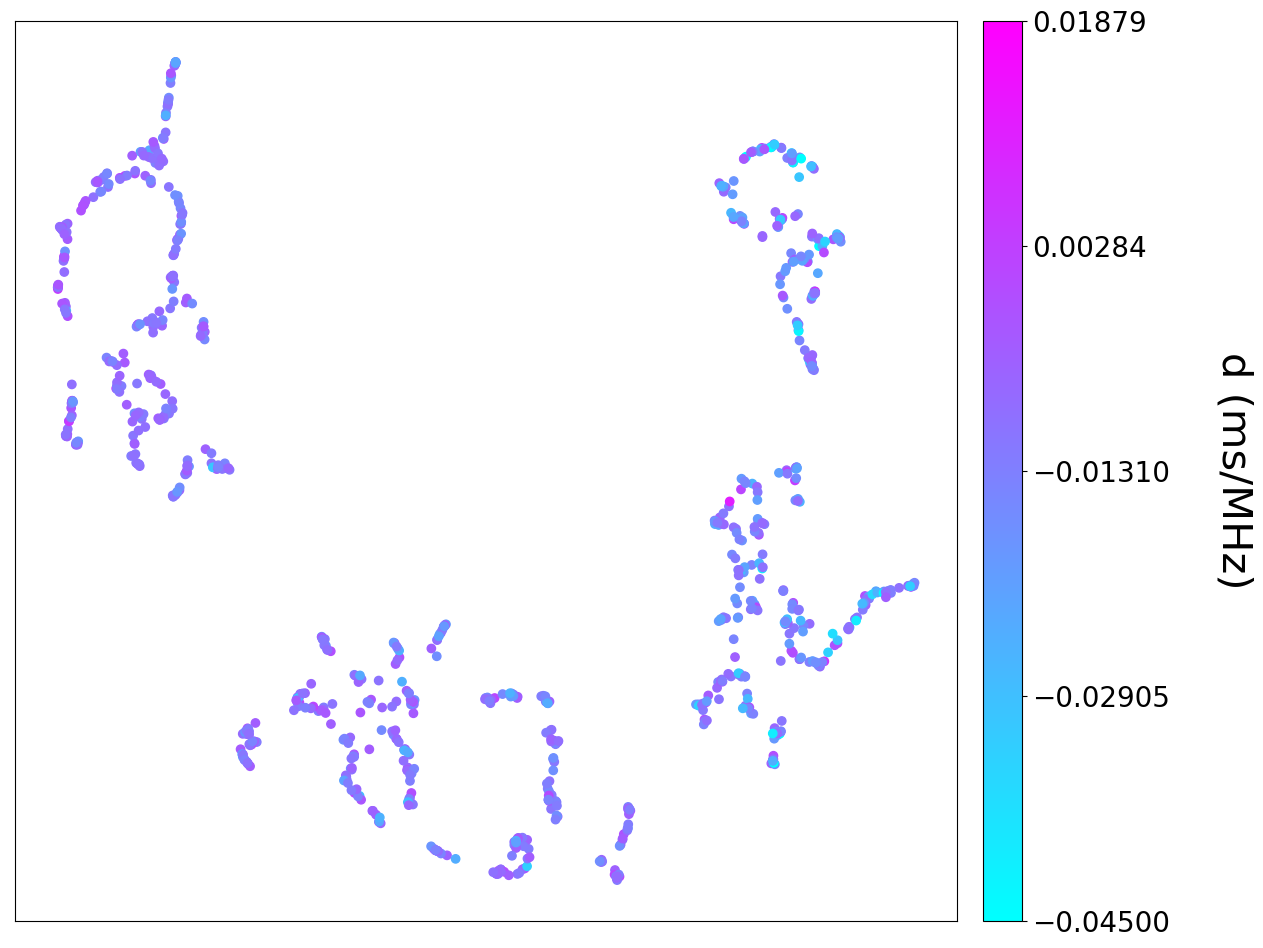}}\par
                \end{multicols}
                \caption{Linear Temporal Drift colouring of the clustering results for Fig. \ref{d5} {\ttfamily n\_neighbors = 5}, Fig. \ref{d6} {\ttfamily n\_neighbors = 6}, Fig. \ref{d7} {\ttfamily n\_neighbors = 7}, and Fig. \ref{d9} {\ttfamily n\_neighbors = 9}. 
                }
            \label{A_6}
            \end{figure*}
         \begin{figure*}
                \centering
                \begin{multicols}{2}
                    \subcaptionbox{\label{f5}}{\includegraphics[width=1\linewidth, height=0.8\columnwidth]{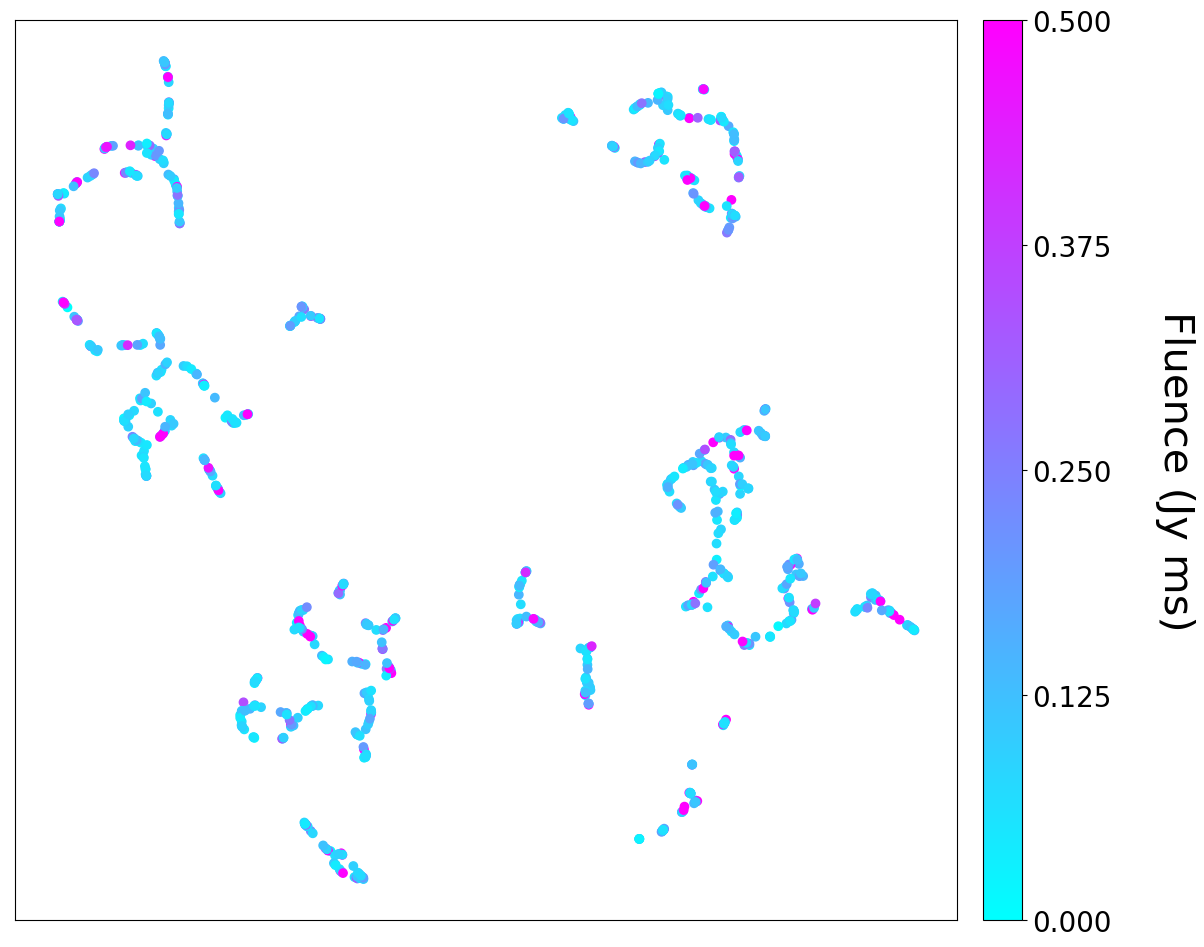}}\par 
                    \subcaptionbox{\label{f6}}{\includegraphics[width=1\linewidth, height=0.8\columnwidth]{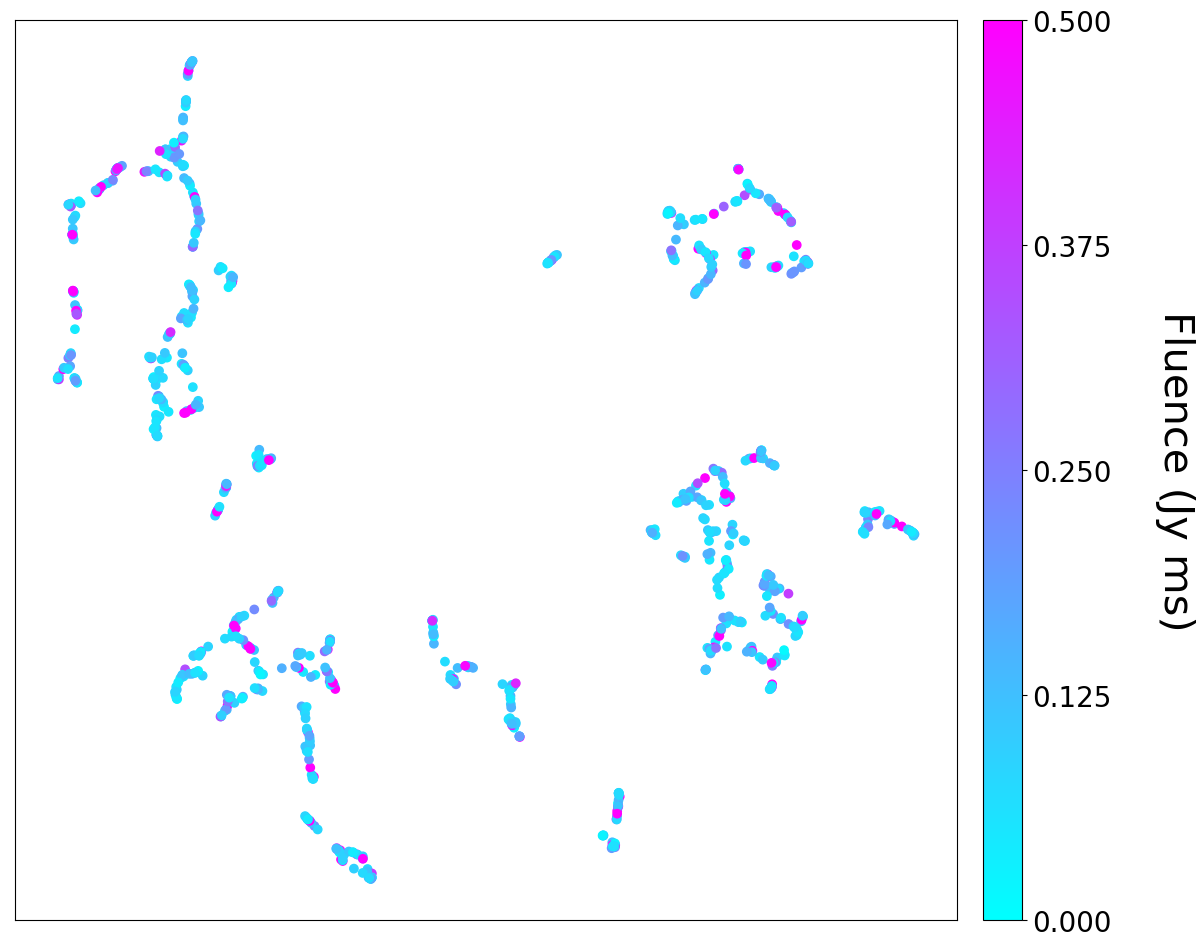}}\par 
                \end{multicols}
                \begin{multicols}{2}
                    \subcaptionbox{\label{f7}}{\includegraphics[width=1\linewidth, height=0.8\columnwidth]{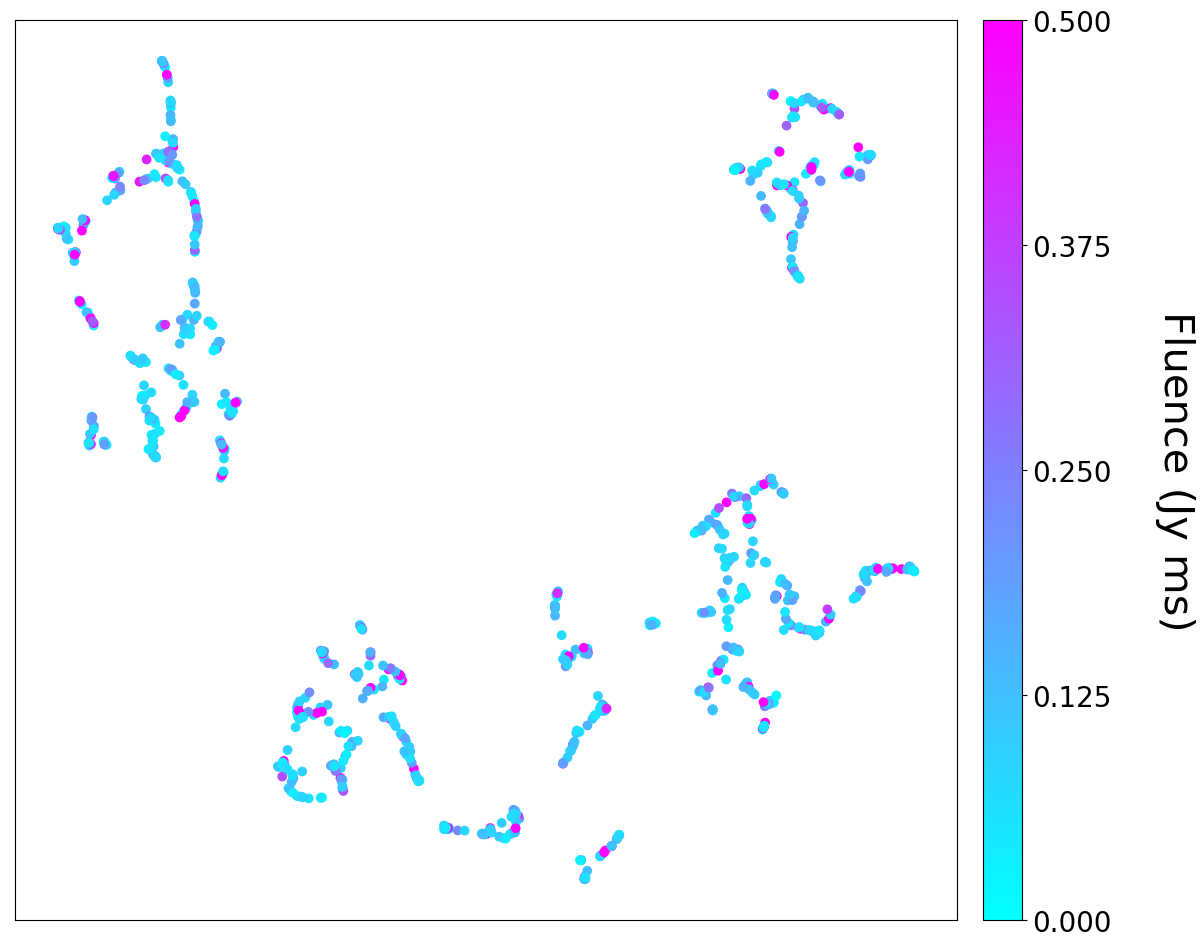}}\par
                    \subcaptionbox{\label{f9}}{\includegraphics[width=1\linewidth, height=0.8\columnwidth]{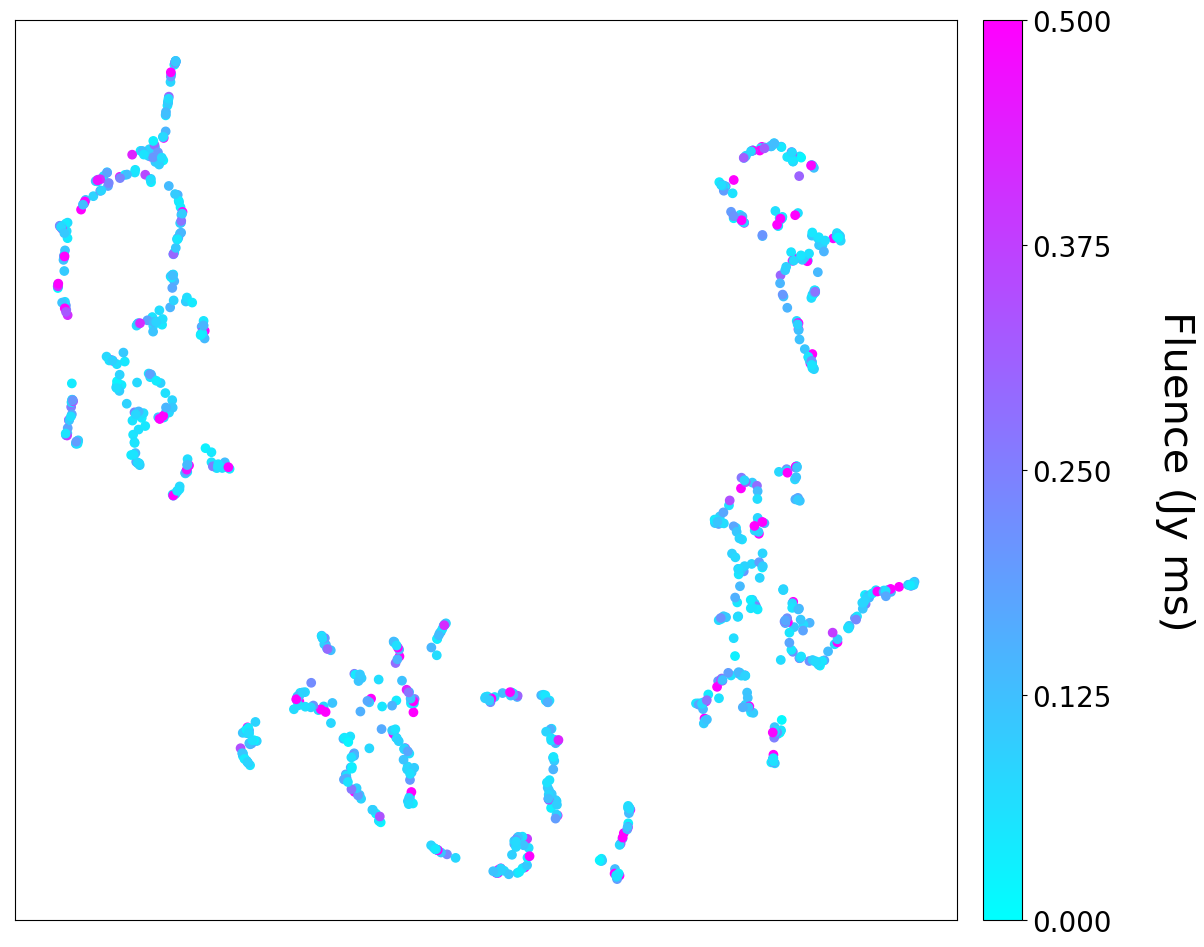}}\par
                \end{multicols}
                \caption{Fluence colouring of the clustering results for Fig. \ref{f5} {\ttfamily n\_neighbors = 5}, Fig. \ref{f6} {\ttfamily n\_neighbors = 6}, Fig. \ref{f7} {\ttfamily n\_neighbors = 7}, and Fig. \ref{f9} {\ttfamily n\_neighbors = 9}. 
                }
            \label{A_7}
            \end{figure*}
            
         \begin{figure*}
                \centering
                \begin{multicols}{2}
                    \subcaptionbox{\label{s5}}{\includegraphics[width=1\linewidth, height=0.8\columnwidth]{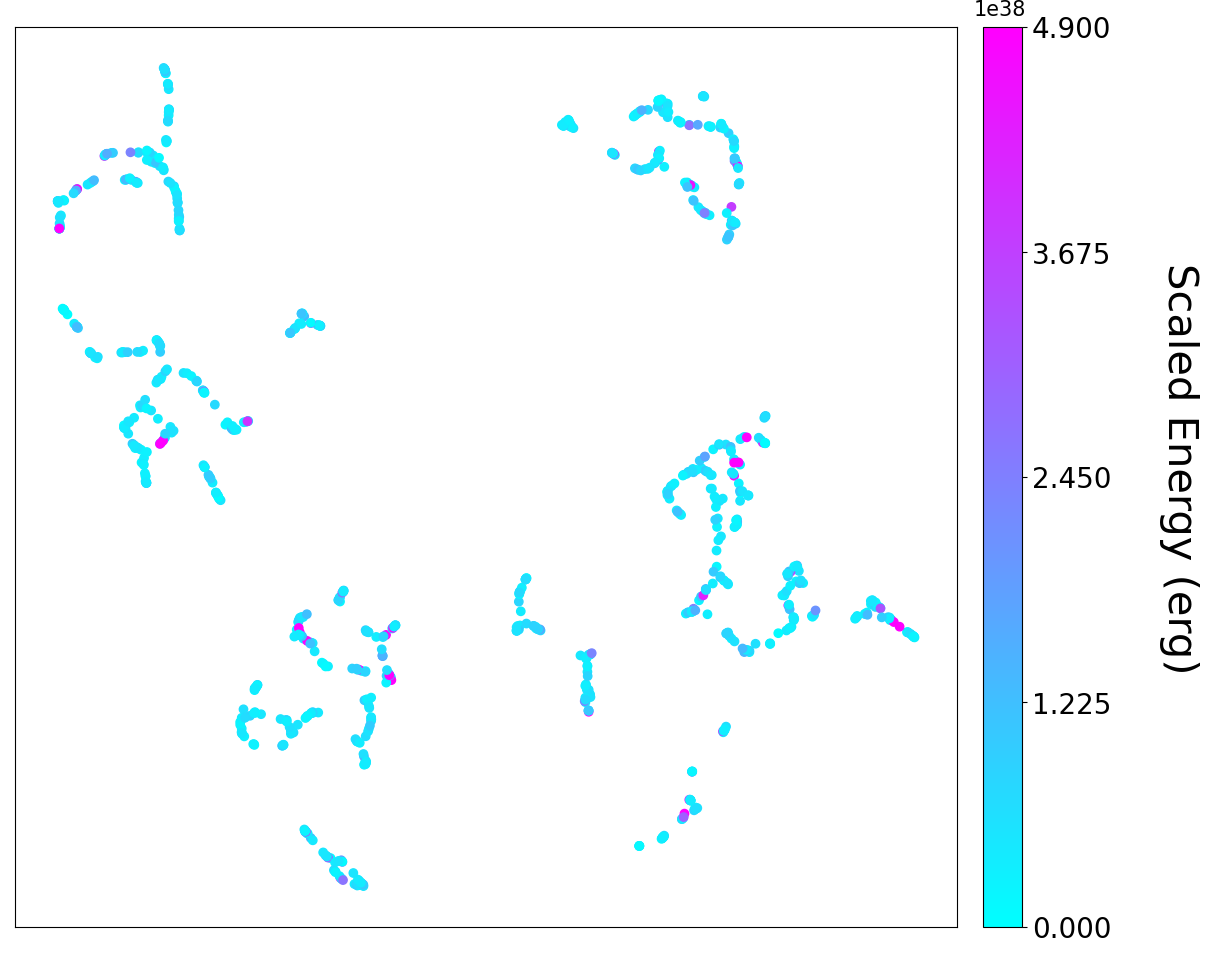}}\par 
                    \subcaptionbox{\label{s6}}{\includegraphics[width=1\linewidth, height=0.8\columnwidth]{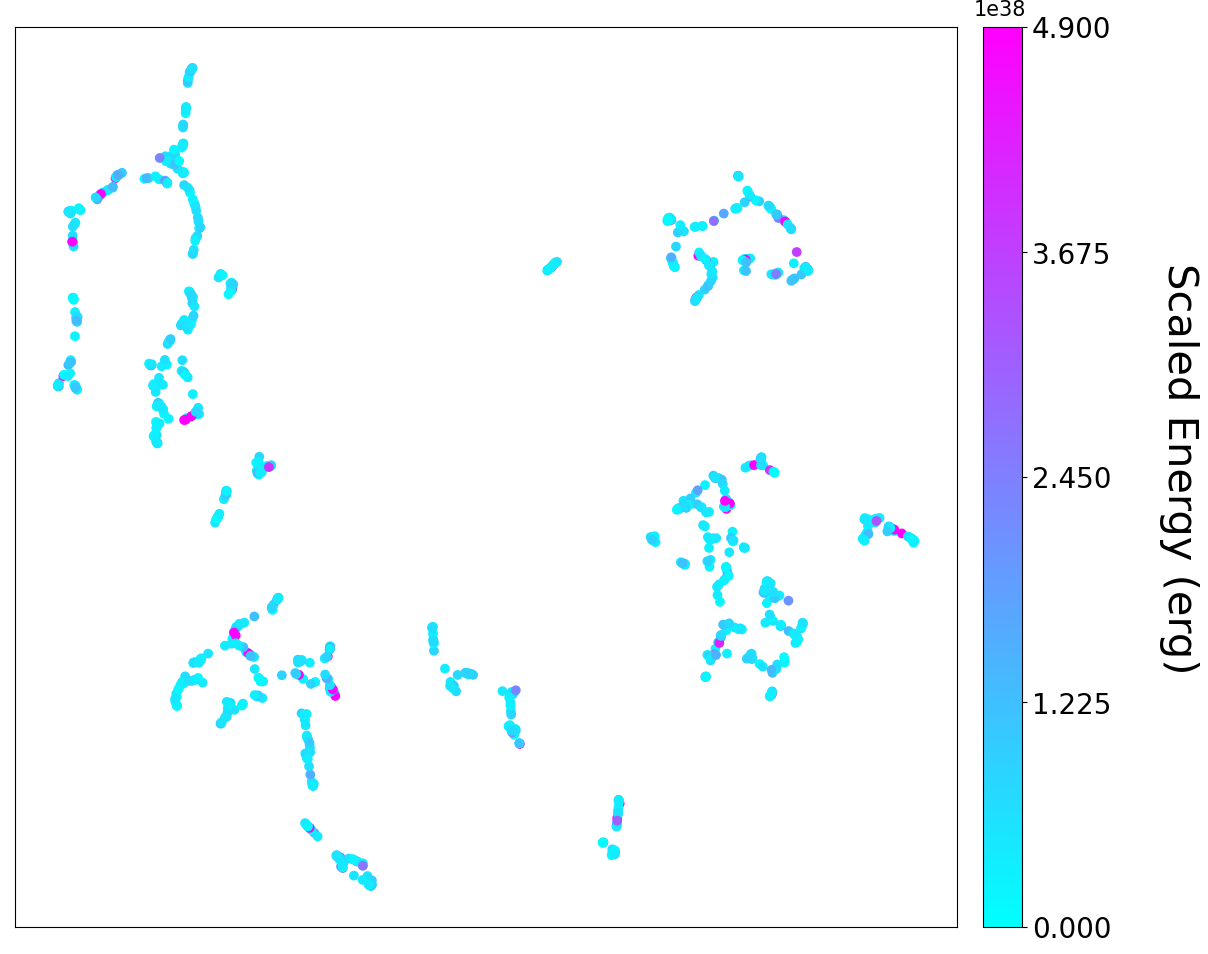}}\par 
                \end{multicols}
                \begin{multicols}{2}
                    \subcaptionbox{\label{s7}}{\includegraphics[width=1\linewidth, height=0.8\columnwidth]{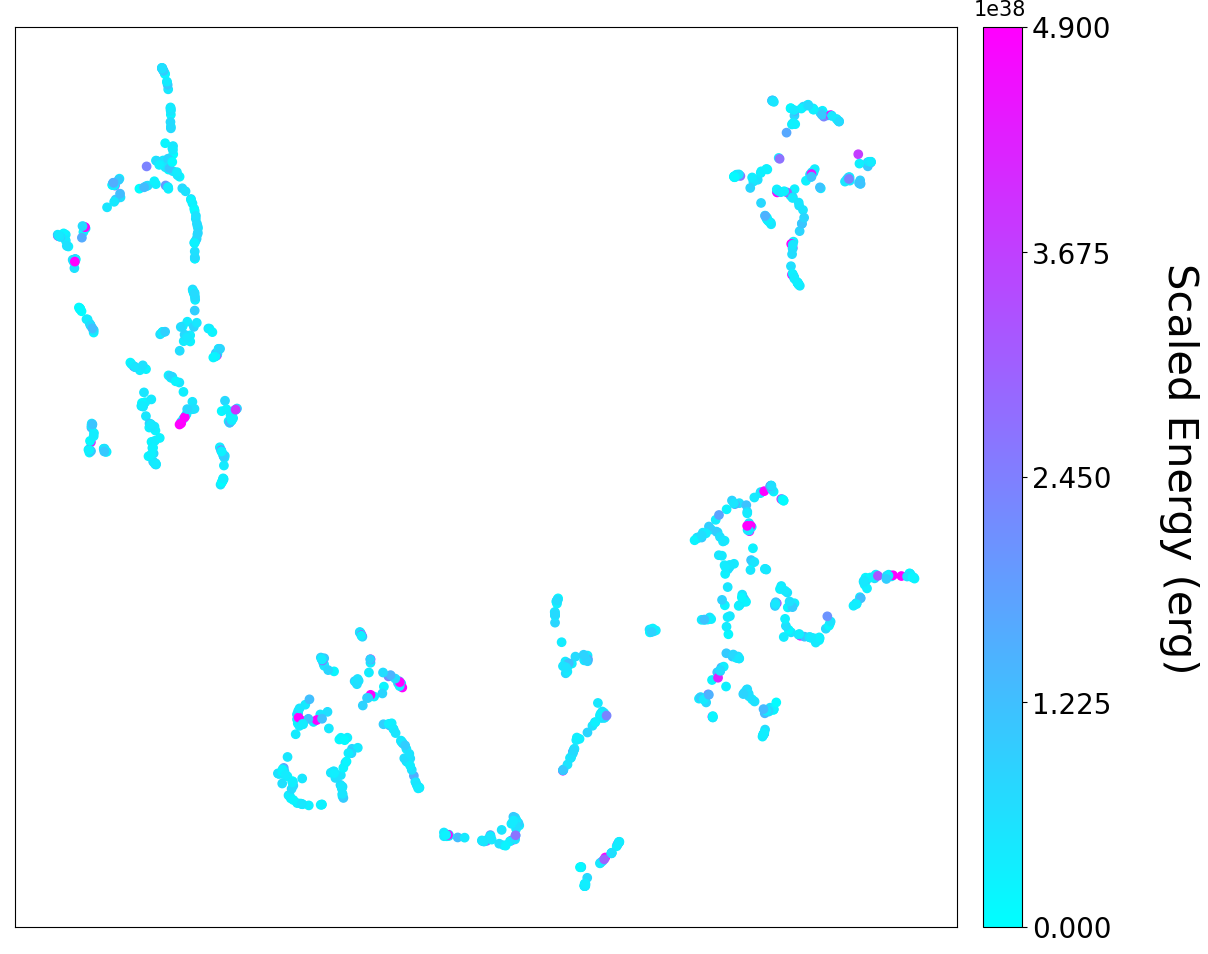}}\par
                    \subcaptionbox{\label{s9}}{\includegraphics[width=1\linewidth, height=0.8\columnwidth]{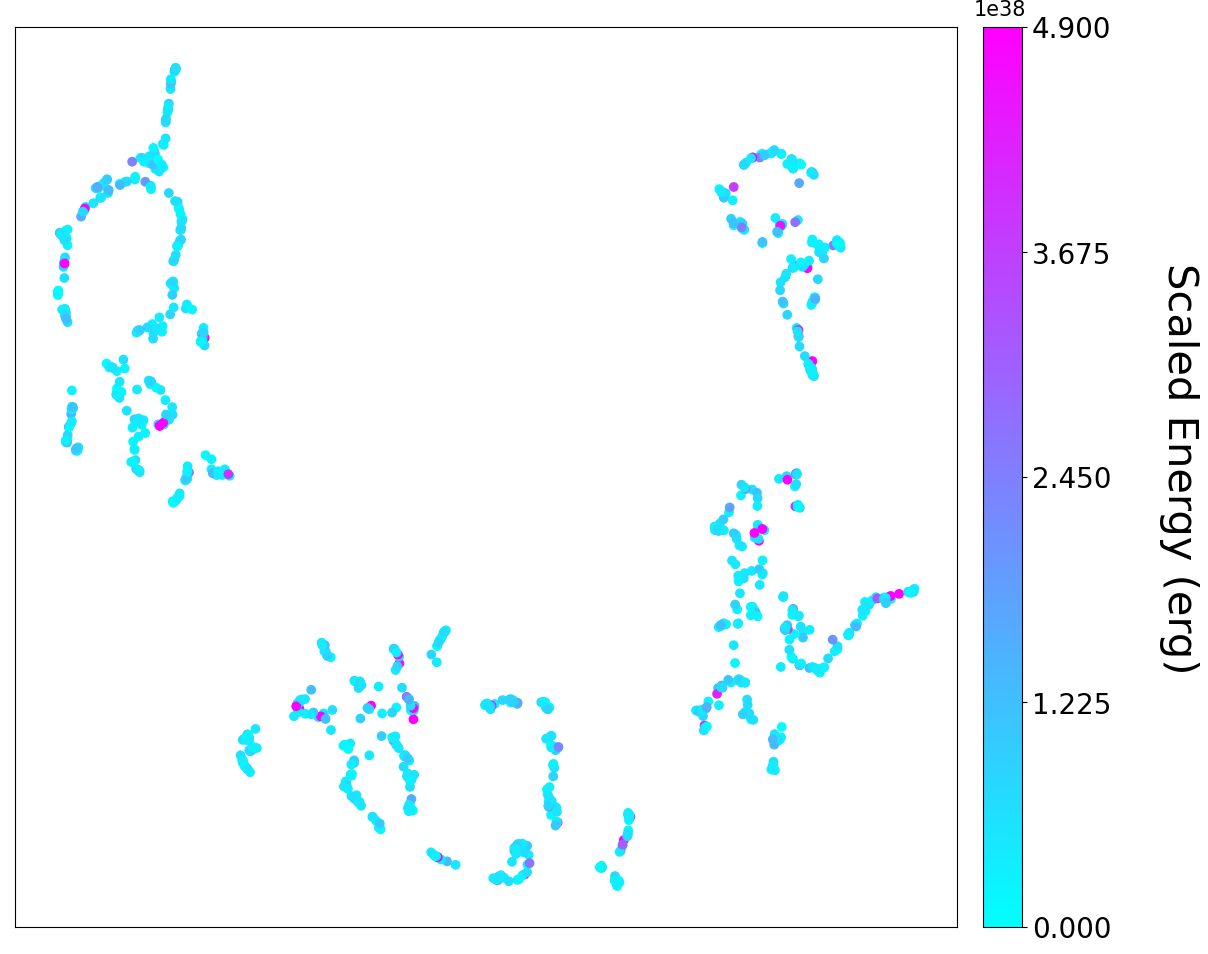}}\par
                \end{multicols}
                \caption{Scaled Energy colouring of the clustering results for Fig. \ref{s5} {\ttfamily n\_neighbors = 5}, Fig. \ref{s6} {\ttfamily n\_neighbors = 6}, Fig. \ref{s7} {\ttfamily n\_neighbors = 7}, and Fig. \ref{s9} {\ttfamily n\_neighbors = 9}. 
                }
            \label{A_8}
            \end{figure*}

            \begin{figure*}
                \centering
                \begin{multicols}{2}
                    \subcaptionbox{\label{t5}}{\includegraphics[width=1\linewidth, height=0.8\columnwidth]{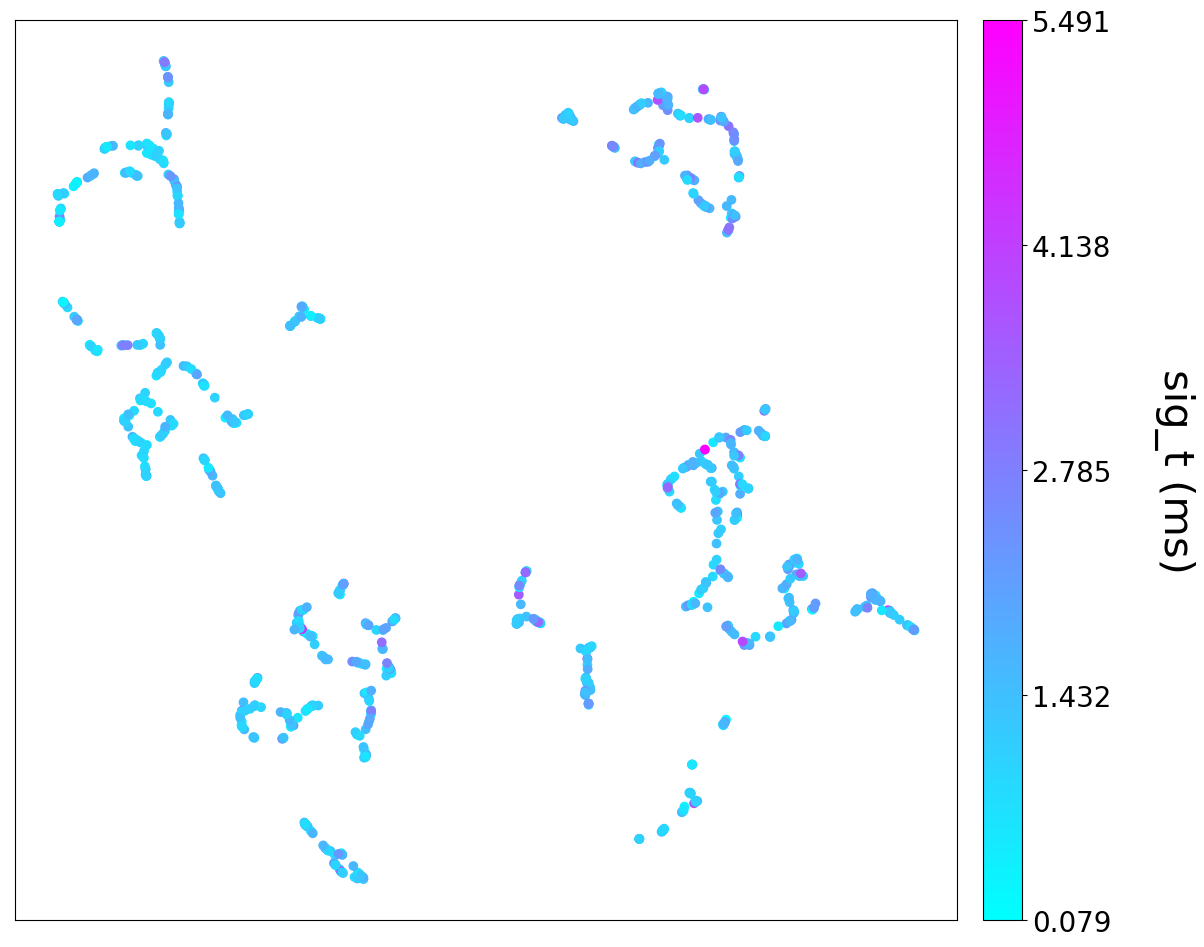}}\par 
                    \subcaptionbox{\label{t6}}{\includegraphics[width=1\linewidth, height=0.8\columnwidth]{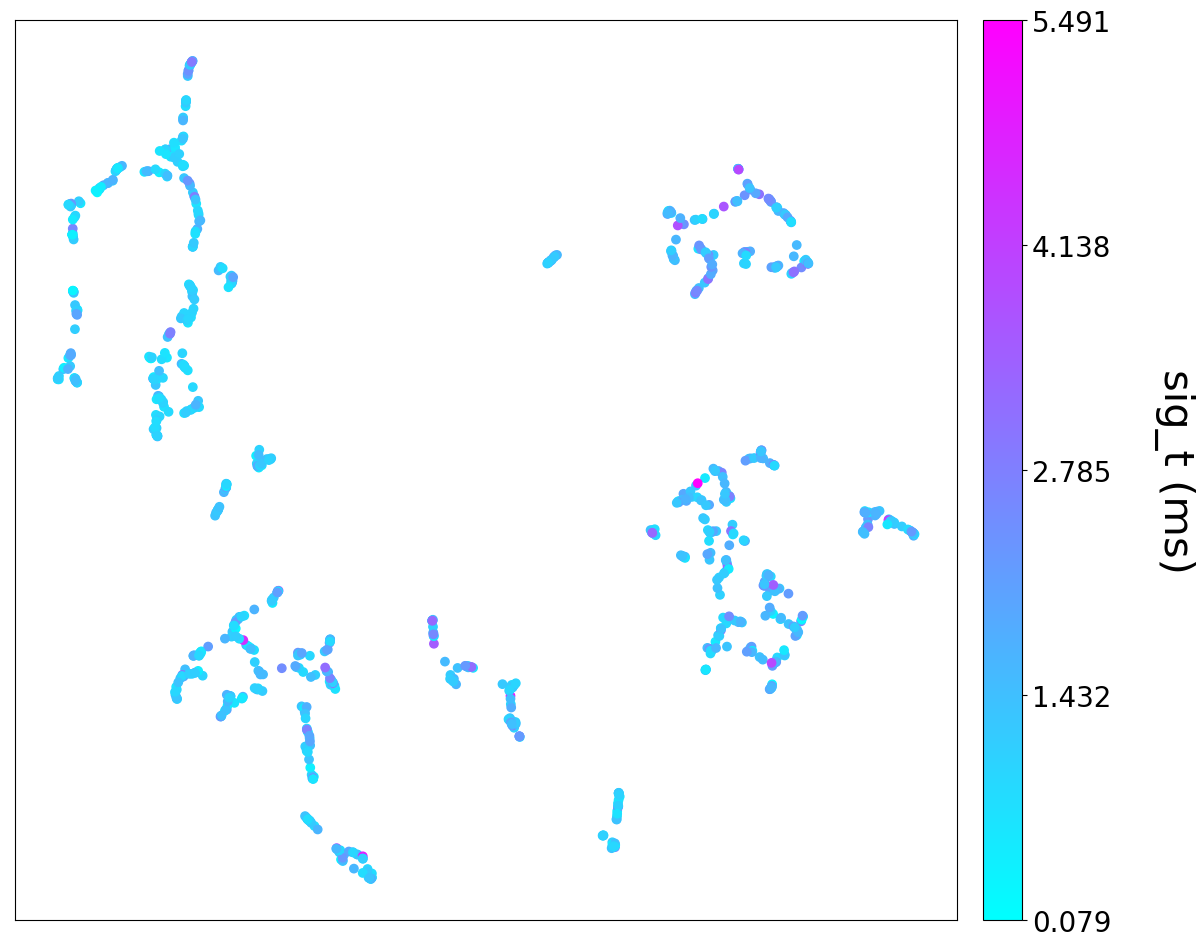}}\par 
                \end{multicols}
                \begin{multicols}{2}
                    \subcaptionbox{\label{t7}}{\includegraphics[width=1\linewidth, height=0.8\columnwidth]{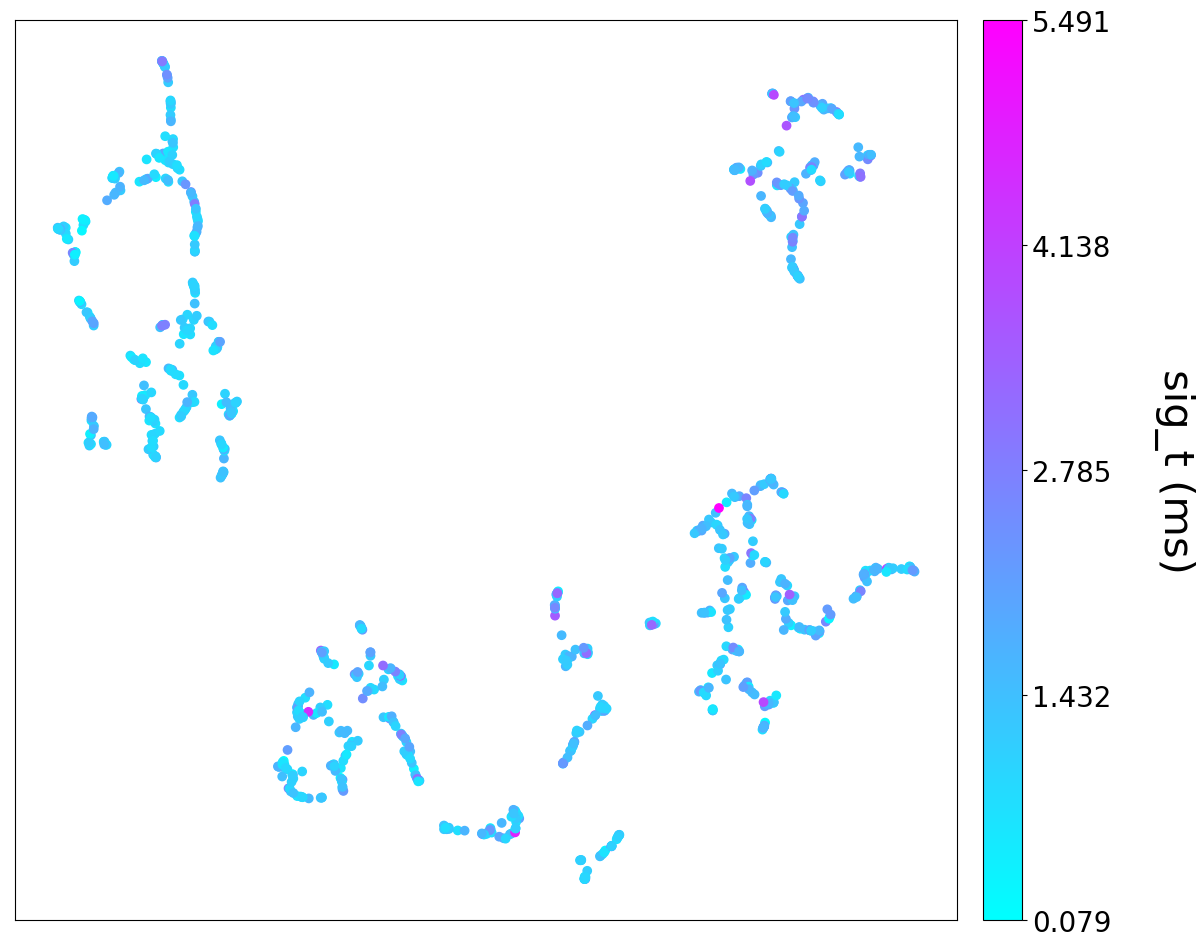}}\par
                    \subcaptionbox{\label{t9}}{\includegraphics[width=1\linewidth, height=0.8\columnwidth]{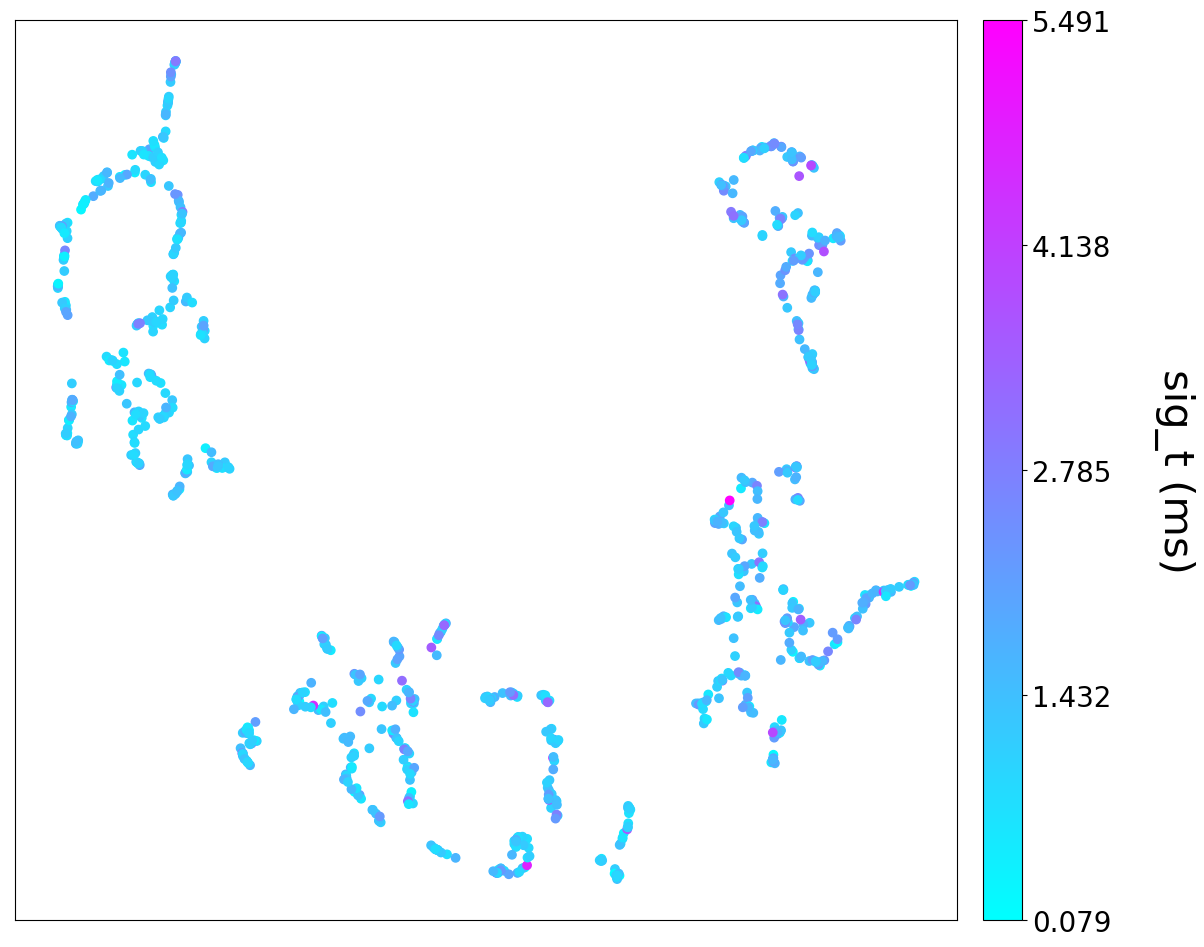}}\par
                \end{multicols}
                \caption{Time Duration colouring of the clustering results for Fig. \ref{t5} {\ttfamily n\_neighbors = 5}, Fig. \ref{t6} {\ttfamily n\_neighbors = 6}, Fig. \ref{t7} {\ttfamily n\_neighbors = 7}, and Fig. \ref{t9} {\ttfamily n\_neighbors = 9}. 
                }
            \label{A_9}
            \end{figure*}
         \begin{figure*}
                \centering
                \begin{multicols}{2}
                    \subcaptionbox{\label{a5_his}}{\includegraphics[width=1\linewidth, height=0.8\columnwidth]{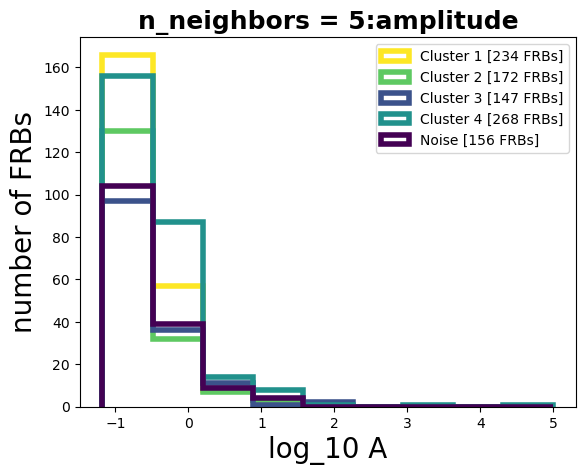}}\par 
                    \subcaptionbox{\label{a6_his}}{\includegraphics[width=1\linewidth, height=0.8\columnwidth]{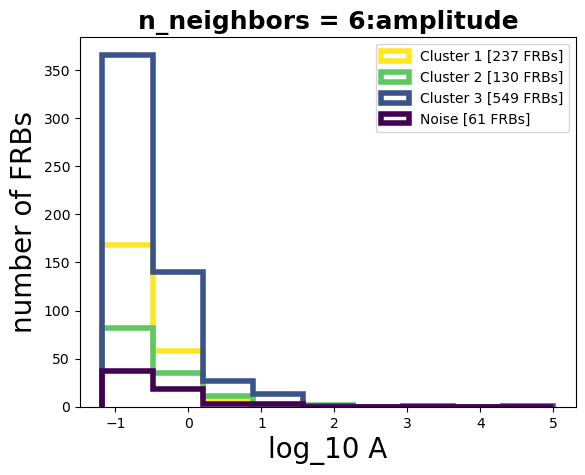}}\par 
                \end{multicols}
                \begin{multicols}{2}
                    \subcaptionbox{\label{a7_his}}{\includegraphics[width=1\linewidth, height=0.8\columnwidth]{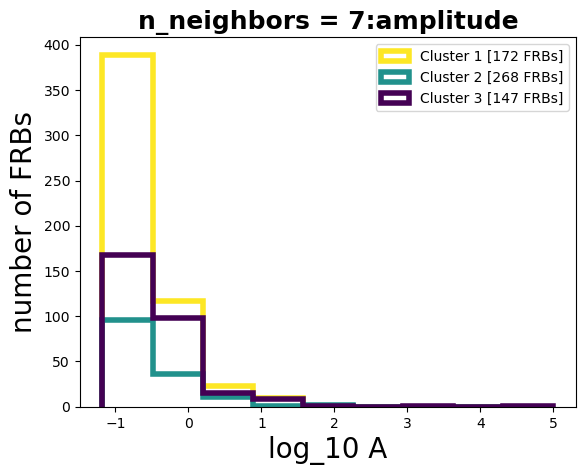}}\par
                    \subcaptionbox{\label{a9_z_his}}{\includegraphics[width=1\linewidth, height=0.8\columnwidth]{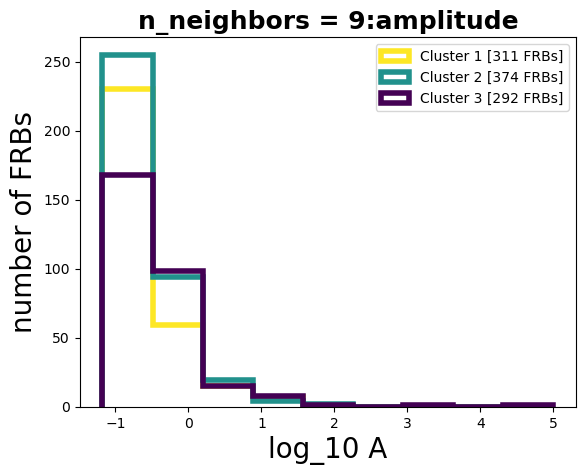}}\par
                \end{multicols}   
                \caption{Histograms for Amplitude with different {\ttfamily n\_neighbors}. }
            \label{his_a}
            \end{figure*}

         \begin{figure*}
                \centering
                \begin{multicols}{2}
                    \subcaptionbox{\label{b5_his}}{\includegraphics[width=1\linewidth, height=0.8\columnwidth]{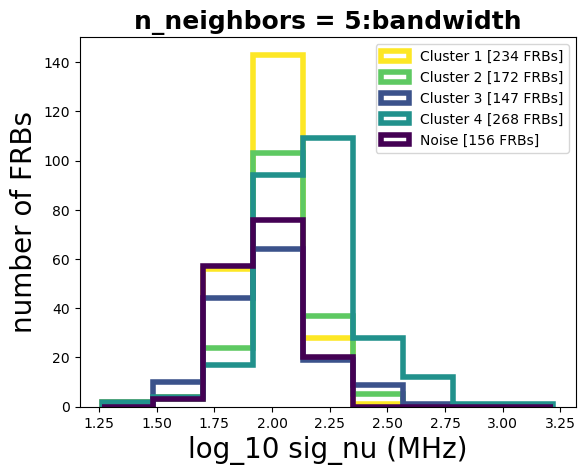}}\par 
                    \subcaptionbox{\label{b6_his}}{\includegraphics[width=1\linewidth, height=0.8\columnwidth]{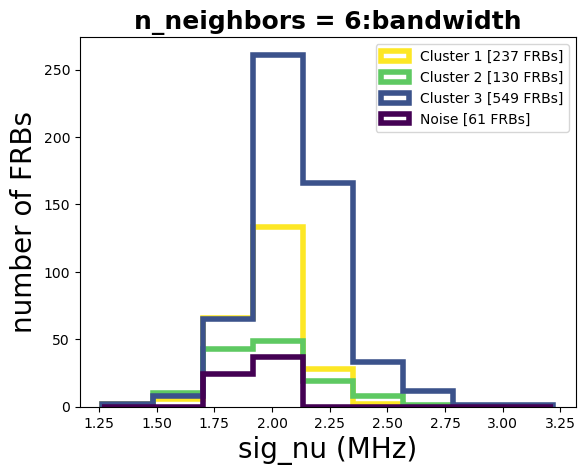}}\par 
                \end{multicols}
                \begin{multicols}{2}
                    \subcaptionbox{\label{b7_his}}{\includegraphics[width=1\linewidth, height=0.8\columnwidth]{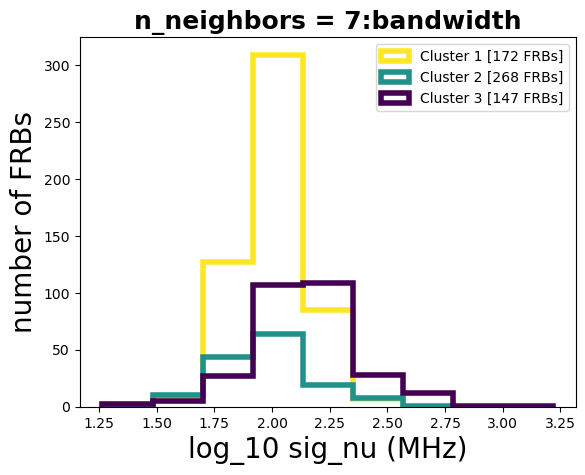}}\par
                    \subcaptionbox{\label{b9_his}}{\includegraphics[width=1\linewidth, height=0.8\columnwidth]{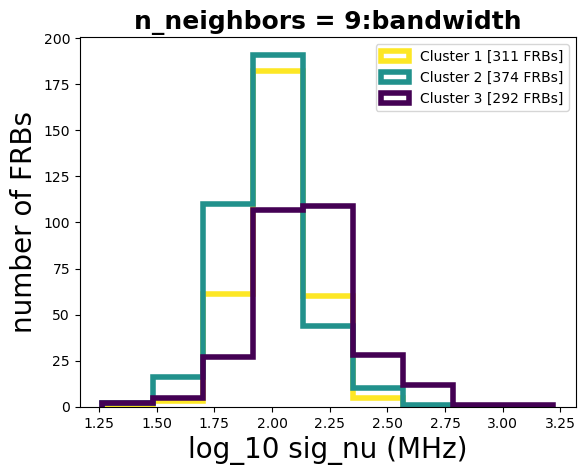}}\par
                \end{multicols}   
                \caption{Histograms for Bandwidth with different {\ttfamily n\_neighbors}.}
            \label{his_b}
            \end{figure*}

            \begin{figure*}
            \centering
            \begin{multicols}{2}
                    \subcaptionbox{\label{c5_his}}{\includegraphics[width=1\linewidth, height=0.8\columnwidth]{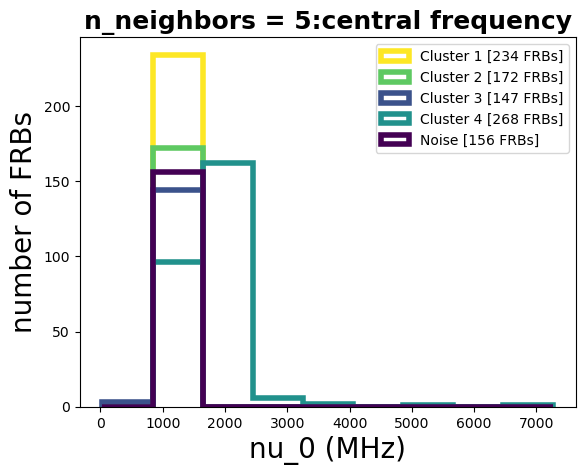}}\par 
                    \subcaptionbox{\label{central_freqency_1}}{\includegraphics[width=1\linewidth, height=0.8\columnwidth]{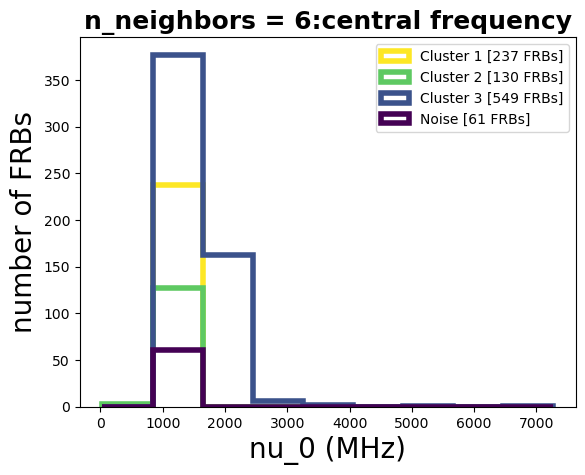}}\par 
            \end{multicols}
            \begin{multicols}{2}
                    \subcaptionbox{\label{c7_his}}{\includegraphics[width=1\linewidth, height=0.8\columnwidth]{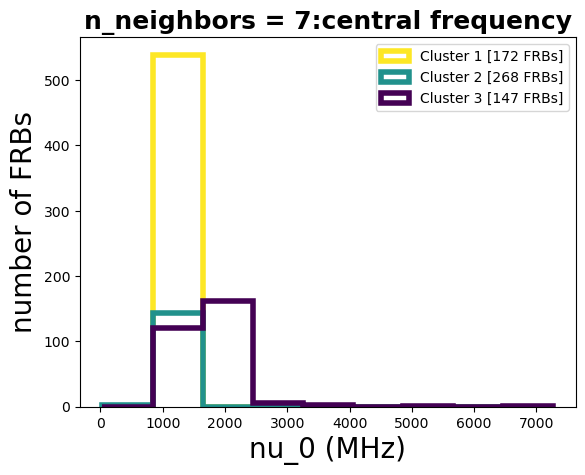}}\par
                    \subcaptionbox{\label{c9_his}}{\includegraphics[width=1\linewidth, height=0.8\columnwidth]{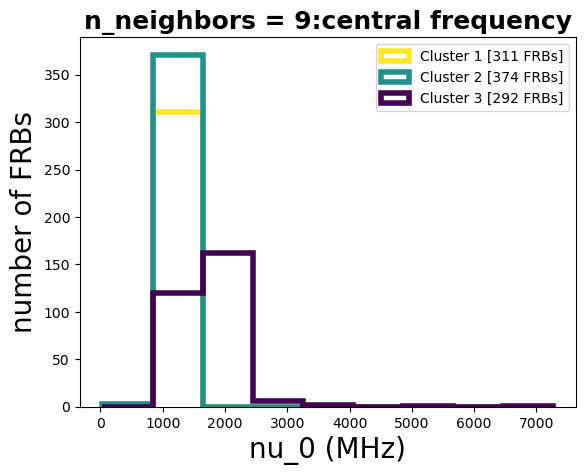}}\par
            \end{multicols}   
            \caption{Histograms for Central frequency with different {\ttfamily n\_neighbors}. }
            \label{his_c}
            \end{figure*}
            \begin{figure*}
            \centering
            \begin{multicols}{2}
                    \subcaptionbox{\label{d5_his}}{\includegraphics[width=1\linewidth, height=0.8\columnwidth]{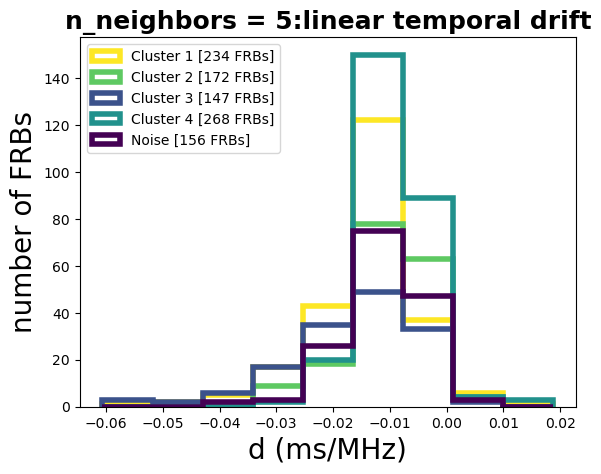}}\par 
                    \subcaptionbox{\label{d6_his}}{\includegraphics[width=1\linewidth, height=0.8\columnwidth]{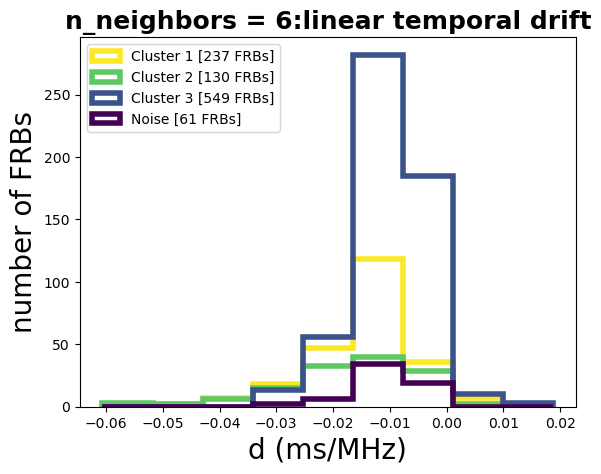}}\par 
            \end{multicols}
            \begin{multicols}{2}
                    \subcaptionbox{\label{d7_his}}{\includegraphics[width=1\linewidth, height=0.8\columnwidth]{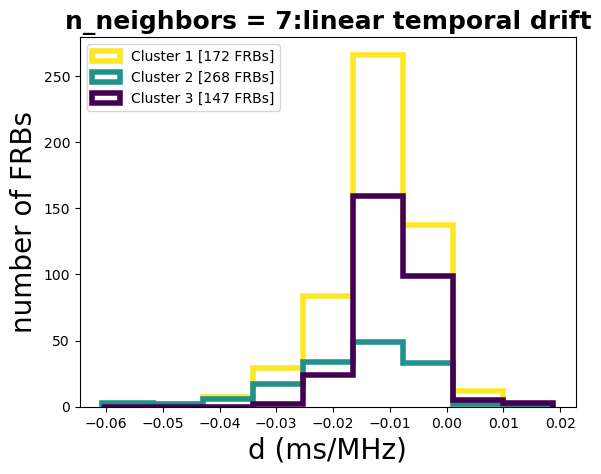}}\par
                    \subcaptionbox{\label{d9_his}}{\includegraphics[width=1\linewidth, height=0.8\columnwidth]{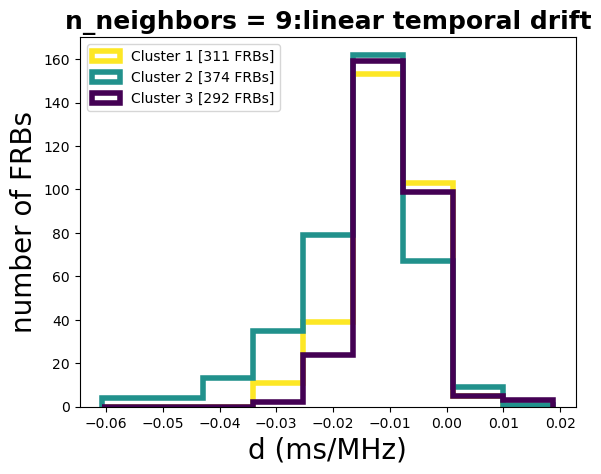}}\par
            \end{multicols}   
            \caption{Histograms for Linear temporal drift with different {\ttfamily n\_neighbors}. 
            }
            \label{his_d}
            \end{figure*}
            \begin{figure*}
            \centering
            \begin{multicols}{2}
                    \subcaptionbox{\label{f5_his}}{\includegraphics[width=1\linewidth, height=0.8\columnwidth]{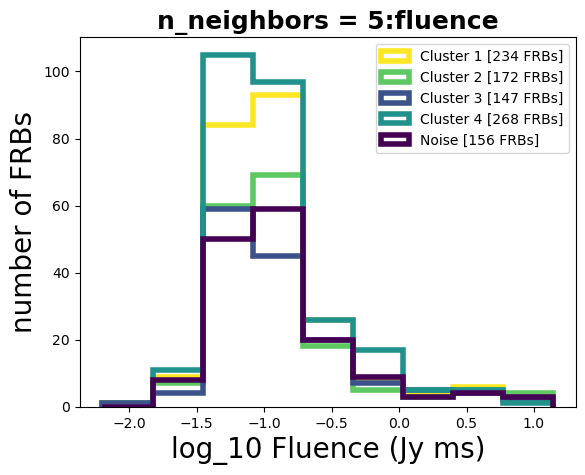}}\par 
                    \subcaptionbox{\label{c6_his}}{\includegraphics[width=1\linewidth, height=0.8\columnwidth]{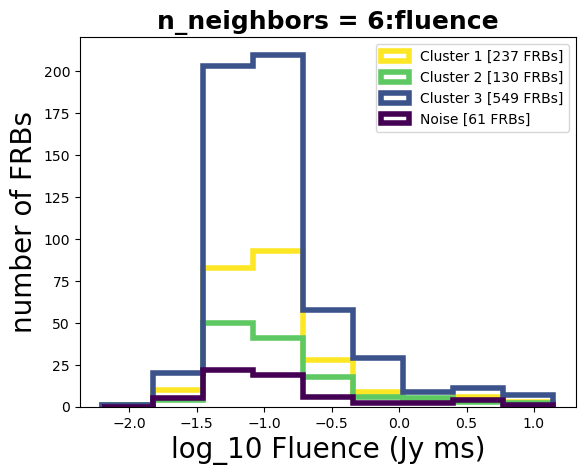}}\par 
            \end{multicols}
            \begin{multicols}{2}
                    \subcaptionbox{\label{f7_his}}{\includegraphics[width=1\linewidth, height=0.8\columnwidth]{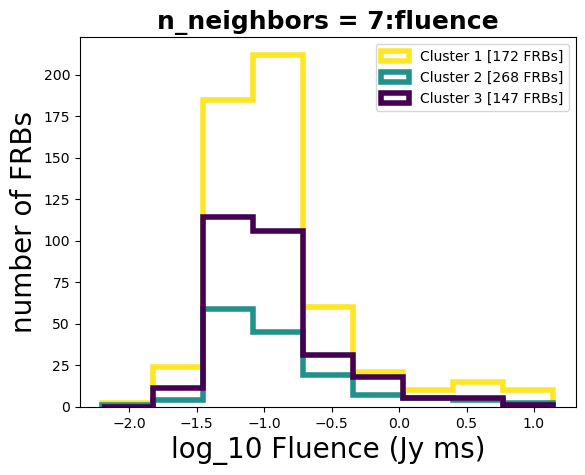}}\par
                    \subcaptionbox{\label{f9_his}}{\includegraphics[width=1\linewidth, height=0.8\columnwidth]{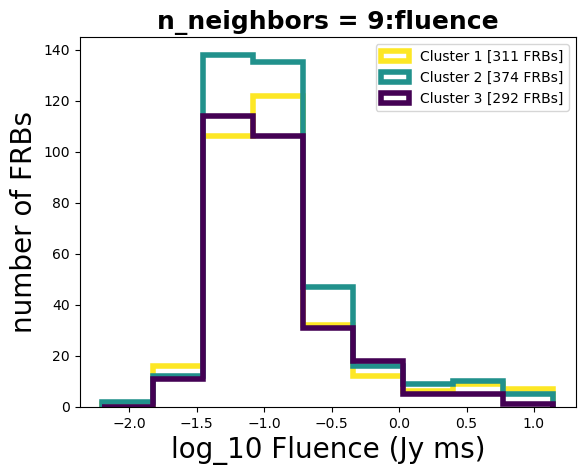}}\par
            \end{multicols}   
            \caption{Histograms for Fluence with different {\ttfamily n\_neighbors}.}
            \label{his_f}
            \end{figure*}
            \begin{figure*}
            \centering
            \begin{multicols}{2}
                    \subcaptionbox{\label{s5_his}}{\includegraphics[width=1\linewidth, height=0.8\columnwidth]{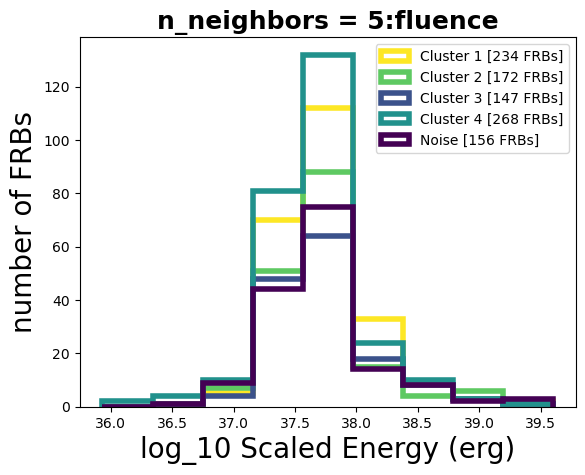}}\par 
                    \subcaptionbox{\label{s6_his}}{\includegraphics[width=1\linewidth, height=0.8\columnwidth]{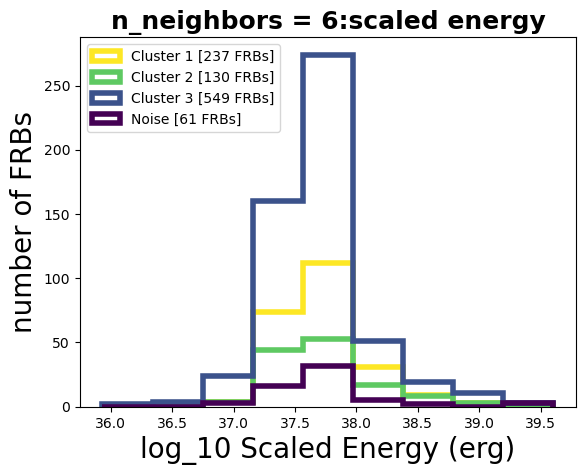}}\par 
            \end{multicols}
            \begin{multicols}{2}
                    \subcaptionbox{\label{s7_his}}{\includegraphics[width=1\linewidth, height=0.8\columnwidth]{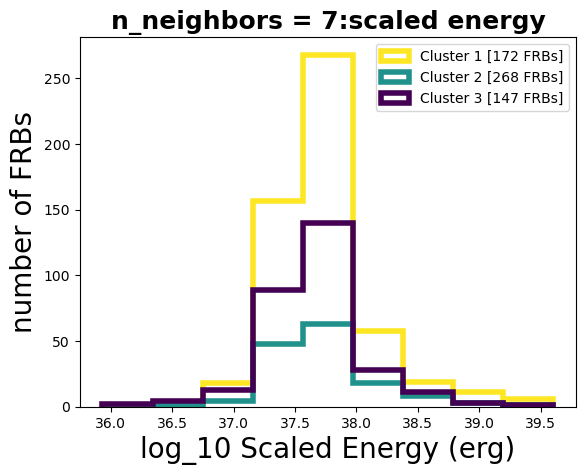}}\par
                    \subcaptionbox{\label{s9_his}}{\includegraphics[width=1\linewidth, height=0.8\columnwidth]{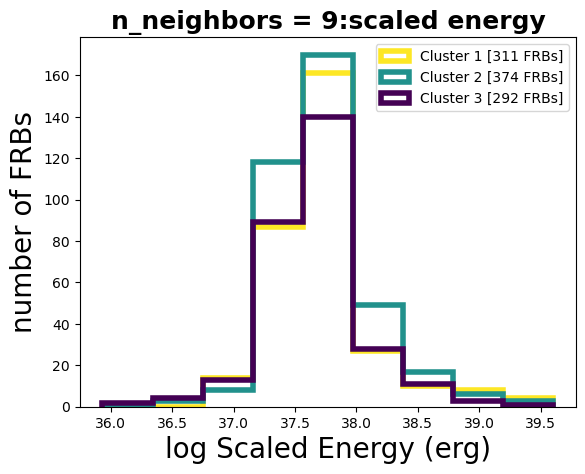}}\par
            \end{multicols}   
            \caption{Histograms for Scaled energy with different {\ttfamily n\_neighbors}. }
            \label{his_s}
            \end{figure*}
            \begin{figure*}
            \centering
            \begin{multicols}{2}
                    \subcaptionbox{\label{t5_his}}{\includegraphics[width=1\linewidth, height=0.8\columnwidth]{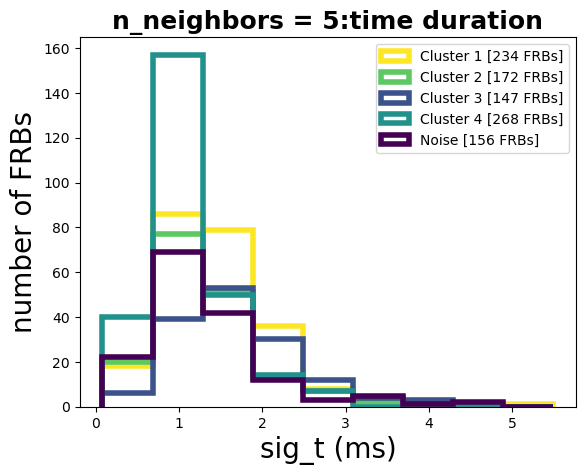}}\par 
                    \subcaptionbox{\label{t6_his}}{\includegraphics[width=1\linewidth, height=0.8\columnwidth]{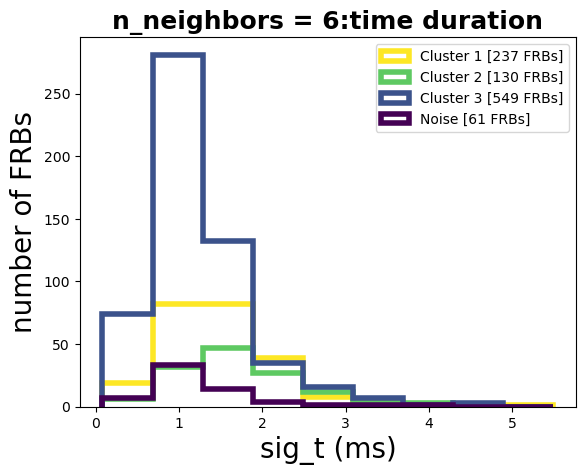}}\par 
            \end{multicols}
            \begin{multicols}{2}
                    \subcaptionbox{\label{t7_his}}{\includegraphics[width=1\linewidth, height=0.8\columnwidth]{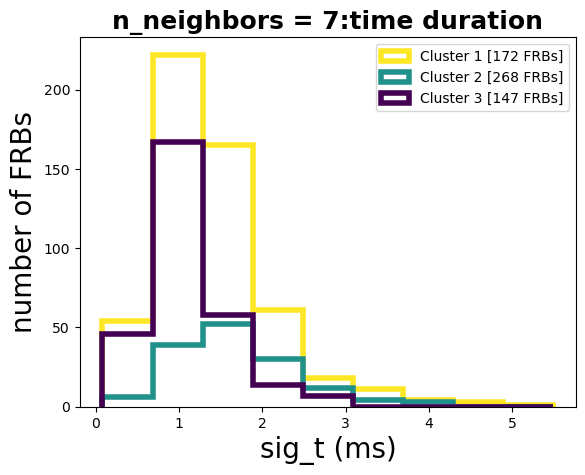}}\par
                    \subcaptionbox{\label{t9_his}}{\includegraphics[width=1\linewidth, height=0.8\columnwidth]{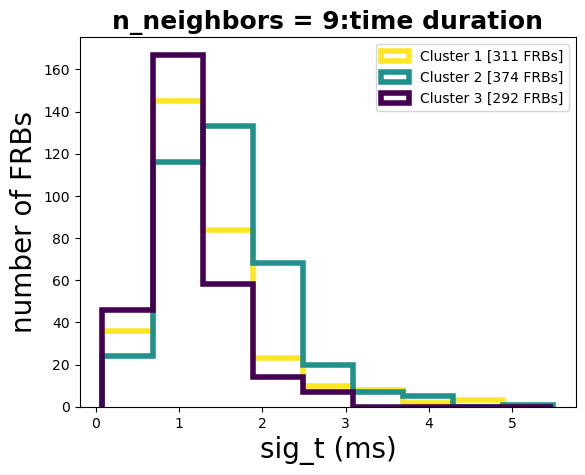}}\par
            \end{multicols}   
            \caption{Histograms for Time duration with different {\ttfamily n\_neighbors}. }
            \label{his_t}
            \end{figure*}

\end{document}